\documentclass[titlepage,12pt]{article}
\usepackage[T1]{fontenc}
\usepackage{times,latexsym,graphicx,wrapfig,url}
\usepackage{epsfig,lineno,bm}
\usepackage{epsfig,bm}

\usepackage{amsmath}

\usepackage{rotating}
\usepackage{amstext}

\usepackage[dvipsnames]{xcolor}

\newcommand{\etal}{{\em et al.}}

\newcommand{\brem}{bremsstrahlung} 
\newcommand{\breml}{\brem{}-like}

\newcommand{\pbo}  {PbWO$_4$}
\newcommand{\moll} {M{\o}ller}

\begin{document}
\widowpenalty=100000
\title{\vspace{-2.0in}
{\bf \Large{A Direct Detection Search for Hidden Sector New Particles in the 3 - 60 MeV Mass Range}
\vspace{-0.4in}
}}

\author{ \normalsize A. Ahmidouch, S.~Davis, A.~Gasparian (spokesperson, contact), \\
\normalsize T.~J.~Hague (co-spokesperson), S. Mtingwa\\ 
 {\em \normalsize North Carolina A\&T State University, Greensboro, NC 27411}\\*[0.05in]
 \normalsize C.~Ayerbe-Gayoso, H.~ Bhatt, B.~Devkota, J. Dunne,\\
 \normalsize D.~Dutta (co-spokesperson), L.~El Fassi, A. Karki, P. Mohanmurthy\\
 {\em \normalsize Mississippi State University, Mississippi State, MS 39762} \\*[0.05in] 
 \normalsize  C.~Peng (co-spokesperson)\\
 {\em \normalsize Argonne National Lab, Lemont, IL 60439} \\*[0.05in] 
 \normalsize  S. Ali, X. Bai, J. Boyd, B. Dharmasena, V. Gamage, S. Jeffas, S. Jian\\  
 \normalsize N. Liyanage (co-spokesperson), H. Nguyen, A. Rathnayake\\
 {\em \normalsize University of Virginia, Charlottesville, VA 22904} \\*[0.05in] 
 \normalsize  M. Khandaker\\
 {\em \normalsize Energy Systems, Davis, CA 95616} \\*[0.05in] 
 \normalsize D. Byer, H. Gao (co-spokesperson), C. Howell, B. Karki, V. Khachatryan, G. Matousek,\\ \normalsize E. van Nieuwenhuizen, A. Smith, B. Yu, Z. Zhao, J. Zhou \\
 {\em \normalsize Duke University, Durham, NC 27708}
 \\*[0.05in] 
 \normalsize A. Shahinyan \\
 {\em \normalsize Yerevan Physics Institute, Yerevan, Armenia}\\*[0.05in] 
 \normalsize K.~Gnanvo, D. Higinbotham, V. Kubarovsky,\\ 
 \normalsize  R. Paremuzyan (co-spokesperson), E. Pasyuk, S. Stepanyan, H. Avakian\\
 {\em \normalsize Thomas Jefferson National Accelerator Facility, Newport News, VA 23606} \\*[0.05in]
 \normalsize A. Bianconi, G. Costantini, G. Gosta, M. Leali, S. Migliorati, L. Venturelli\\
{\em \normalsize Dipartimento di Ingegneria dell'Informazione, Università di Brescia, Italy }\\
{\em \normalsize and Isituto Nazionale di Fisica Nucleare, sezione di Pavia, Italy}\\*[0.05in] 
\normalsize V. Mascagna \\
{\em \normalsize DiSAT, Università dell'Insubria, Como, Italy
and INFN, sezione di Pavia, Italy}\\*[0.05in] 
\normalsize M. De Napoli\\
{\em \normalsize Isituto Nazionale di Fisica Nucleare, Sezione di Catania, Italy}\\*[0.05in] 
\normalsize M. Battaglieri, R.~De~Vita \\
{\em \normalsize Isituto Nazionale di Fisica Nucleare, Sezione di Genova, Genova, Italy}\\*[0.05in]
\normalsize I. Larin, R. Miskimen \\
{\em \normalsize University of Massachusetts, Amherst, MA 01003}\\*[0.05in]
\normalsize P. L. Cole \\
{\em \normalsize Lamar University, Beaumont, Texas 77710}\\*[0.05in]
 \normalsize  J. Rittenhouse~West \\
 {\em \normalsize Lawrence Berkeley National Laboratory, Berkeley, CA 94720}\vspace{-1.0in}
} 
\date{}
\maketitle
\section*{Abstract}
In our quest to understand the nature of dark matter and discover its non-gravitational interactions with ordinary matter, we propose an experiment using a \pbo ~calorimeter to search for or set new limits on the production rate of i) hidden sector particles in the $3 - 60$ MeV mass range via their $e^+e^-$ decay (or $\gamma\gamma$ decay with limited tracking), and ii) the hypothetical X17 particle, claimed in multiple recent experiments. The search for these particles is motivated by new hidden sector models and dark matter candidates introduced to account for a variety of experimental and observational puzzles: the small-scale structure puzzle in cosmological simulations, anomalies such as the 4.2$\sigma$ disagreement between experiments and the standard model prediction for the muon anomalous magnetic moment, and the excess of $e^+e^-$ pairs from the $^8$Be M1 and $^4$He nuclear transitions to their ground states observed by the ATOMKI group. In these models, the $1 - 100$ MeV mass range is particularly well-motivated and the lower part of this range still remains unexplored. Our proposed direct detection experiment will use a magnetic-spectrometer-free setup (the PRad apparatus) to detect all three final state particles in the visible decay of a hidden sector particle allowing for an effective control of the background and will cover the proposed mass range in a single setting. The use of the well-demonstrated PRad setup allows for an essentially ready-to-run and uniquely cost-effective search for hidden sector particles in the $3 - 60$ MeV mass range with a sensitivity of 8.9$\times$10$^{-8}$ - 5.8$\times$10$^{-9}$  to $\epsilon^2$, the square of the kinetic mixing interaction constant between hidden and visible sectors.\\
This updated proposal includes our response to the PAC49 comments and a new simulation with improved generators that was used to estimate the projected sensitivity of the experiment.
\newpage

\section{Executive Summary}
We request 60 PAC days to perform a direct detection search for hidden sector particles in the $3 - 60$ MeV mass range using the magnetic-spectrometer-free PRad setup in Hall-B. This experiment will exploit the well-demonstrated PRad setup to perform a ready-to-run and cost-effective search. This proposed search experiment is timely given the recent agreement between measurements of the muon $(g-2)$ anomaly at BNL and FNAL as well as the $17$ MeV particle proposed to account for the excess of $e^+e^-$ pairs found in a nuclear transition in $^8$Be from one of its $1^+$ resonance to its ground state and the electromagnetically forbidden M0 transition in $^4$He. In particular, the $3 - 60$ MeV mass range remains relatively unexplored amplifying the urgency. 

The experiment will use 2.2 GeV and 3.3 GeV CW electron beams, with a current of $50 - 100$~nA, on a $1~\mu$m Ta foil placed in front of the PRad setup. All three final state particles - the scattered electron and the $e^{+}e^{-}$ from the visible decay of the hidden sector particle - will be detected in the \pbo ~part of the HyCal calorimeter. A pair of GEM chambers will be used to suppress the neutral background and events not originating from the target. This technique will help effectively suppress the background from the Bethe-Heitler process and provide a sensitivity of 8.9$\times$10$^{-8}$ - 5.8$\times$10$^{-9}$ to $\epsilon^2$, the square of the kinetic mixing interaction coupling constant. The $\epsilon^2 -  m_{X}$ parameter space covered by this proposed experiment as shown below will help fill the void left by current, ongoing and other planned searches, thereby helping to validate or place limits on hidden sector models. A new simulation with improved generators is used to estimate the projected sensitivity for this proposal that was conditionally approved by PAC49.
\begin{figure}[hbt!]
  \centerline{\includegraphics[width=0.6\textwidth]{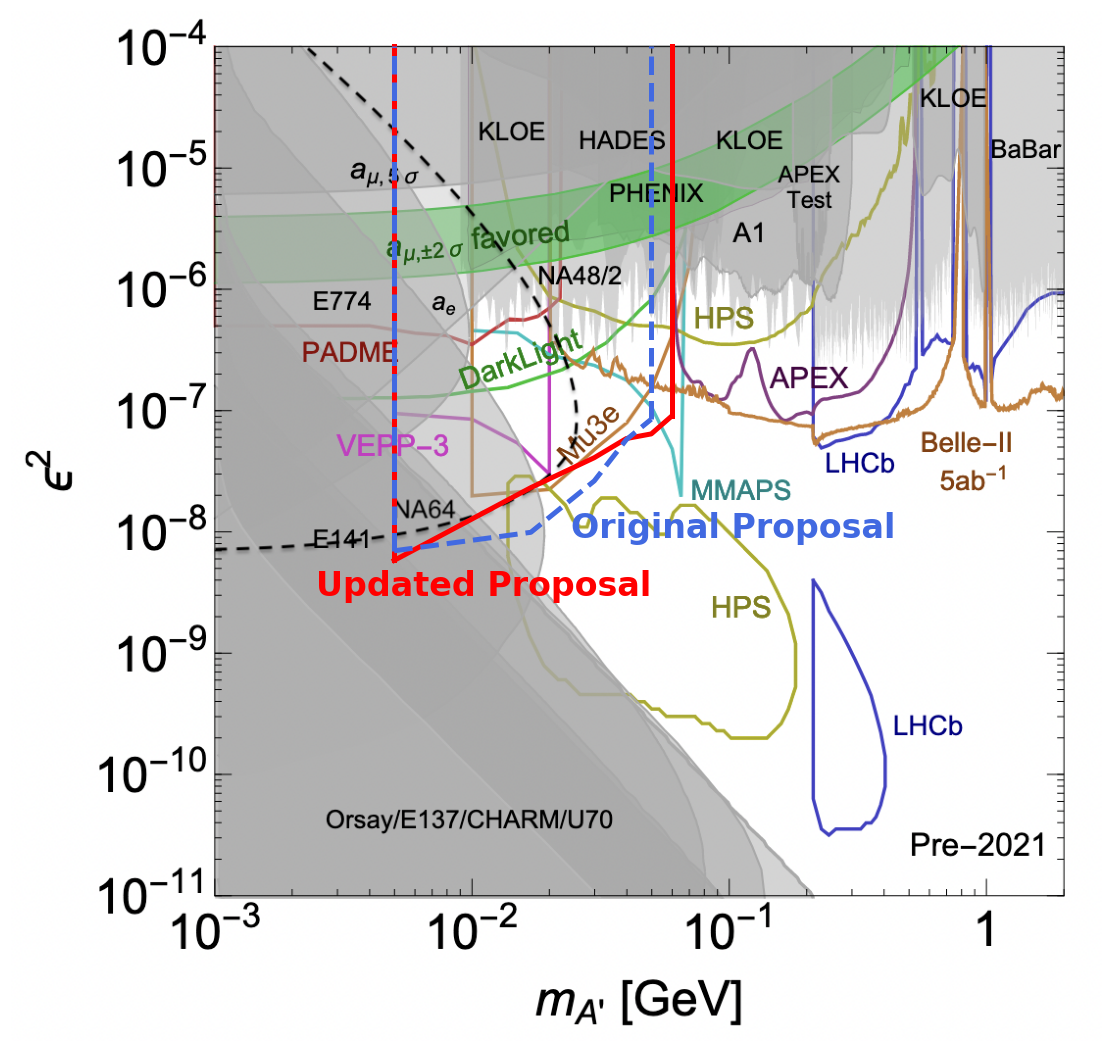}}
  \caption{Projected coverage of the $\epsilon^2 - m_{X}$ parameter space by this proposal is shown by the thick red lines for the combined statistics of the two beam energies. Adapted from Ref.~\cite{Alexander:2016hsr}.}
  \label{fig:proj_ps2}
\end{figure}

\newpage

\section*{Summary of Response to PAC49 Comments}
\subsection*{Overview}
The experiment of ``A Direct Detection Search for Hidden Sector New Particles in the 3 - 60 MeV Mass Range" (C12-21-003) was conditionally approved by PAC49  with a C2 status. This update addresses the issues raised in the PAC49 report which is followed by the updated full proposal.
\subsection*{Response}
{\bf (1)} A comprehensive simulation of the experiment was carried out using the Geant3 and Geant4 simulation packages developed for the PRad experiment which included the detector geometry and all materials of the experimental setup.  The simulation included light propagation and digitization of the simulated events. We consider the Geant4 simulation to be a more complete simulation of the experiment, with the Geant3 simulation serving as a crosscheck. The reconstructed invariant mass spectrum for backgrounds from the Geant3 and Geant4 simulations are compared and found to be consistent with each other (see Section 6).\\

An alternate simulation was performed using the MadGraph5 event generator that was used by the HPS collaboration. The background events generated by MadGraph5 were fed into the two Geant simulation packages that were used to trace them through the experimental setup
and detection of events.  The invariant mass spectrum of the resulting simulated background events for the Geant3 and Geant4 simulations were compared and found to be consistent with each other. However, the Geant and Madgraph5 background generators seem to disagree in the mass range of interest. It is expected that Madgraph5 is the better generator for the Bethe-Heitler radiative trident background. Therefore, we have created a hybrid generator by combining the Madgraph5  and the Geant4 (excluding Bethe-Heitler) generators to utilize the best features of the two generators. The shape of the background was also found to be similar to the background using similar event selection of PRad carbon and hydrogen data (see Appendix A and B). The results from the hybrid simulation was used to estimate the epsilon {it vs.} mass sensitivity.\\

\noindent
{\bf (2)} This experiment will also detect $\gamma \gamma$ events, albeit, with significantly lower invariant mass resolution. The optimization of the setup is somewhat orthogonal for the $e^{+}e^-$ and $\gamma \gamma$ channels and are therefore not discussed is this proposal.  However, the $\gamma \gamma$ channel is a relevant visible decay channel for hidden sector particles and it will also be sensitive to additional possibilities such as the proposed QED mesons~\cite{qed_meson1, qed_meson2}. The $\gamma \gamma$ data will certainly be analyzed. If a signal were to be observed, a new experiment with improved ability to detect  $\gamma \gamma$ events could be designed. For example the tracking detectors would need a thin converter added in front of them. This would improve the $\gamma \gamma$ invariant mass resolution at the cost of the $e^{+}e^-$ invariant mass resolution.\\

\noindent
{\bf(3)} The proposal has been carefully revised to ensure that it always clearly states that the experiment is a search for a hidden sector particle in the 3-60 MeV range and not a dark matter search.\\

\newpage
\section{Introduction}
The remarkable fact that $\sim$~85\% of the matter in the Universe is of unknown origin - termed dark matter (DM) - is inferred from astronomical measurements over a wide range of distance and time scales, from the Milky Way to the largest cosmological structures and as early as Big Bang Nucleosynthesis to today. Yet all of the evidence thus far is gravitational, providing no direct information about the identity of dark matter. Consequently, the investigation into the nature of DM, from its origin to its composition and how it interacts with other forces both old and new, is one of the grand challenges in fundamental science. There are many candidate theories for dark matter and dark sector mediators that span an enormous mass range, from $10^{-22}$ eV to at least 100 times the mass of the sun. 
Several recent observations and anomalies have required new dark matter interactions and candidates, such as hidden sector dark matter (HSDM)~\cite{Alexander:2016hsr} models, that point to the $1 - 100$ MeV/c$^2$ mass range as well-motivated for high priority searches~\cite{Battaglieri:2017aum}. 
In this proposal, we describe an experiment that will search in the 3 - 60 MeV mass region for heretofore unobserved hidden sector particles. 
This experiment will utilize the bremsstrahlung-like production of a hidden sector boson that subsequently decays to a $e^+e^-$ or $\gamma\gamma$ pair (i.e. visible decays) which will be detected in the PRad setup with minimal modifications.

The availability of a high duty factor, high luminosity electron beam at Jefferson Lab provides an ideal setup to search for MeV-scale dark mediators with small coupling constants. 
The well tested PRad setup in Hall B will be used in this experiment to reach our physics goals.
Using a magnetic-spectrometer-free setup allows the experiment to be sensitive to the full mass range in a single experimental setting, thus eliminating systematic uncertainties associated with field mapping or moving the spectrometer. The detection of all three final state particles in the \pbo calorimeter along with tracking with GEM chambers allows for an effective control of the backgrounds. Moreover, it provides an essentially ready-to-run and uniquely cost effective search for hidden-sector particles in the 3 - 60 MeV mass range.

\section{Physics Motivation}
\label{sec:motivation}
On large distance scales, the structure of the Universe as  inferred from cosmological data is consistent with dark matter particles that are cold, collision-less, and interact with ordinary matter purely via gravity~\cite{Tulin:2017sss}. Decades of cosmological data have converged on the standard model of cosmology, dubbed $\Lambda \rm CDM$, with cold dark matter (CDM) as a crucial ingredient ~\cite{Bachall:99cdm} along with dark energy (the cosmological constant of general relativity, $\Lambda$) and ordinary matter.  Weakly interacting massive particles (WIMPs) have been one of the primary DM candidates~\cite{jungman96}, in part due to the ``WIMP miracle'' in which weak-scale DM masses and couplings combine into cross-sections to give the correct abundance of DM to ordinary matter \cite{Feng:2010gw}. Consequently, the experimental search for the particle nature of DM focused on WIMPs. While WIMP dark matter remains well-motivated, significant parameter space for both WIMPs
and models that realize WIMP dark matter have been explored and ruled out by recent searches~\cite{Battaglieri:2017aum}. To date, the strongest bound on the WIMP-nucleon spin-independent elastic cross-section was made by the XENON1T Experiment~\cite{Aprile:2018x1t}, which excludes cross-section values down to 4$\times$10~$^{-47}$~cm$^2$. Clearly, new models and 
candidates are needed. Meanwhile, there are several astronomical observations and experimental anomalies that suggest the existence of a new MeV-scale force-carrier connecting dark matter to standard model (SM) particles. Spurred on by these developments, new theoretical frameworks have been developed that are natural generalizations of the WIMP DM idea but include interactions through new forces rather than just SM forces while retaining the ability to fit the observational constraints~\cite{Alexander:2016hsr}.  

In typical models, the new force carrier is a $\rm U(1)_X$ gauge boson, or dark photon, henceforth referred to as $X$.
The $X$ can couple to SM matter via electromagnetic interactions. In the simplest schemes these couplings arise from a kinetic mixing interaction, $\frac{\epsilon}{2}F^Y_{\mu\nu}F^{\prime\mu\nu}$, where $F^Y_{\mu\nu}$ is the SM hypercharge field strength, $F^{\prime\mu\nu} = \left[\partial_\mu,X\nu\right]$ is the dark gauge boson field, and $\epsilon$ is the dimensionless coupling constant of the $X$ to SM matter. The $X$ can acquire mass through the Higgs mechanism and a massive $X$ will induce a charge on SM particles proportional to $\epsilon$
~\cite{Holdom:1985ag,Bjorken:2009mm}. 
In these models the MeV to GeV-scale $X$ masses are found to be particularly well-motivated~\cite{Bjorken:2009mm}.

Recently, there has also been an increased interest in an $X$ that does not couple proportionally to electric charge.
Rather, the coupling is tied to flavor of quarks or leptons.
One such example is the protophobic X17 proposed in Ref.~\cite{Feng:2016jff}.
If an $X$ were to have flavor-dependent couplings the parameter space for the $X$ is more open than previously thought \cite{Feng:2022inv}.

Here we will discuss a few primary motivators for dark mediator searches in the sub-GeV mass range of the proposed experiment, and a more detailed discussion of dark sector searches can be found in Ref.~\cite{Battaglieri:2017aum}.

\begin{figure}[!htb]
    \centering
    \includegraphics[width=0.65\textwidth]{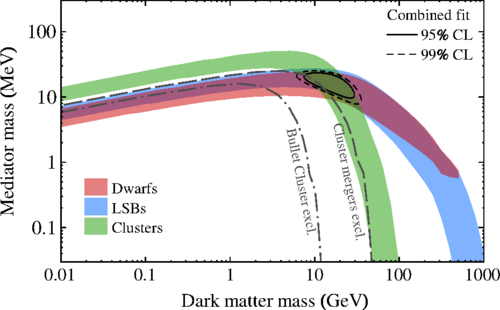}
    \caption{Parameter space for the dark photon model of self-interactions (with  $\alpha^{\prime}= \alpha_{EM}$), preferred by dwarfs (red), LSB spiral galaxies (blue), and clusters (green), each at 95\% C.L. The combined 95\% (99\%) region is shown by the solid (dashed) contours. The estimated Bullet Cluster excluded region lies below the dot-dashed curve and the ensemble merging cluster excluded region below the long-dashed curve. Reproduced from Ref.~\cite{Kaplinghat:2016sss}.}
    \label{fig:sss1}
\end{figure}
\subsection{Small Scale Structures in Astrophysical Observations}

Since the 1990s, a series of astrophysical observations at scales smaller than the virial radius of galaxies - collectively called small-scale structure - have posed a challenge to the results from simulations of traditional weakly interacting CDM models~\cite{Tulin:2017sss}. Dark matter models with significant self-interactions, such as hidden sector dark matter, are able to account for this small-scale puzzle while retaining the ability of CDM to describe large-scale structure. In order to be compatible with
both small-scale and large-scale observations, the single particle mediated self-interactions of dark matter must be velocity dependent
and the favored mediator mass that is consistent with all astrophysical observations lies in the $\sim$ 1 - 100 MeV mass range, as shown in Fig.~\ref{fig:sss1}. Thus, in light of the observational evidence for small scale structure, it is critical to explore the $\sim$ 1 - 100 MeV mass range for hidden sector mediator candidates.

\subsection{Muon Anomalous Magnetic Moment}
\begin{figure}[hbt!]
    \centering
    \includegraphics[width=0.75\textwidth]{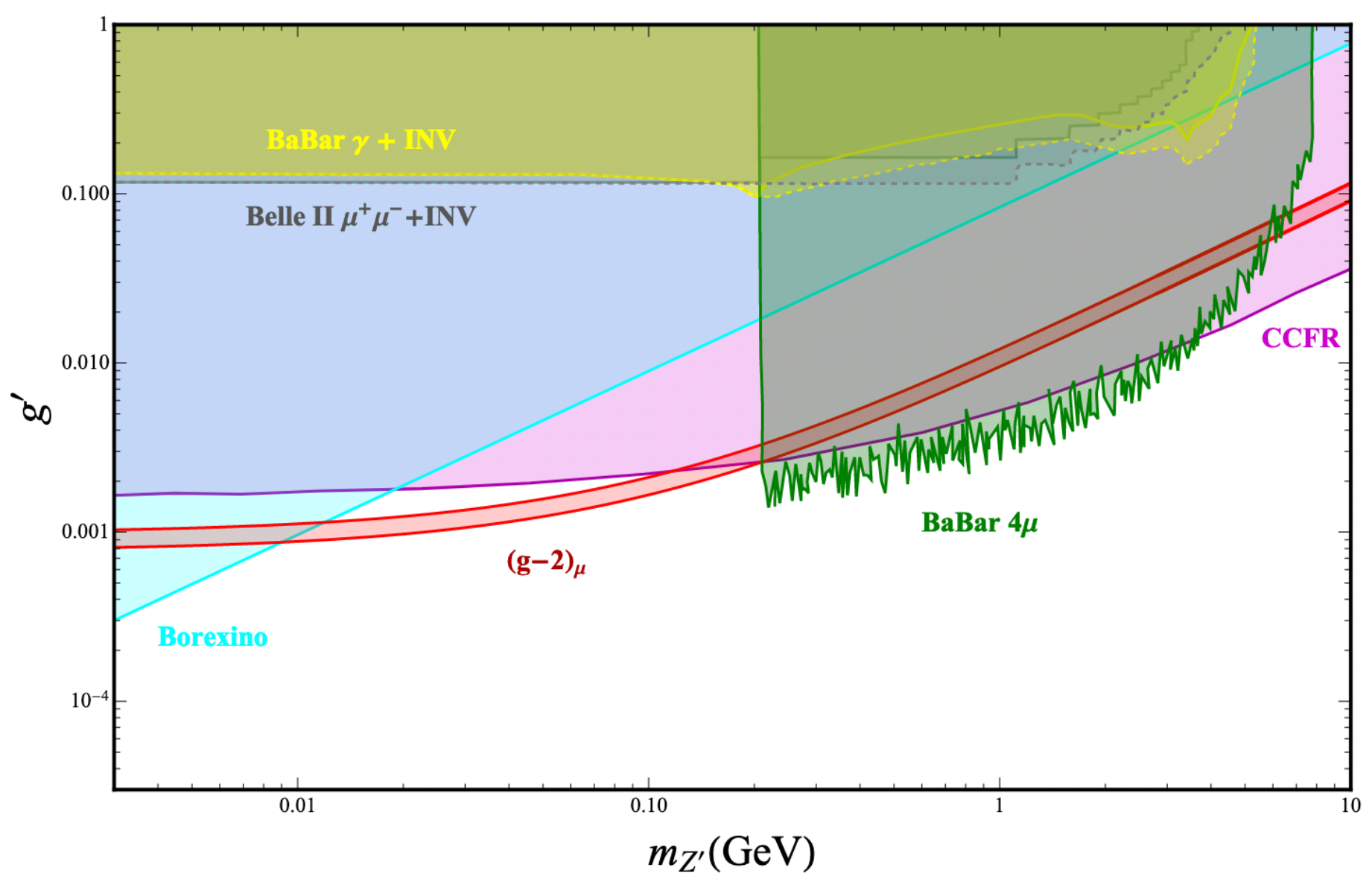}
    \caption{Allowed region to explain the$(g-2)_{\mu}$ anomaly and exclusion regions on the Z$^{\prime}$ mass and gauge coupling constant $g^{\prime}$ from different experiments: e$^{+}$e$^{-}$ colliders BABAR and Belle II; neutrino-electron scattering from Borexino,  and neutrino trident production from the CCFR collaboration.  These limits exclude most of the parameter space except for the region contained in the interval 10~MeV$<m_{Z^\prime}<$~200~MeV. 
    This corresponds to a $X$ mass of 
    5~MeV~$< m_{X} <$~100~MeV. 
    Reproduced from Ref.~\cite{Arcadi:2021yyr}.}
    \label{fig:g-2}
\end{figure}

The Muon $g-2$ collaboration has recently reported their measurement of the muon magnetic moment at Fermi National Accelerator Laboratory (FNAL) which is consistent with their previous measurement at Brookhaven National Laboratory (BNL)~\cite{Albahri:2021kmg,Bennett:2006fi]}.
These two results, with their uncertainties combined, show a $4.2\sigma$ deviation from the SM prediction.
There are many proposed solutions to this discrepancy, here we will focus on potential MeV-scale bosons leading to loop corrections that can account for the muons'  anomalous magnetic moment. For example the scalar Higgs doublet model which adds a second scalar doublet to the SM, resulting in a scalar sector with two Higgs doublets~\cite{Arcadi:2021yyr}. The  muon  couples  directly  to  the  new Z$^{'}$ gauge boson which can  provide the  required  contribution  to  the  muon  $g-2$. In fact, the Z$^{'}$ can be made very light with a sufficiently small coupling constant $g^{'}$ and can address the $(g-2)_{\mu}$ anomaly with Z$^{'}$ mass roughly in the range of 10 -200~MeV~\cite{Arcadi:2021yyr}, as shown in Fig.~\ref{fig:g-2}. In this model the DM is introduced via the addition of a vector-like fermion $X$ which undergoes s-channel mediated annihilations into SM particles. The DM relic abundance constrains the allowed phase space to near the Z$^{'}$ resonance, which implies $m_X \sim m_{Z'}/2$.   If the Z$^{'}$ mass  should  lie  around 10 - 200~MeV  to  explain the $(g-2)_{\mu}$ the $X$ mass must lie in the mass range $m_{X} \approx$~5 - 100~MeV~\cite{Arcadi:2021yyr}.  

The $X$ can also directly contribute  to  the $(g-2)_{\mu}$. The contributions of an $X$ to the $(g-2)_{\mu}$ scales as $1/m_{X}^2$, thus a smaller mass will have a comparably larger contribution than a higher mass particle. There have been numerous publications proposing dark photons in the MeV-range to, at a minimum, partially account for the muon anomalous magnetic moment~\cite{Arcadi:2021yyr,Athron:2021iuf,Borah:2021jzu,Ge:2021cjz}.
It is critical to search the MeV-scale region for a potential $X$ to account for the observed muon $(g-2)$ anomaly.

\subsection{ATOMKI Beryllium Anomaly}
\label{sec:atomki}
A 1996 experiment~\cite{deBoer1996}, using the 2.5 MV Van de Graaff accelerator at Institut f{\"u}r Kernphysik of the University of Frankfurt, noted a $4.5\sigma$ deviation from expectation in the angular distribution of $e^+e^-$ from Internal Pair Conversion (IPC) of the $^8$Be M1 resonance.
Analysis and Simulations of the signal that was seen led to the conclusion that a neutral boson of mass 9 MeV was a possible explanation that could not be ruled out by existing constraints~\cite{deBoer1997,deBoer:2001sjo}.

A 2015 experiment at the ATOMKI 5 MV Van de Graaff accelerator sought to repeat this measurement to further study the reported anomaly~\cite{Krasznahorkay:2015iga}.
This experiment again saw an excess of $e^+e^-$ pairs beyond the expectation of IPC.
A subsequent analysis of these results has shown that the $6.8\sigma$ anomaly is consistent with a new particle with a mass of 16.84 MeV, dubbed X17. 
A followup experiment by the ATOMKI group was conducted on the 20.01 MeV $0^-\to0^+$ transition in $^4$He.
The 2019 preprint, and subsequent 2021 publication, of these results reports an $e^+e^-$ excess consistent with the so-called X17 particle~\cite{Krasznahorkay:2019lyl,Krasznahorkay:2021joi}.
\begin{figure}[hbt!]
    \centering
    \includegraphics[width=0.65\textwidth]{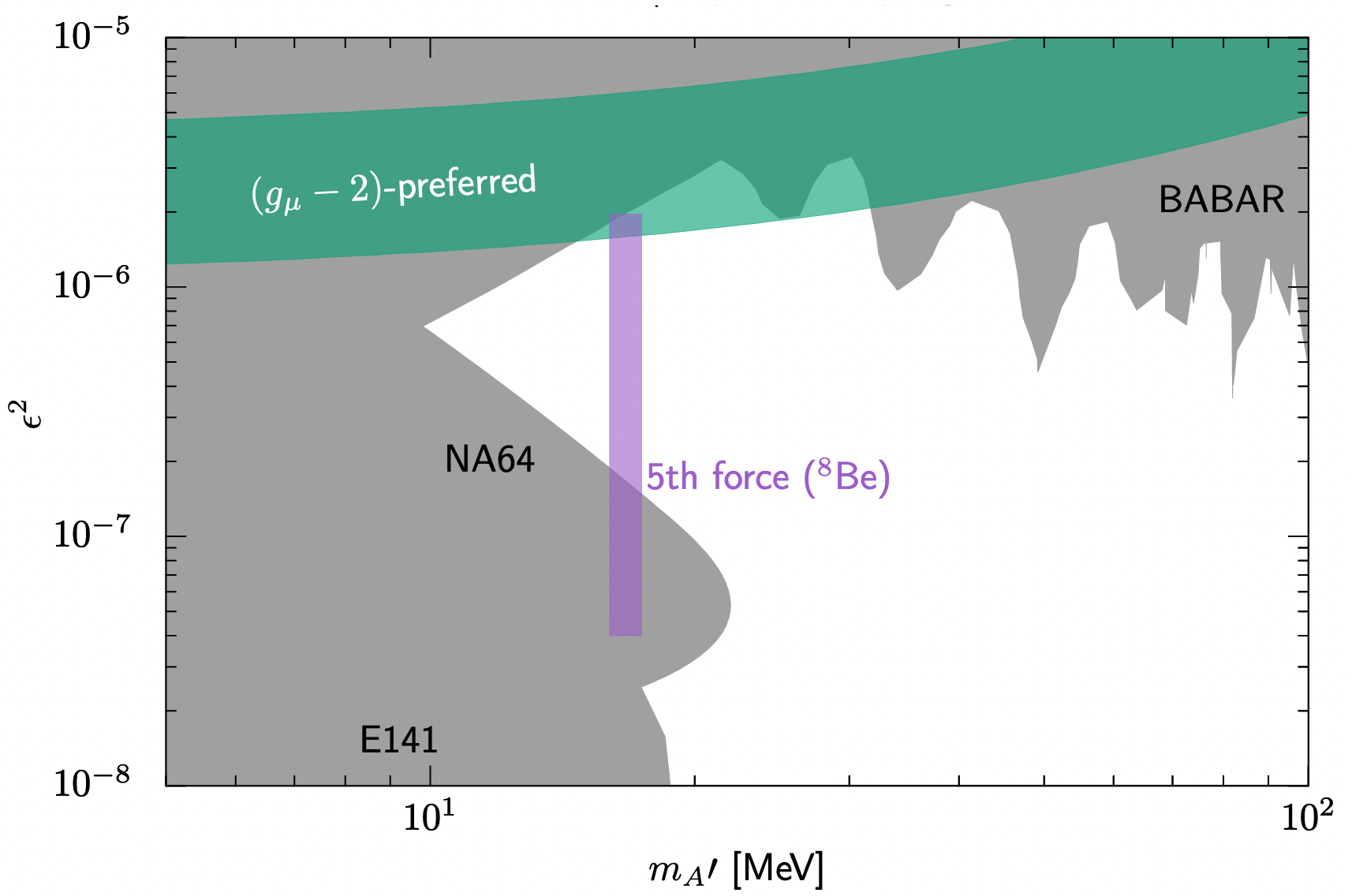}
    \caption{Current constraints on a fifth force explanation of the $^8$Be anomaly.  The vertical axis is the leptonic coupling strength relative to $\alpha_{QED}$, with horizontal axis the mass of the mediator. Excluded regions, in gray, are taken from measurements that depend solely on leptonic interactions. Dark photon exclusions via hadronic measurements are not shown.
    Reproduced from Ref.~\cite{darklight:2020pp}.}
    \label{fig:be_anom}
\end{figure}

Feng \etal~\cite{Feng:2016jff} analyzed this signal against existing constraints.
The proposed explanation is that the signal is from the decay of a protophobic gauge boson that mediates a fifth force with a length scale of 12~fm.
This explanation can also possibly explain the muon anomalous magnetic moment and an excess of $\pi^0\to e^+e^-$ decays. The current constraints from leptonic production mechanisms, where the effective coupling to a new force-carrier is proportional to electric charge, are shown in Fig.~\ref{fig:be_anom}. For more generic fifth-force models with  quark flavor-dependent couplings~\cite{Feng:2016jff}, a much wider parameter space with multiple couplings must be considered.

The fifth force based explanations are being challenged by recent reanalyses~\cite{Zhang:2020ukq} and the observed discrepancies could be the result of as-yet-unidentified nuclear reactions or experimental effects. Nonetheless, these results have garnered a lot of attention, and must be independently validated with the highest urgency.




\subsection{Current Constraints}
\begin{figure}[!hbt]
\centerline{
\includegraphics[width=0.9\textwidth]{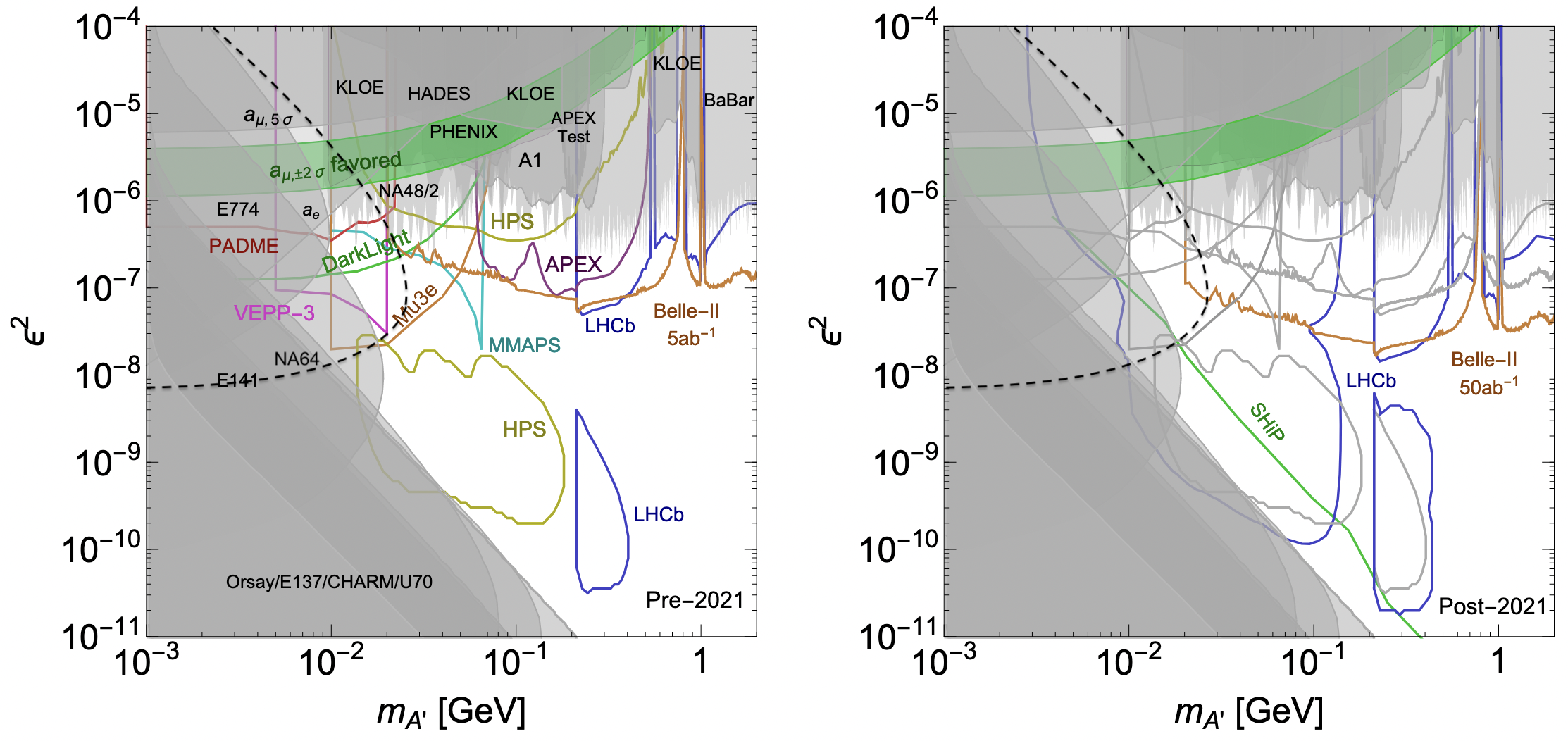}}
\caption{Sensitivity to $X$ (sometime called $A^{\prime}$) for exclusive experiments seeking visible decay modes $X \rightarrow l^{+}l^{-}$. Left:Experiments  capable  of  delivering  results  soon.   Shaded  regions  show existing bounds.  Colored regions are experiments that are being prepared or are currently underway. The green band shows 2$\sigma$ region in which an $X$ can explain the discrepancy between the  calculated  and  measured  value  for  the  muon $(g-2)$. The black dashed line is the exclusion from the recently reported NA64 experiment. Right:Longer  term  prospects  beyond 2021 for experimental sensitivity.  All projections on left plot are repeated in gray here. Reproduced from Ref.~\cite{Alexander:2016hsr}.}
\label{fig:cc}
\end{figure}

There are many constraints from previous experiments on the parameter space for a dark photon as shown in Fig.~\ref{fig:cc} reproduced from Ref.~\cite{Alexander:2016hsr}. The parameter space of mass $m_X$ and the square of the coupling $\epsilon^2$ has been constrained from two sides by existing data. The low mass and small-$\epsilon^2$ parameter space has been excluded primarily by previous beam-dump experiments~\cite{bdx1,bdx2,bdx3} and the energy loss in supernovas from observations of the neutrinos they generate~\cite{nova_cooling}. The constraints from the 2020 results of the NA64 experiment measuring $e^{+}e^{-}$ decays~\cite{Banerjee:2019hmi} have been added to Fig.~\ref{fig:cc} as the black dashed line. In the NA64 experiment 8.4$\times$10$^{10}$ 150 GeV electrons were incident on a $\approx 30-40$ (depending on the run) radiation length (r.l.) active target-tungsten-calorimeter. The active target also served as a dump for the recoil electrons as well as the secondary particles emitted by the electron beam before the production of $X$. The subsequent decay of $X$ to $e^{+}e^{-}$ would appear as a bump in the spectrum of events producing two showers, one in the active target and another one in a second downstream calorimeter, with the total energy of the two showers almost equal to the beam energy. The experiment described in this proposal has a radically different design with $\sim$ 2.7 $\times$ 10$^{16}$ electrons incident on a 2.5$\times$ 10$^{-4}$ r.l. target and a high resolution calorimeter and tracking detectors that will individually detect all three particles in the final state making it a direct detection experiment. The systematic differences between the two methods are further discussed in Sec.~\ref{sec:oexpt}.

The larger-$\epsilon^2$ values~(10$^{-6}$ - 10$^{-4}$) are excluded for a broad range of masses by searches at B-factories~\cite{bfactory} and lepton $g-2$ measurements~\cite{g2_meas}. With these constraints, there still exists a gap in the MeV-scale region with coupling constant $\epsilon^2$ in the range of $10^{-6}-10^{-10}$. Several experimental efforts such as DarkLight~\cite{darklight:2020pp}, MMAPS~\cite{mmaps:2017}  and Mu3e~\cite{mu3e} are being prepared and several others such as the HPS~\cite{hps} and APEX~\cite{apex} experiments at JLab are currently underway to cover parts of this gap as described in Sec.~\ref{sec:oexpt} and shown in Fig.~\ref{fig:cc} (left). While the gap in the parameter space is small, it is a high-priority search window as well-motivated models described above can exist within this previously unexplored region.
This gap opens up even more when considering the possibility of a flavor-dependent coupling, as any constraints derived from proton or muon beams would then be irrelevant to an electron beam search. Based on the region of parameter space that is still unexplored and at the same time having healthy overlap with other searches, we have designed an experiment to search the $3-60$ MeV mass range as described in the next section.

\section{Experimental Method}
\label{sec:method}
\begin{figure}[hbt!]
    \centering
    \includegraphics[width=0.35\textwidth]{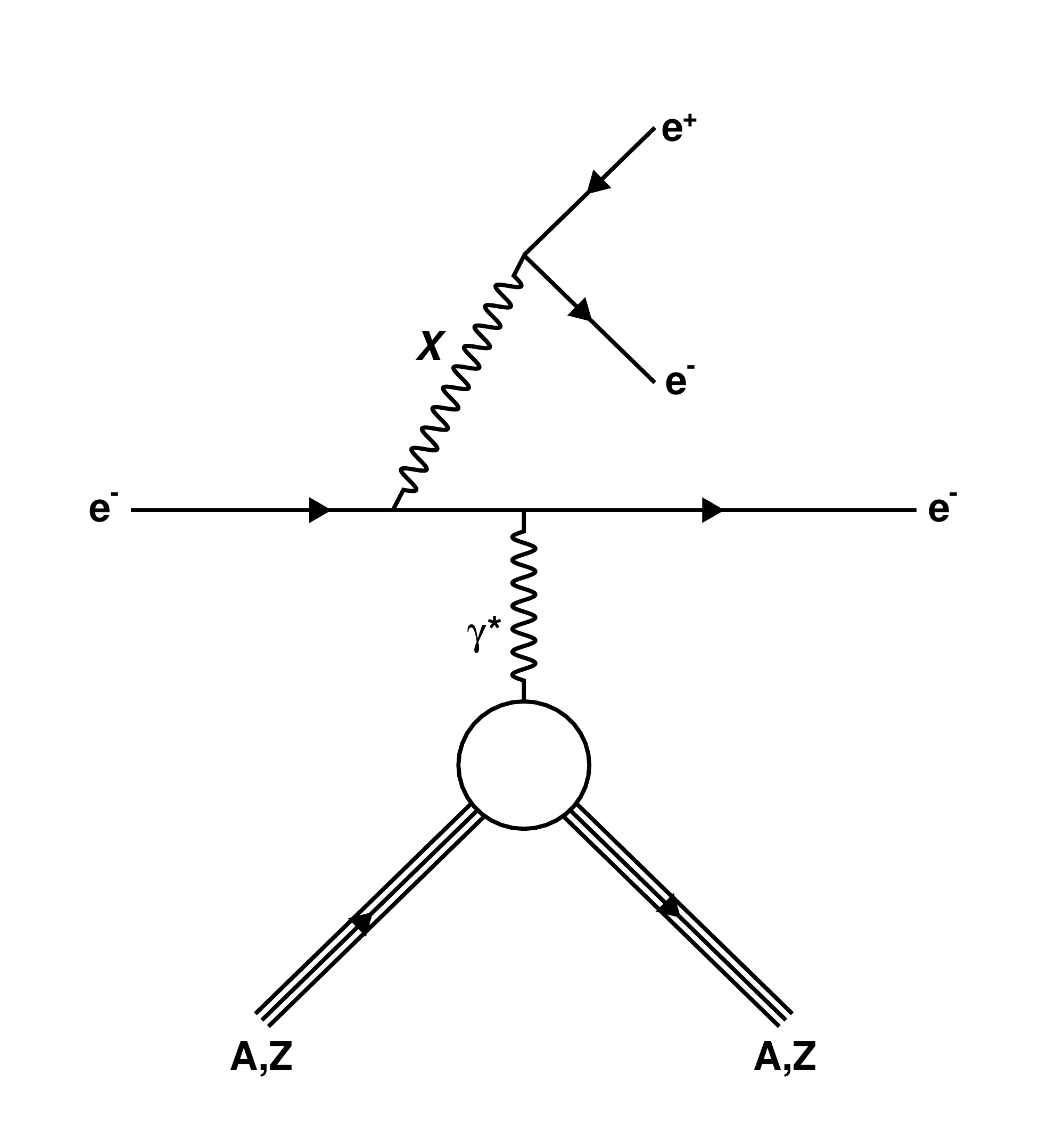}
    \includegraphics[width=0.35\textwidth]{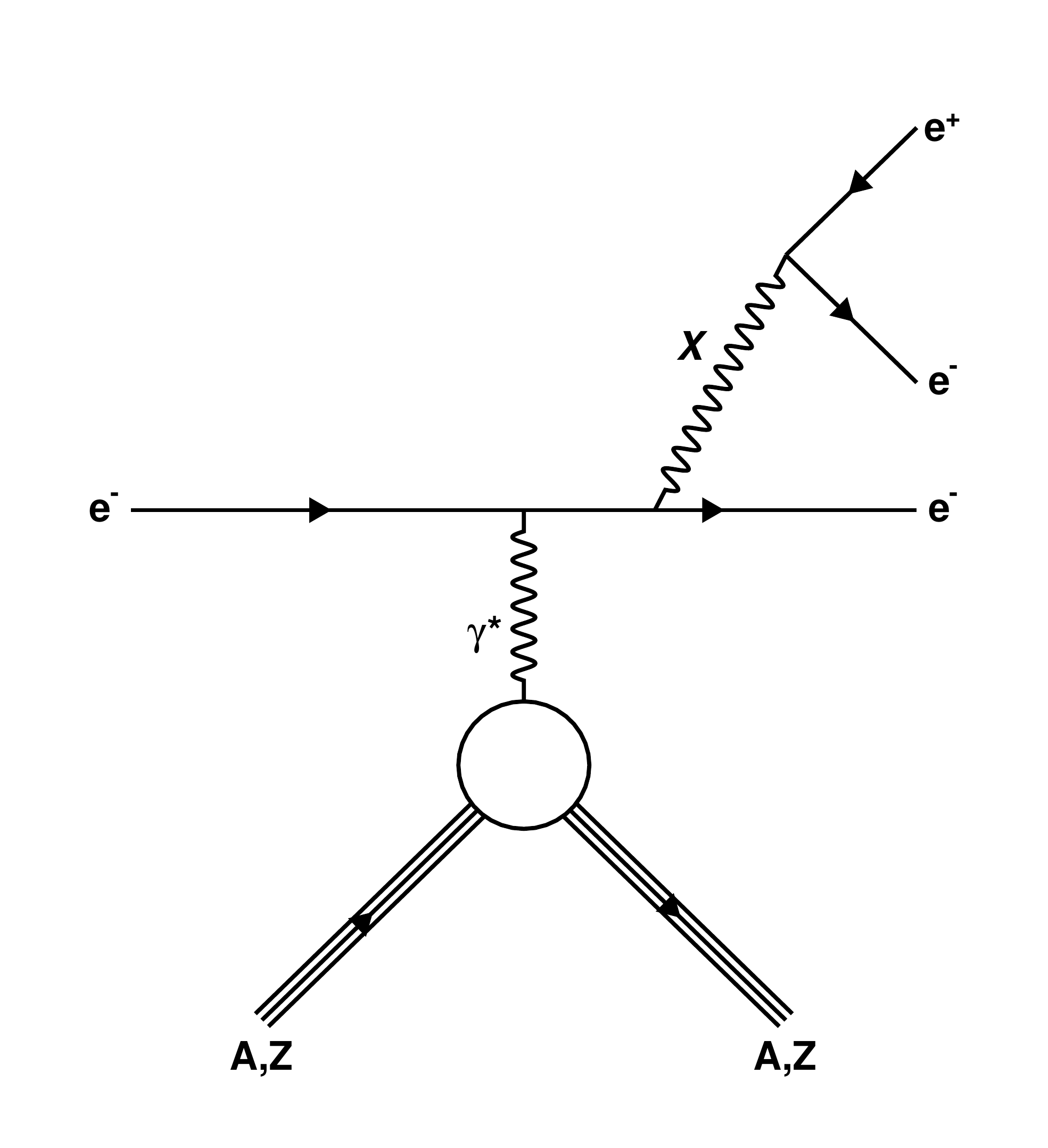}
    \caption{Bremsstrahlung-like production of a hidden sector force carrier $X$ from electron scattering. The left diagram shows production from the incoming electron and the right diagram shows production from the scattered electron.}
    \label{fig:bremsstrahlung-feynman}
\end{figure}

\begin{figure}[hbt!]
    \centering
    \includegraphics[width=0.3\textwidth]{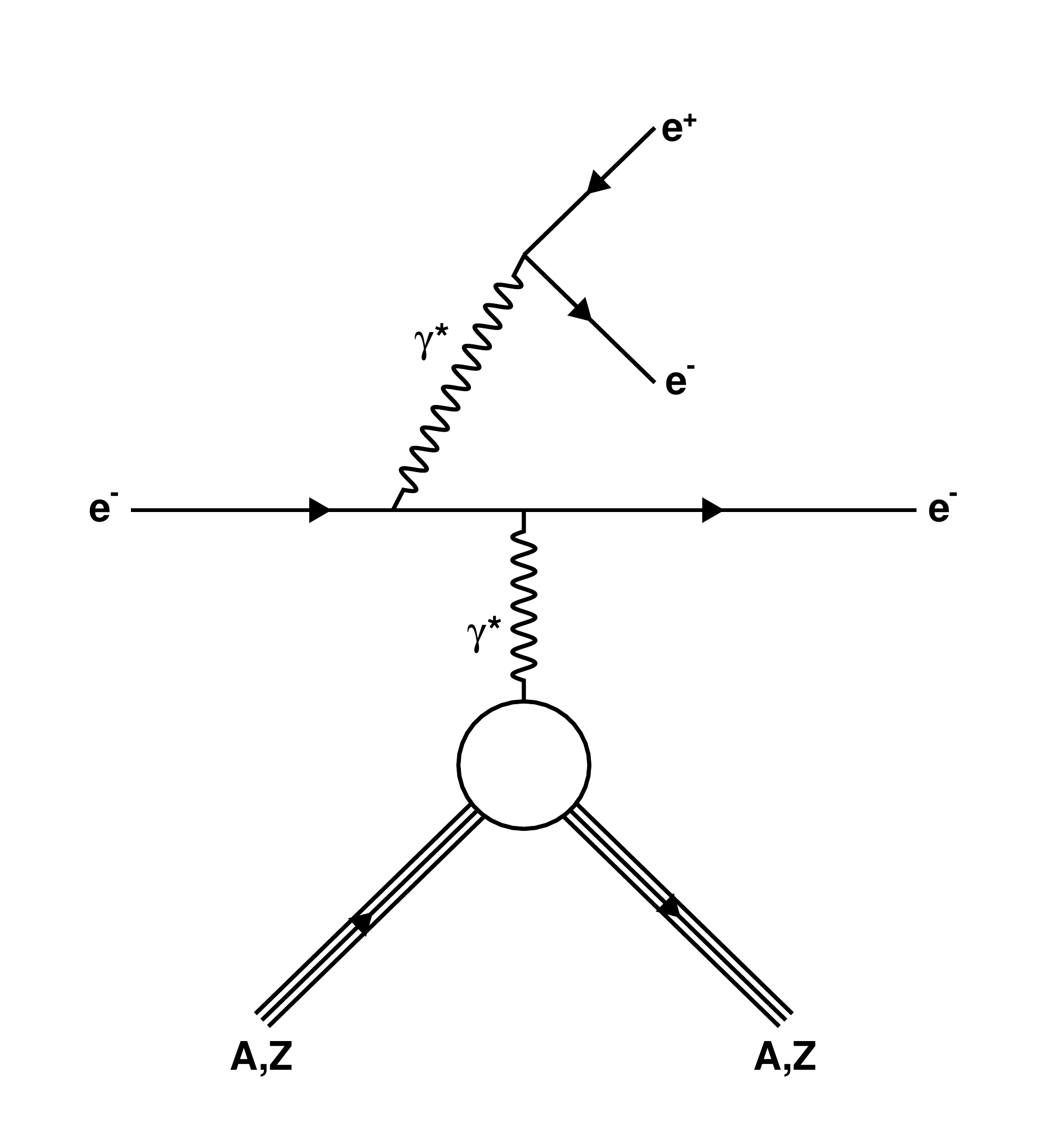}    \includegraphics[width=0.3\textwidth]{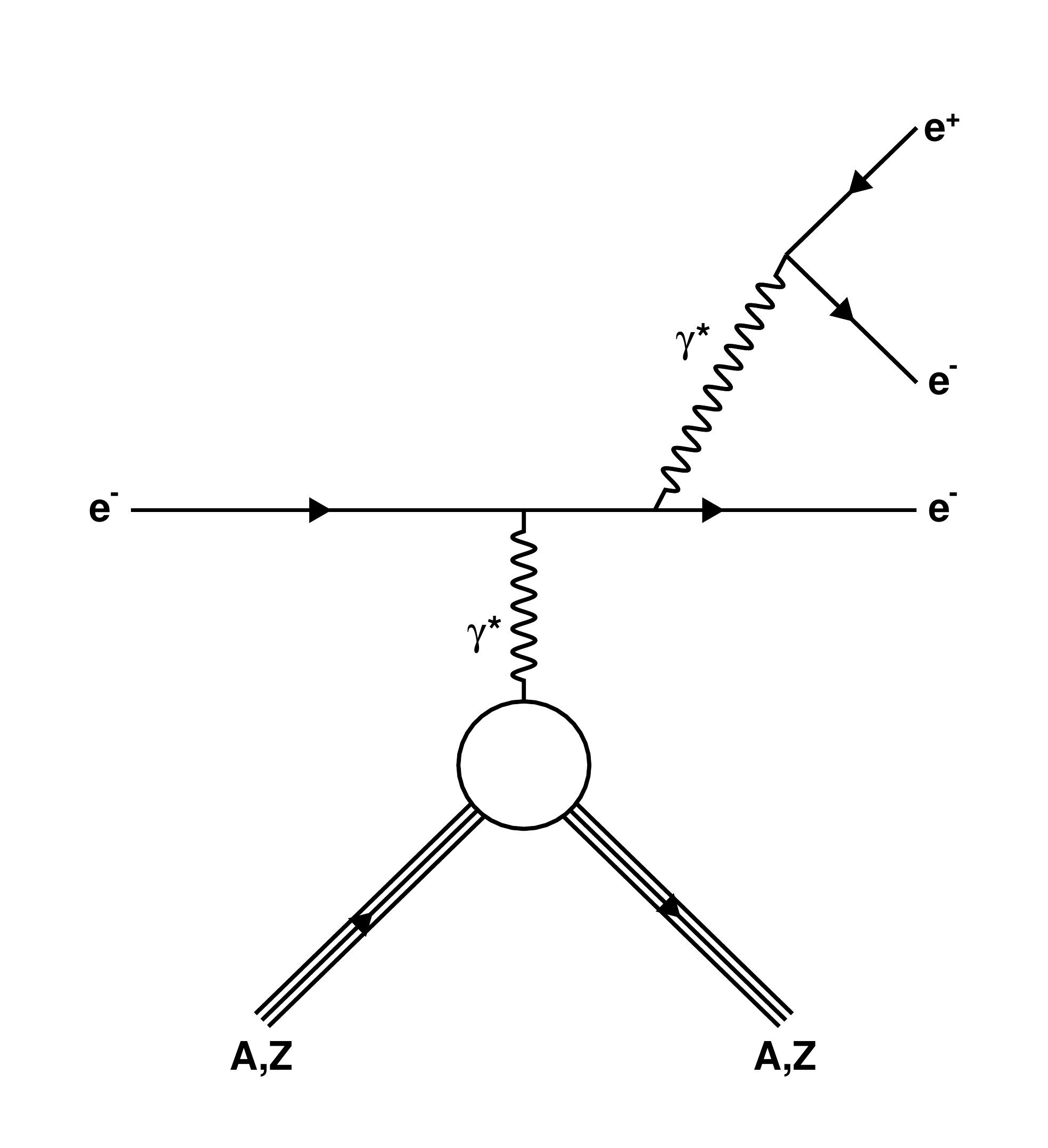}
    \includegraphics[width=0.3\textwidth]{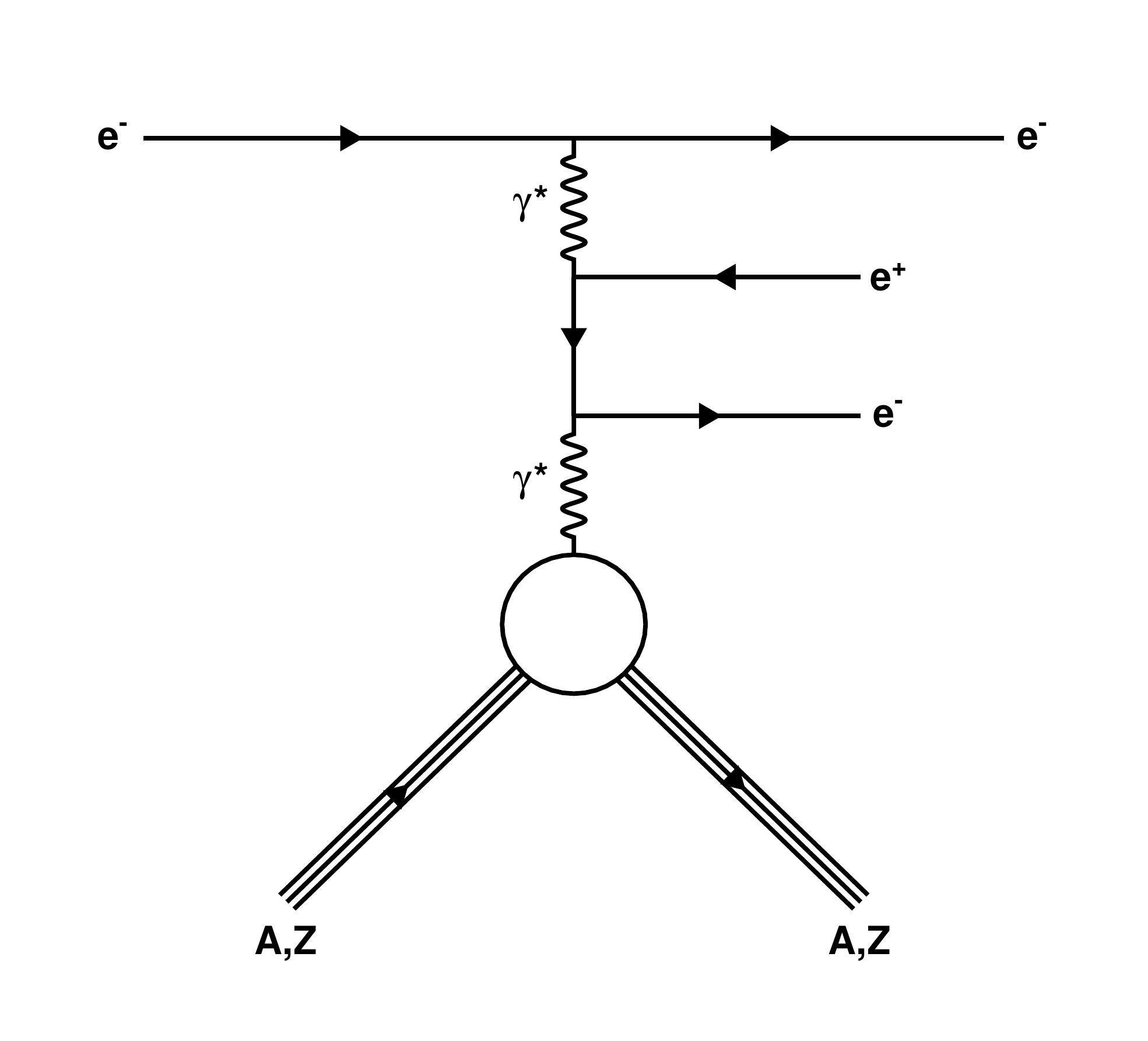}
    \caption{The QED background from radiative (left and middle) and Bethe-Heitler (right) process.}
    \label{fig:background-feynman}
\end{figure}

This experiment will focus on the bremsstrahlung-like production of hidden sector particles from the initial electron or the scattered electron (both shown in Fig.\ref{fig:bremsstrahlung-feynman}), in the $3-60$ MeV mass range.
When searching for new particles, it is of the utmost importance to minimize backgrounds in order to prevent false ``bumps'' in the mass spectrum. The primary QED background in this experiment (see Fig.~\ref{fig:background-feynman}) are from the radiative pair production which is an irreducible background and the Bethe-Heitler trident reactions which can be kinematically suppressed.

The cross section for bremsstrahlung-like production of an $X$ can be estimated within the Weizs{\"a}cker-Williams approximation~\cite{Bjorken:2009mm} giving a total production rate of:

\begin{equation}
\label{eq:rate}
 N_X \sim N_e \mathcal{C} T \epsilon^2 \frac{m_e^2}{m_X^2},   
\end{equation} 
where $N_e$ is the number of incident electrons, $T$ is the target thickness, $m_X$ is the mass of the produced $X$, and $\epsilon^2$ is the square of the dimensionless coupling constant of the $X$ to SM matter. For $T<<1$ and $m_X$ in the range of this experiment, the factor $\mathcal{C} \sim$ 5 (it is related to the effective photon flux, atomic screening, and nuclear size effects). This implies that the total $X$ production rate is suppressed relative to photon bremsstrahlung by $\sim \epsilon^2 \frac{m_e^2}{m_X^2}$~\cite{Bjorken:2009mm}. Moreover, the bremsstrahlung produced $X$ is sharply peaked at $y = \frac{E_{X}}{E_{beam}} \approx$ 1, i.e. when an $X$ is produced, it carries nearly the entire beam energy. Further, the emission of $X$ is dominantly co-linear with the beam with a cut-off emission angle ($\theta_X$) that is smaller than the opening angle of the decay products i.e. $\theta_{X} < m_{X}/E_{beam}$~\cite{Bjorken:2009mm}. In the limit of small $m_X/yE_{beam}$, the lab-frame opening angles $\theta_{\pm}$ and energies $E_{\pm}$ of the $X$ decay products are given by~\cite{Bjorken:2009mm};
\begin{equation}
\tan{(\theta_{\pm})} = \pm \frac{1}{\gamma} \sqrt{\frac{1 \mp \cos{(\theta_{CM})}}{1 \pm \cos{(\theta_{CM})}}} + \tan{(\theta_{X})}
\end{equation}
and
\begin{equation}
E_{\pm} = \frac{yE_{beam}}{2}\left(1 \pm \theta_{CM} \right), 
\end{equation}
where $\gamma = yE_{beam}/m_X$ and $\theta_{CM}$ is the the emission angle of the forward decay product relative to the direction of $X$ in its rest frame. At the mass range probed in this experiment the recoiling electron largely balances the recoil of the $X$ and the energy $E_R$ and the angle $\theta_R$ of the recoiling electron in the laboratory frame are given by~\cite{Bjorken:2009mm};
\begin{equation}
E_R = (1-y)E_{beam} \approx m_X 
\end{equation}
and
\begin{equation}
\tan{(\theta_R)} \approx \sqrt{\frac{m_X}{E_{beam}}}\left(1 + \frac{m_X}{2E_{beam}} + ... \right) ,
\end{equation}
indicating the relatively wide angle of the recoiling electron relative to the $X$ decay products. These kinematics characteristics of $X$ production can be used to suppress the background from the Bethe-Heitler trident reaction which can otherwise be prohibitively large.  
Once again using the Weizs{\"a}cker-Williams approximation it is found that Bethe-Heitler production is peaked at small $y$, it has $\cos{(\theta_{CM})} \rightarrow 1$ and it is co-linear to the photon. Therefore, the kinematic characteristics of the Bethe-Heitler production is quite different from the signal (from $X$ production) and from radiative backgrounds, which  are peaked at large values of $y$ and vary slowly with $\cos{(\theta_{CM})}$. This implies that requiring $y >$ 1 - $\delta$, with $\delta$ near or below its median value $\bar{\delta} = {\mbox{max}}(m_X/E_{beam}, m_e/m_X)$, keeps a large fraction of the signal while suppressing the Bethe-Heitler background by a factor of $\delta$. Similarly, since the signal from $X$ production is relatively flat in $\cos{(\theta_{CM})}$ the Bethe-Heitler background can be suppressed by constraining the cosine of the opening angles of the two decay products to be near unity~\cite{Bjorken:2009mm}.


The proposed experiment will use a ``bump hunt'' technique in the direct detection search of heretofore unknown MeV mass particles. The experiment is designed based on the kinematic constraints described above. A 2.2 and 3.3 GeV CW electron beam from CEBAF will be incident on a retractable ultra-thin target consisting of a 1~$\mu$m Tantalum foil. The scattered particles will traverse the 7.5 m long flight path in a vacuum chamber consisting of the PRad target chamber (or an appropriate diameter beam pipe) and the PRad vacuum chamber en route to a pair of common ionization volume GEM chambers coupled to the HyCal EM calorimeter. 
 
 All 3 cluster events with individual cluster energy within (0.02-0.85)$\times E_{beam}$ and with the sum of total energy deposited $E_{sum} >$ 0.7$\times E_{beam}$ will be recorded and examined for ``bumps'' in the the $M_{e^{+} e^{-}}$ invariant mass spectrum reconstructed from these events. This will allow for the $X$-particles production by virtual photons over a wide energy range in the forward solid angle coverage of the \pbo{} calorimeter. The capability of detecting events produced in a wide energy and angle range, in a single experimental setting, is one of the important features of our experiment which stands in contrast to all other magnetic spectrometer methods. The tracking provided by the pair of GEM chambers 
will be used to suppress background events from the large area window at the exit of the vacuum chamber. The GEM chambers will also be used to veto any neutral particles. Only the \pbo{} part of the HyCal calorimeter will be used in the experiment. The lower resolution of the lead-glass part makes it unusable in this experiment. The experimental method discussed here applies directly to any spin-1 boson in the hidden sector that decay directly to a lepton pair. 
The experiment is designed with several key features to provide a clean signal of any unknown particles that may exist in the 3 - 60 MeV/c$^2$   mass range:
\begin{itemize}
    \item The experiment will detect the scattered electron along with the decay products of the hidden sector boson (all three final states are detected - i.e. it is a direct detection experiment).
    \begin{itemize}
        \item The full energy of the event can be reconstructed, reducing kinematic mimicking.
        \item It can be verified that the dark photon and the recoil electron are co-planar.
    \end{itemize}
    \item With the use of two GEMs spaced 10~cm apart, charged decays and recoil electrons can be tracked to ensure that they do not originated from the vacuum chamber exit window. The GEM detectors also help suppress the neutral background.
    
    \item The use of two beam energies (2.2 GeV and 3.3 GeV) will ensure that any background processes that mimic a bump can be identified.
\end{itemize}
This optimized design can be implemented with the PRad setup with very minor modifications, making it a ready-to-run and uniquely cost-effective search for hidden-sector particles in the 3 - 60 MeV mass range.

\section{Experimental Setup}
\begin{figure}[!hbt]
\centerline{
\includegraphics[width=1.0\textwidth]{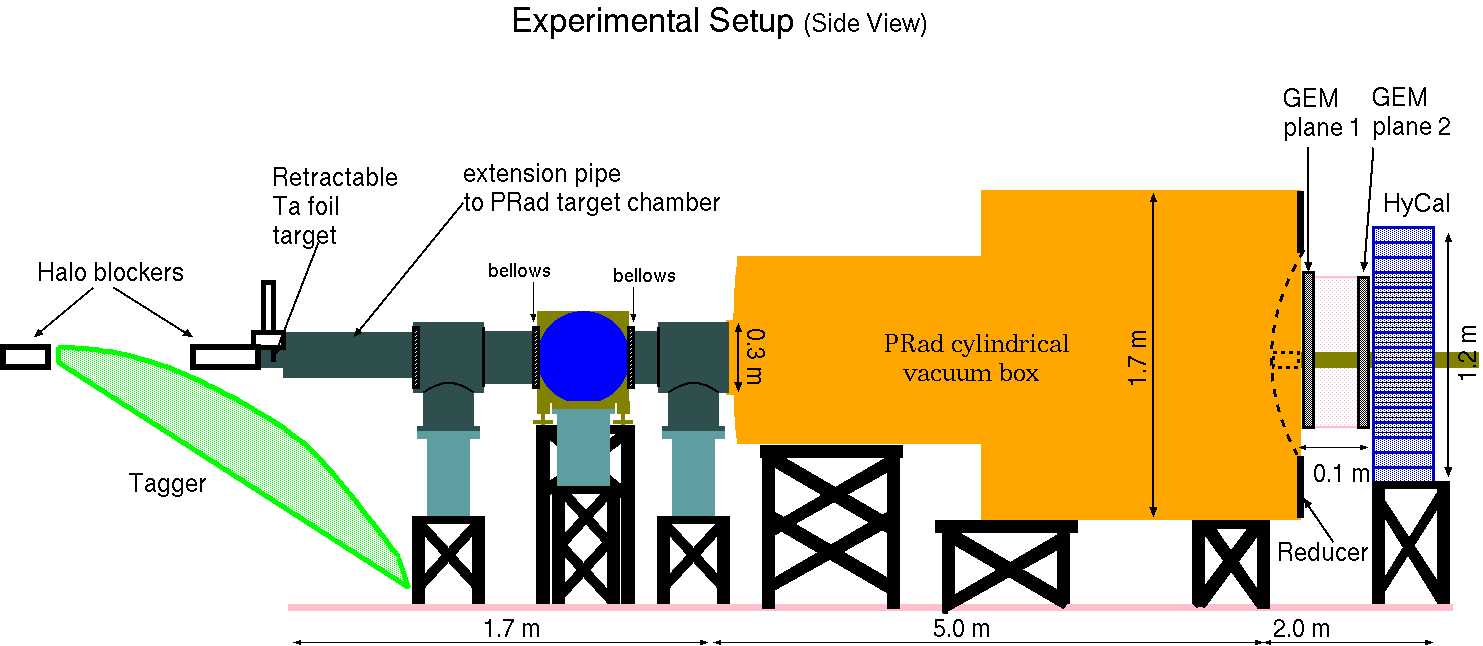}}
\caption{Schematic of the experimental setup.}
\label{fig:expt_setup}
\end{figure}

The proposed experiment plans to reuse the PRad setup (shown in Fig.~\ref{fig:expt_setup}) but with a Tantalum foil target placed 7.5 m upstream of the calorimeter. Only the high resolution \pbo ~crystal part of the electromagnetic calorimeter will be used together with a new flash ADC (FADC) based readout system for the calorimeter. 
Just as in the PRad experiment, the scattered electrons will travel through the 5~m long vacuum chamber with a  thin window to minimize multiple scattering and backgrounds. The vacuum chamber matches the geometrical acceptance of the calorimeter. An extension piece will be added to the upstream end to couple the PRad target chamber to the super-harp which will now hold the target foil. A reducer ring will be attached to the downstream exit of the PRad vacuum chamber and a new 1~mm thick Al exit window will be used such that it matches the \pbo ~portion of the calorimeter. Two layers of GEM detectors  will add a modest tracking capability to help reduce the photon background and to reduce the background originating from the vacuum chamber exit window.

The elements of the experimental apparatus along the beamline are as follows:
\begin{itemize}
\item Ta foil target held inside the super-harp enclosure.
\item The PRad target chamber attached to the super-harp via an extension piece and coupled to the two stage, large area vacuum chamber with a single thin 1 mm Al. window at the calorimeter end.
The PRad target chamber is proposed to be used in this experiment only to be able run in sequence with the PRad-II experiment. If the experiment runs as a stand alone installation, an appropriate diameter beam pipe would be more practical.
\item A pair of GEM detector planes, separated by about 10~cm for coordinate measurement as well as tracking.
\item
High resolution \pbo ~crystal calorimeter (the Pb-glass part of the HyCal will not be used) with a FADC based readout.
\end{itemize}

\subsection{Electron beam}
\begin{table}[hbt!]
\caption{Beam parameters for the proposed experiment}
\label{tab:beampar}
\vspace{-0.25cm}
\center{
\begin{tabular}{c|c|c|c|c|c} \hline \hline
  Energy & Current & Polarization & Size & Position stability & Beam halo \\
  (GeV)  &   (nA)  &   (\%) &  (mm)  &  (mm) &  \\ \hline
  &         &        &        &       &  \\ [-11pt]
  2.2 & 50 & Non & $<$ 0.1 & $\leq$ 0.1 & $\sim$ 10$^{-6}$ \\
  3.3  &   100 & Non  & $<$ 0.1 & $\leq$ 0.1 & $\sim$ 10$^{-6}$ \\ \hline \hline  
\end{tabular}
}
\end{table}
\begin{figure}[htb!]
  \centerline{
    \includegraphics[width=0.5\textwidth]{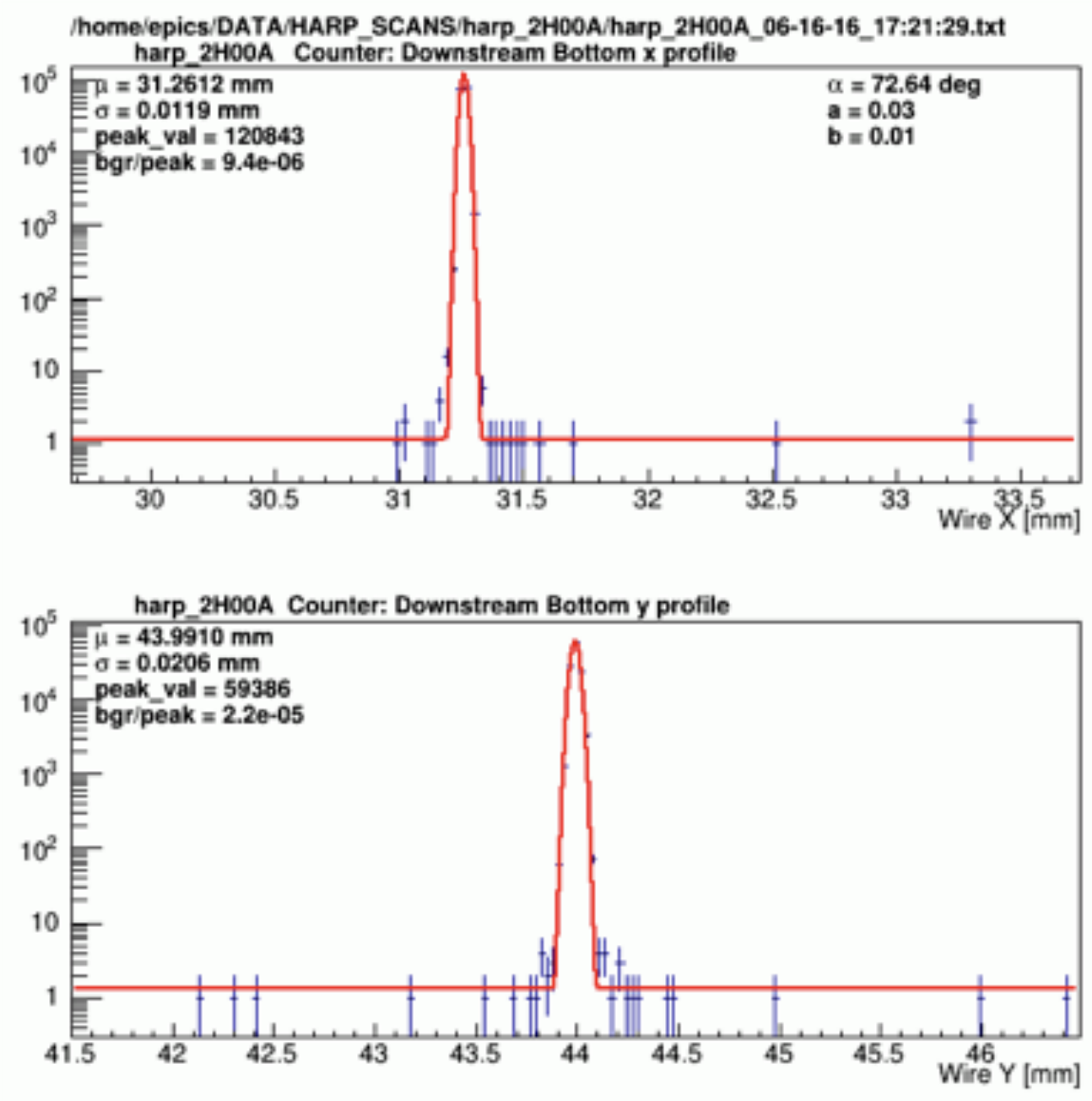}}
  \caption{Typical beam profile during the PRad experiment, showing a beam size of $\sigma_x=$~0.01~mm and $\sigma_y=$~0.02~mm.}
  \label{fig:beamprof}
\end{figure}
\begin{figure}[htb!]
  \centerline{
    \includegraphics[width=0.9\textwidth]{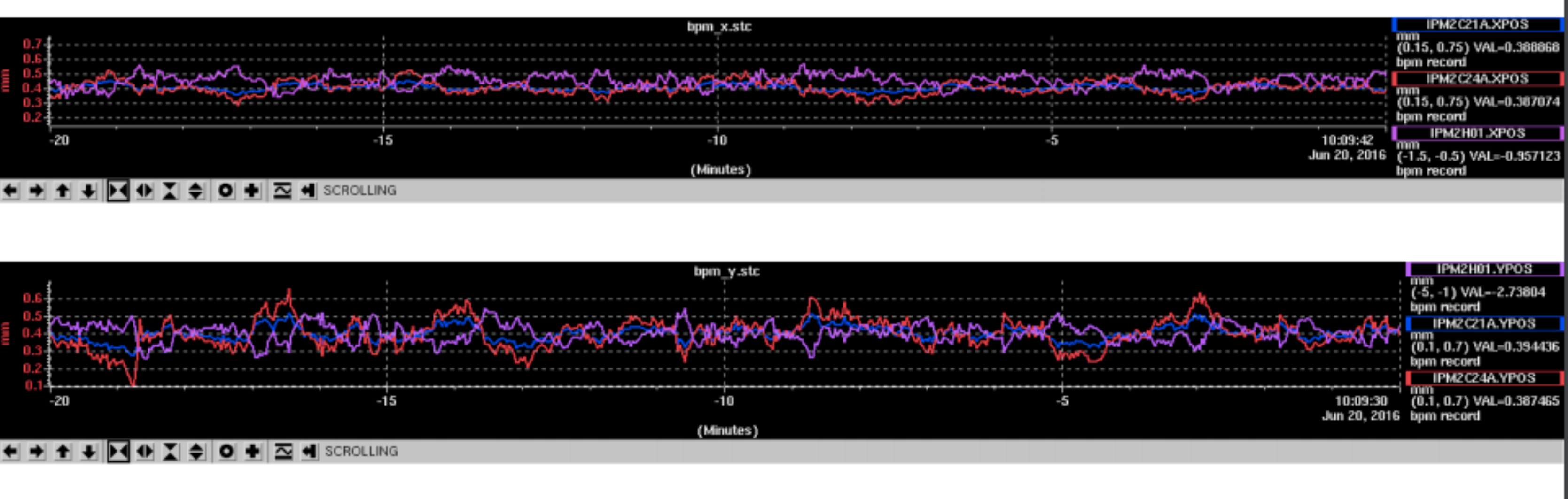}}
  \caption{Beam X,Y position stability ($\simeq \pm$ 0.1~mm) during the PRad experiment.}
  \label{fig:beampos}
\end{figure}
We propose to use the CEBAF beam at two incident beam energies $E_0 =$ 2.2 and 3.3 GeV for this experiment. The beam requirements are listed in Table~\ref{tab:beampar}. All of these
requirements, or even better, were achieved during the PRad experiment. 
A typical beam profile during the PRad experiment is shown in Fig.~\ref{fig:beamprof} and the beam X, Y position stability was  $\simeq \pm$ 0.1~mm as  shown in Fig.~\ref{fig:beampos}.

\subsection{Target}
\label{sec:target}
\begin{table}[hbt!]
\caption{Material properties}
\label{tab:targ_prop}
\vspace{-0.25cm}
\center{
\begin{tabular}{c|c|c|c|c|c} \hline \hline
  Material & Z &Melting Point & Density & Tensile strength & Min. foil thickness\\
   &  &($^{\circ}$C)  &   (g/cc) &  (1000 psi)  & $\mu$ m \\ \hline
  &         &        &        &       &  \\ [-11pt]
  Ta & 73& 2996 & 16.6 & 35-75 & 0.5 \\
  W  & 74 & 3410 & 19.3  & 100-500 & 2.5 \\ \hline \hline  
\end{tabular}
}
\end{table}
A 1 $\mu$m thin Ta foil (2.5$\times$10$^{-4}$ r.l.) target will be placed inside the super-harp setup that is part of the PRad setup in Hall-B. The super-harp will be located right after the second beam halo blocker placed right after the Hall-B tagger in the Hall-B beamline. The target foil will be placed on the same ladder that holds the super-harp wires. The mechanism used to insert the super-harp wires into the electron beam will be utilized to insert or retract the target foil from the beam. Tantalum was chosen because foils as thin as 0.5$\mu$m are available commercially. Properties of Ta and W are compared in Table ~\ref{tab:targ_prop}. Multiple target foils will be available on the target ladder for redundancy.

\begin{figure}[!ht]
  \centerline{
    \includegraphics[width=0.75\textwidth]{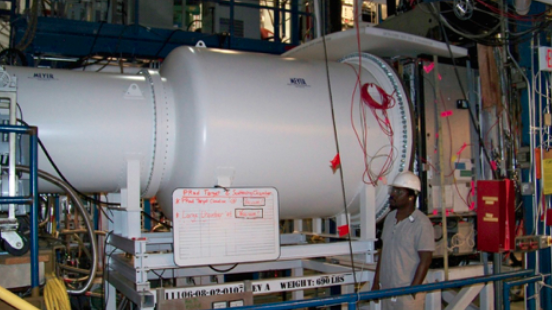}}
  \caption{A photograph of the $\sim$5~m long, two stage vacuum chamber used during the PRad experiment.}
  \label{fig:vacphoto}
\end{figure}
\subsection{Large volume vacuum chamber}
\begin{figure}[!ht]
  \centerline{\includegraphics[width=0.28\textwidth]{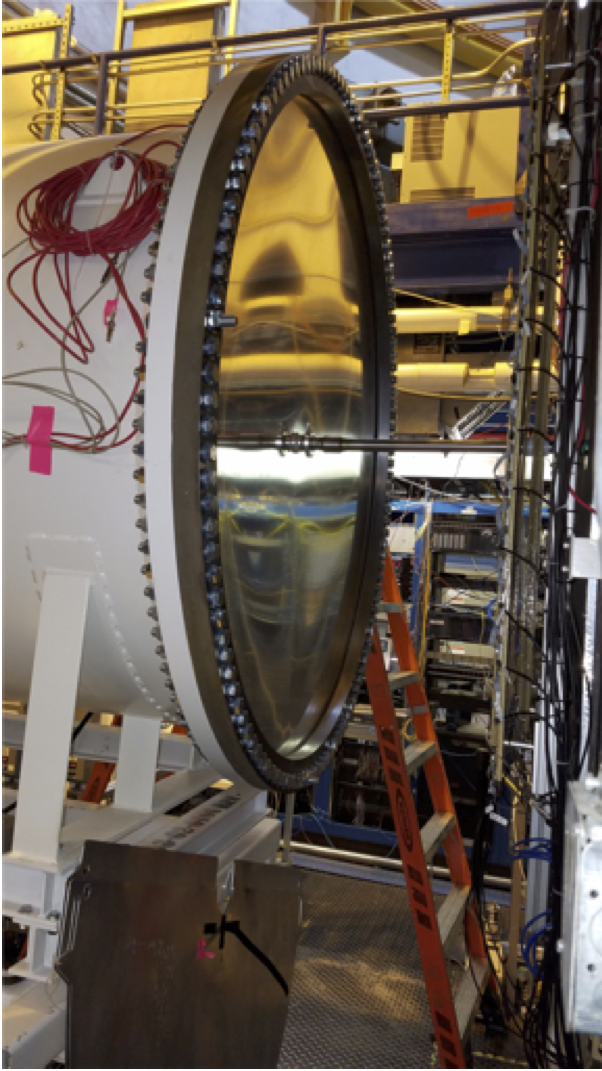}\hspace{5ex}\includegraphics[width=0.7\textwidth]{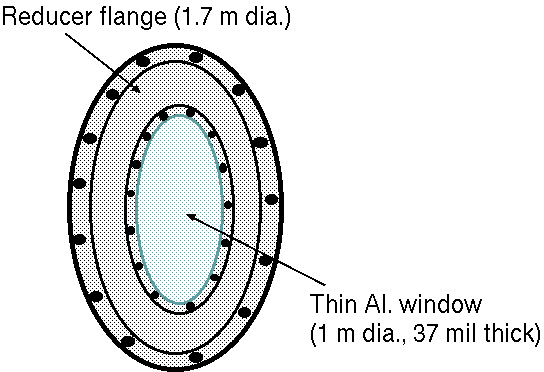}}
  \caption{A photograph of the 1.7~m diameter thin window
  at one end of the vacuum chamber (left). A sketch of the reducer flange and $1.0$~m diameter, $37$ mil thick Al. window (right).}
  \label{fig:vwin}
\end{figure}

For the PRad experiment a new large $\sim$ 5~m long, two stage vacuum chamber was designed and built. A photograph of the vacuum chamber is shown in Fig.~\ref{fig:vacphoto}. We will reuse this vacuum chamber with a new reducer flange such that a single $1.0$~m diameter, $37$ mil thick Al. window at the downstream end of the vacuum chamber, just before the GEM detector, can be used. Thinner window material will also be explored.
A photograph of the current 1.7~m diameter, 2~mm thick window is shown in
Fig.~\ref{fig:vwin} (left) along with a sketch of the reducer flange (right) which will allow a smaller, thinner window to be used for this proposed experiment.
A 2 inch diameter beam pipe will be
attached using a compression fitting to the center of the thin window. This vacuum chamber along with the PRad target chamber (without the target cell)
and a new extension piece connecting it to the super-harp will ensure that the electron beam does not encounter any additional material,
other than the target foil, all the way down to the Hall-B beam dump. The vacuum chamber will also help minimize multiple scattering of the charged particles en route to the detectors. 

\subsection{High resolution forward calorimeter}
\begin{figure}[!ht]
\centerline{
\includegraphics[width=0.75\textwidth]{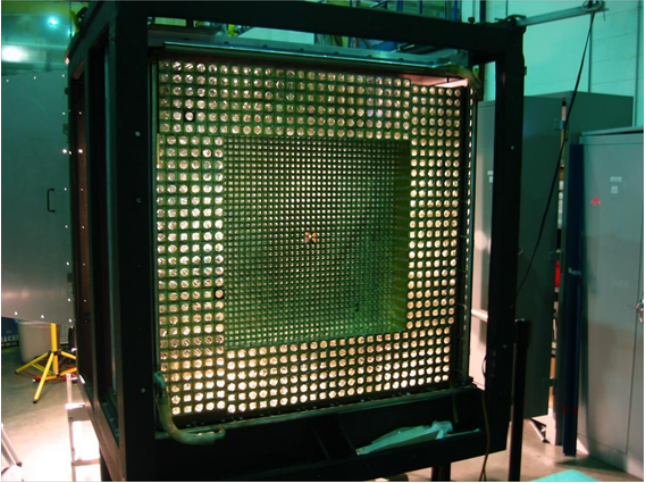}}
\caption{The PrimEx HyCal calorimeter with all modules of the high
performance \pbo ~crystals in place before installation of the light monitoring system.}
\label{fig:hycal}
\end{figure}
The scattered electrons  and electron-positron pair in this precision experiment will be detected with a high resolution 
and high efficiency electromagnetic calorimeter.
In the past decade, lead tungstate (\pbo{}) has became a popular inorganic 
scintillator material for precision compact electromagnetic calorimetry 
in high and medium energy physics experiments ({\em e.g.} CMS and ALICE at the LHC) 
because of its fast decay time, high density and high radiation hardness. 
The performance characteristics of the \pbo ~crystals are well known mostly 
for high energies ($>$10 GeV)~\cite{CMS94} and at energies below 1 
GeV~\cite{Mengel98}.
The PrimEx Collaboration at Jefferson Lab constructed a
novel state-of-the-art multi-channel electromagnetic hybrid (\pbo-lead glass)
calorimeter (HyCal)~\cite{primex-cdr} to perform a high precision (1.5\%) 
measurement of the neutral pion lifetime via the Primakoff effect.
The advantages of using the HyCal calorimeter was also demonstrated in the PRad experiment.

For this experiment we are proposing to use only the \pbo ~part of the calorimeter. A single \pbo ~module is 2.05 $\times$ 2.05 cm$^2$ in cross sectional
area and 18.0 cm in length (20 $X_0$).
The calorimeter consists of 1152 modules arranged in a 34 $\times$ 34
square matrix (69.7 $\times$ 69.7 cm$^2$ in size) with four crystal detectors 
removed from the central part (4.1 $\times$ 4.1 cm$^2$ in size) for passage 
of the incident electron beam. 
As the light yield of the crystal is highly temperature dependent
($\sim$ 2\%/$^{\circ}$C at room temperature), a special frame was developed
and constructed to maintain constant temperature inside of the calorimeter
with a high temperature stability ($\pm$ 0.1 $^{\circ}$C) during the experiments.
Figure~\ref{fig:hycal} shows the assembled PrimEx HyCal calorimeter that was used in the PRad experiment.  
For this experiment the calorimeter will be placed at a distance of about 5.5 m from the target just as in PRad.
\begin{figure}[!ht]
  \centerline{
    \includegraphics[width=0.9\textwidth]{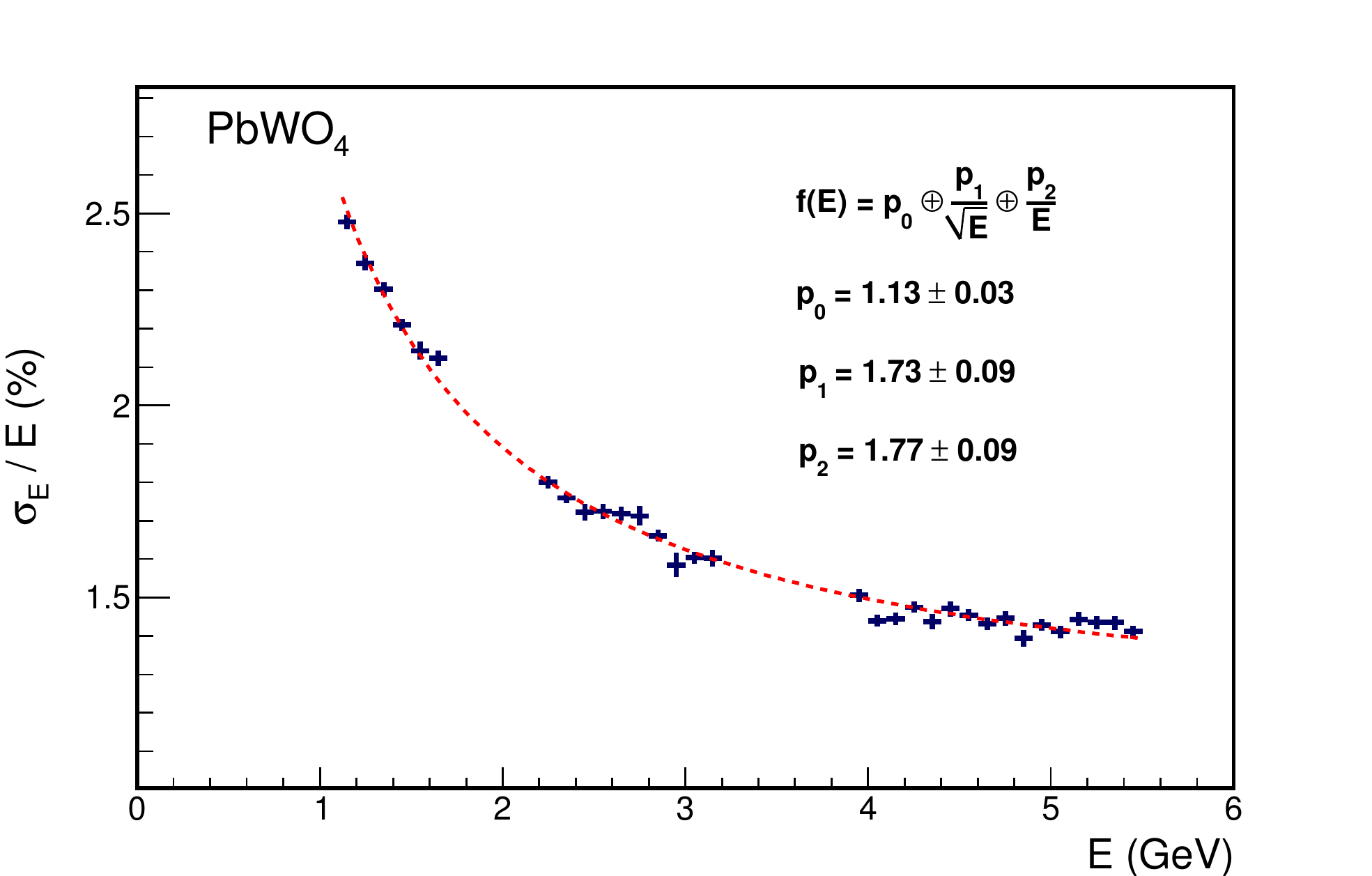}}
  \caption{Energy resolution of the \pbo{} crystal part of the HyCal calorimeter. These data are from the PrimEx experiment.}
  \label{fig:hycal_calib1}
\end{figure}
\begin{figure}[!ht]
  \centerline{
    \includegraphics[width=0.9\textwidth]{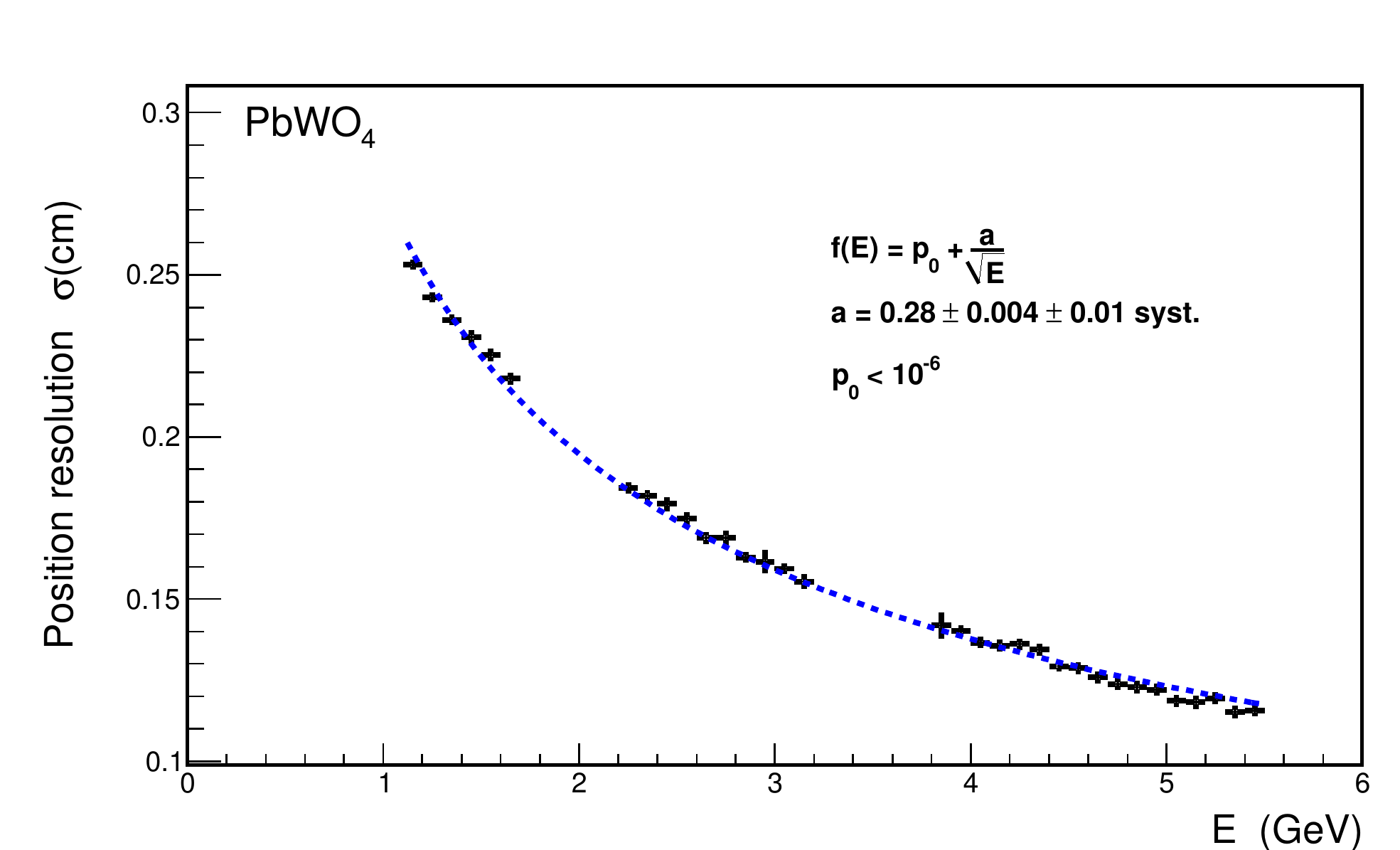}}
  \caption{Position resolution of the \pbo{} crystal part of the HyCal calorimeter. These data are from the PrimEx experiment.}
  \label{fig:hycal_calib2}
\end{figure}
\begin{figure}[!ht]
  \centerline{
\includegraphics[width=0.9\textwidth]{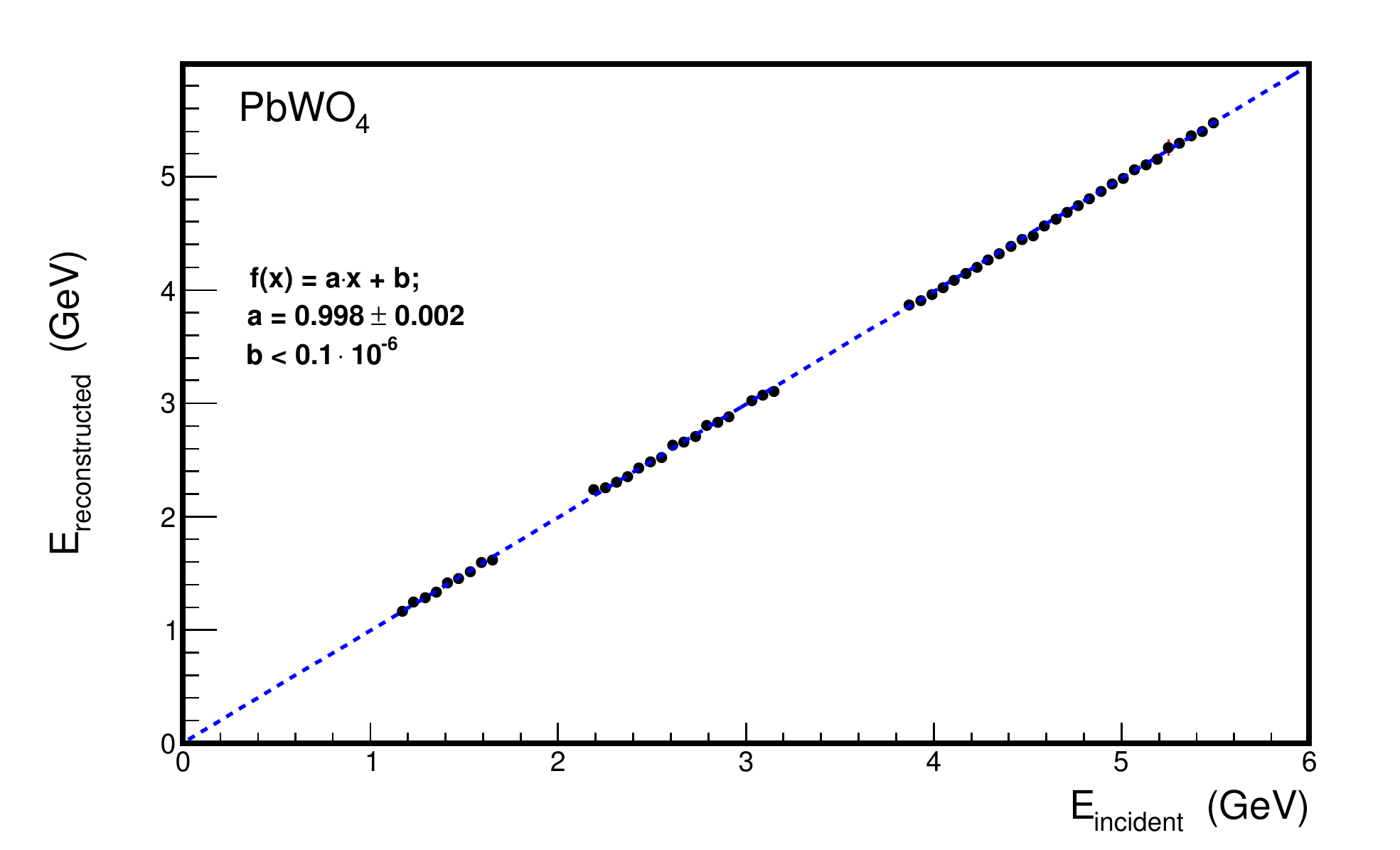}}
  \caption{The linearity of the detector response for the \pbo{} crystal part of the HyCal calorimeter. These data are from the PrimEx experiment.}
  \label{fig:hycal_calib3}
\end{figure}
\begin{figure}[!ht]
    \centering
    \includegraphics[width=0.9\textwidth]{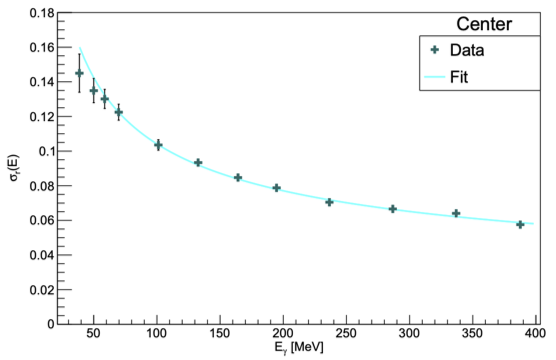}
    \caption{The measured relative resolution of the cluster energy response as function of the incident photon energy $E_{\gamma}$ as reported by a recent Mainz experiment~\cite{mainz_thesis}.}
    \label{fig:lowengres}
\end{figure}

During PrimEx the energy calibration of HyCal was performed by continuously irradiating the
calorimeter with the Hall B tagged photon beam at low intensity ($<$ 100 pA).
 An excellent energy resolution of $\displaystyle \sigma_E/E = 2.6\%/\sqrt{E}$
has been achieved by using a Gaussian fit of the line-shape obtained from the
$6 \times 6$ array.
The impact coordinates of the electrons and photons incident on the crystal
array were determined from the energy deposition of the electromagnetic shower
in several neighboring counters.
Taking into account the photon beam spot size at the calorimeter
($\sigma$=3.0 mm), the overall position resolution reached was
$\sigma_{x,y} = 2.5 ~{\rm mm}/\sqrt{E}$ for the crystal part of the
calorimeter. The calorimeter resolution achieved during the PrimEx experiment and the linearity of the detector response is shown in Figs.~\ref{fig:hycal_calib1}-\ref{fig:hycal_calib3}.

 A very recent measurement, Ref.~\cite{mainz_thesis}, on similar \pbo{} crystal detectors was performed at Mainz down to the $\sim30$ MeV energy range at room temperature (see Fig. ~\ref{fig:lowengres}), demonstrating the excellent 
performance of \pbo{} detectors at MeV energies. Their fit, 
  $\sigma_r(E) = \frac{2.92\%}{\sqrt{E}} +  1.18\%$, 
gives about 4\% resolution when extrapolated to 1 GeV, which is slightly higher than measurements at higher energies carried out during the PrimEx experiment. For example, the 2.7\% resolution at 1 GeV, shown in Fig.~\ref{fig:hycal_calib1} above, was performed with cooled crystals. The authors of the Mainz measurements explain this slight difference with similar arguments about the temperature of the crystals during the measurement.


\subsection{GEM based coordinate detectors}
\label{sec:GEM}

\begin{figure}[!hbt]
\centerline{
\includegraphics[width=0.35\textwidth]{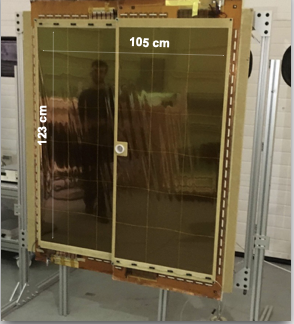} \includegraphics[width=0.45\textwidth]{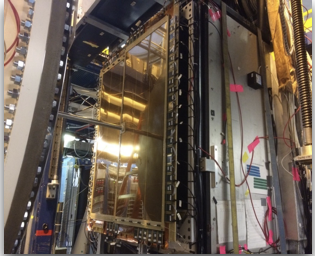}}
\caption{The PRad GEM chambers (left) and the GEM chambers mounted on the HyCal during the experiment (right).}
\label{fig:prad-gems}
\end{figure}

Two planes of 1.0 m $\times$ 1.0 m GEM detector layers separated by 10~cm  wide helium volume will be built by the UVa group for this experiment. These  GEM detector layers will be similar in their basic design to the GEM layer used in the PRad experiment. However, as described below, the GEM detectors for this experiment will be optimized to reduce the material thickness compared to the PRad GEMs. The GEM detectors  have been shown to achieve $\sim$ 70 $\mu$m resolution in the PRad experiment. Similar to the PRad experiment, a custom designed through hole with a 4 cm radius will be  built into  GEM detectors at the center of the  active area for the passage of the beam-line.  The 10 cm thick helium box  between the two GEMs will be built with no front and back windows. The  two GEMs will be mounted on the front and back completing the helium volume. The hole for the beam-line will continue through a pipe in the helium box.  The chamber-tracker box will be mounted to the front face of the HyCal calorimeter using a custom mounting frame.

The material thickness of the two GEMs will be reduced using the following approaches.
\begin{itemize}
    \item Each GEM will be assembled as a double GEM detector with two GEM foils, as opposed to a triple GEM detector. Double GEM detectors have been demonstrated to work successfully in the past.
    \item The GEMs will be operated using a helium based gas mixture as opposed to an argon based gas mixture. 
    \item We will build each GEM without using the honeycomb support layer underneath the bottom readout plane. In a standard GEM detector this support layer holds the slightly elevated pressure inside the detector against the atmosphere. However, in the setup proposed here, the readout layer of each GEM sits against the helium box volume which will be maintained at the same pressure as inside the GEMs. As such,  the support layers could be eliminated in this case reducing the material thickness by over 0.2\% r.l. A large area GEM prototype based on this principle was successfully built and beam tested by the UVa group as part of EIC detector R\&D.
\end{itemize}

With these optimizations the material thickness of single GEM will be reduced to be about 0.3\%, as opposed to about 0.6\% for a standard GEM. With this, the total material thickness for the whole tracker, including the two GEMs and the helium volume, will be approximately 0.6\%.

 In addition to the above improvements, we will explore the possibility of building the GEMs using ultra-thin chromium GEM foils as opposed to standard copper GEM foils. A conventional GEM foil consists of an insulator made of a thin Kapton foil (about 50 $\mu$m) sandwiched between two layers of  copper (each about 5 $\mu$m thick). The new chromium GEM  foils, recently developed at the CERN GEM workshop, have the  copper layers of the foil removed, leaving only a 0.2  $\mu$m   layer of chromium on either side of the Kapton. The material thickness of a GEM module  made of copper-less foils is about 0.2\% radiation lengths. The UVa group has fabricated and successfully operated chromium GEM detectors in beam test runs at Fermilab. 

\begin{figure}[!ht]
  \centerline{
    \includegraphics[height=0.6\textwidth]{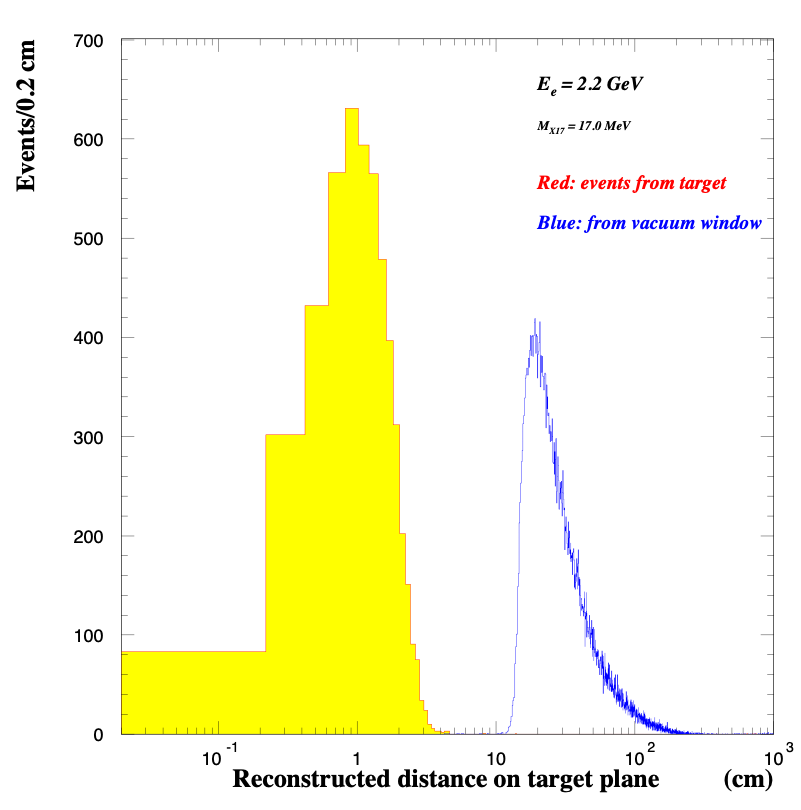}}
  \caption{Events originating from the target and the events originating at the vacuum chamber exit window as reconstructed using tracking information from the GEM detectors.}
  \label{fig:tracking}
\end{figure}

The  GEM detectors will have 2D X-Y strip readout. The readout of the two  GEM layers requires approximately 15 k electronic channels.  This readout for the proposed experiment will be done by using the high-bandwidth optical link based MPD readout system recently   developed  for the SBS program in Hall A.  This system is currently under rigorous resting. This new system uses the APV-25 chip used in the PRad GEM readout. However, the readout of the digitized data is performed over a high-bandwidth optical link to a  Virtual Trigger Processor (VTP)  unit in a CODA DAQ setup. 

Using the GEM layers for tracking the events originating from the target can be distinguished from those originating from the vacuum chamber exit window as shown in Fig.~\ref{fig:tracking}.

\subsection{Electronics, data acquisition, and trigger}
\label{subsec:trigger}

The high resolution calorimeter in this proposed experiment will  have around 1152 channels of charge and timing information.  These  will be readout using the JLab designed and built flash-ADC modules (FADC250),  each with  16 channels. The DAQ system for the calorimeter is thus composed of 72 FADC250 modules that can be held in 5 16-slot VXS crates. The major advantages of the flash-ADC based readout are the simultaneous pedestal measurement (or full waveform in the data stream), sub-nanosecond timing resolution, fast readout speed, and the pipeline mode that allows more sophisticated triggering algorithms such as cluster finding.
Additionally, some VME scalers will be read out and periodically inserted into the data stream.
The total sum will be performed by the new FADC system, however, the completely separate NIM based system for forming the trigger will be left intact. This system using the dynode signals from each shower detector is an integral parts of the HyCal calorimeter which will provide redundancy and extra monitoring.  


\begin{figure}[!ht]
  \centerline{
    \includegraphics[height=0.5\textwidth]{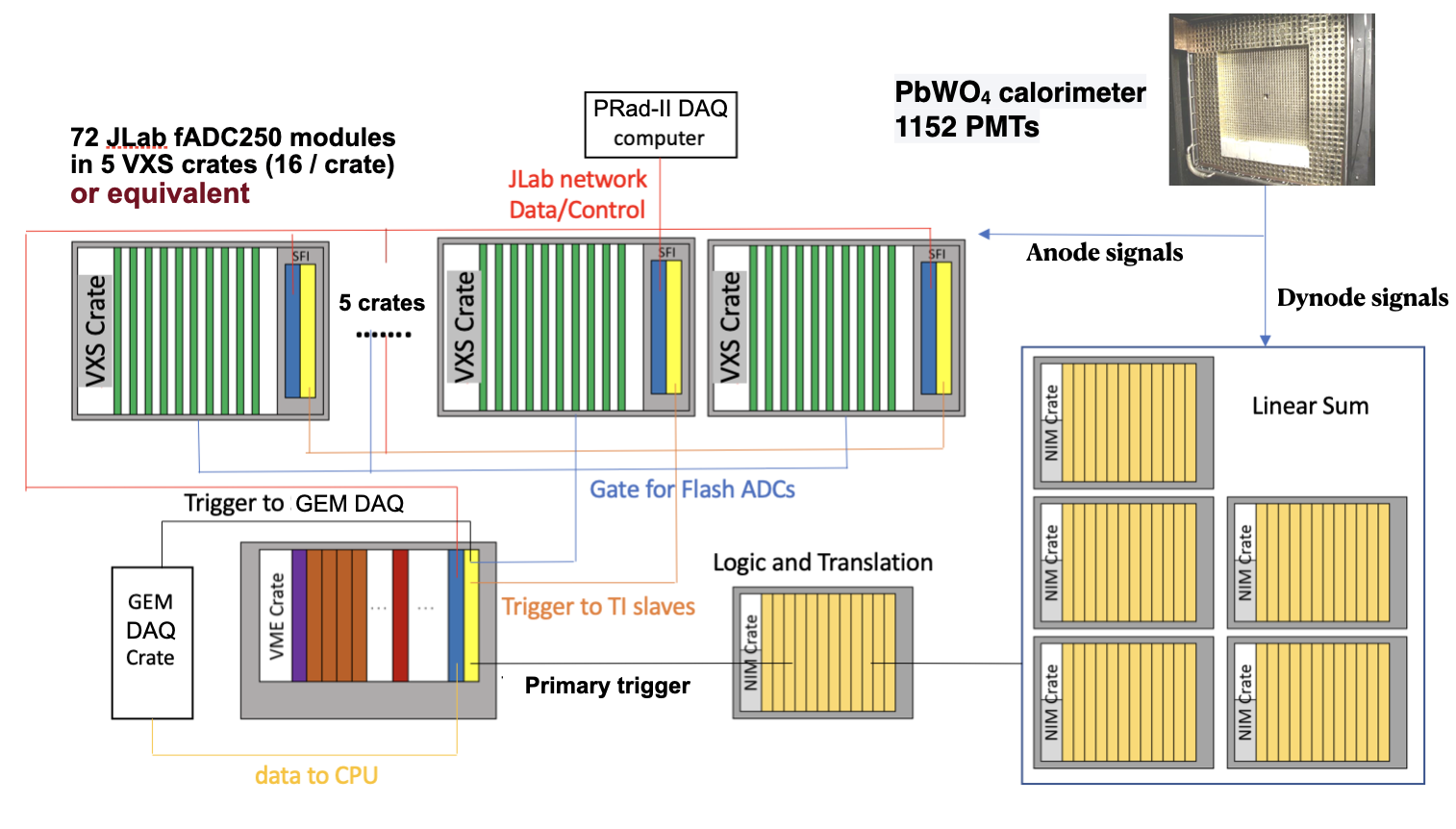}}
  \caption{A modern FADC based DAQ will be used with the primary trigger formed using the sum of all dynode outputs from each of the crystals.}
  \label{fig:daq}
\end{figure}

The DAQ system for the proposed experiment is the standard JLab CODA
based system utilizing the JLab designed Trigger Supervisor.
A big advantage of the CODA/Trigger Supervisor system is the ability
to run in fully buffered mode.
In this mode, events are buffered in the digitization modules themselves
allowing the modules to be ``{\it live}'' while being readout.
This significantly decreases the deadtime of the experiment. 
With the upgraded FADC modules we expect to reach a data-taking rate of about 25 kHz 
events. Such a capability of the DAQ system has already been demonstrated by CLAS12 experiments.


A large fraction of the electronics needed for the DAQ and 
trigger, including the high voltage crates and all necessary cabling for 
the detectors, are available in Hall B from the PRad experiment.

The primary (first-level) trigger will use an analog sum of the dynode outputs from the \pbo{} part of the calorimeter (see Fig.~\ref{fig:daq}).
This trigger will require that the total energy deposited in the \pbo{} crystals is greater than 0.7$ \times $E$_{beam}$.
A second-level trigger will be employed in the FADC through a fast, rudimentary clustering algorithm.
This trigger will check that there are at least three clusters with energy in the range of (0.02 - 0.85) $\times$ E$_{beam}$.
After this trigger, only clusters that meet the energy criteria will be recorded.
The GEM detectors will not be included in the trigger, which will allow for both charged and neutral clusters to be recorded.
A single crystal cluster separation resolution is assumed, as demonstrated in the PrimEx and PRad experiments.
\begin{figure}[hbt!]
  \centerline{
    \includegraphics[height=0.65\textwidth]{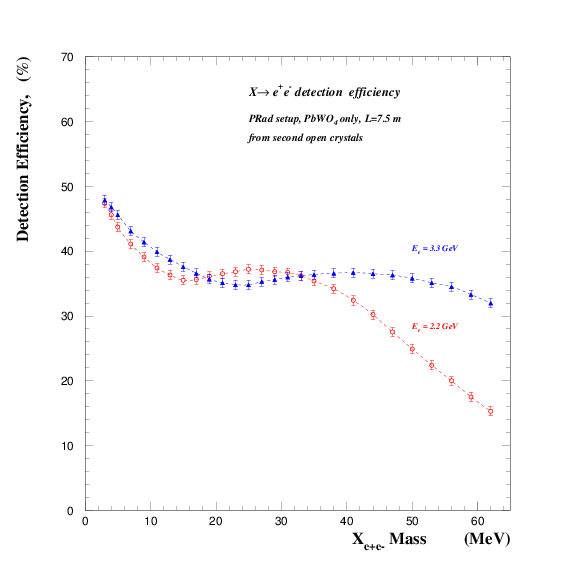}}
  \caption{The efficiency of detecting the scattered electron with energy between (0.03 - 0.7) $\times$ E$_{beam}$ along with the $e^{+}e^{-}$ pair in the \pbo ~calorimeter for E$_{beam}$ = 2.2 and 3.3 GeV. }
  \label{fig:det_eff}
\end{figure}
\begin{figure}[hbt!]
  \centerline{
    \includegraphics[height=0.5\textwidth]{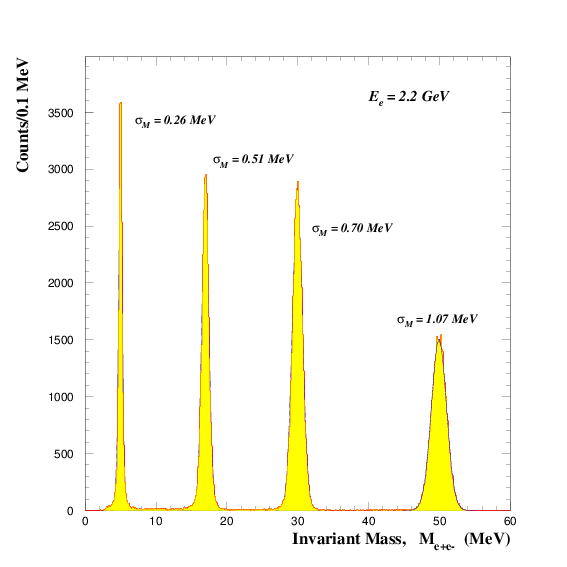}\includegraphics[height=0.5\textwidth]{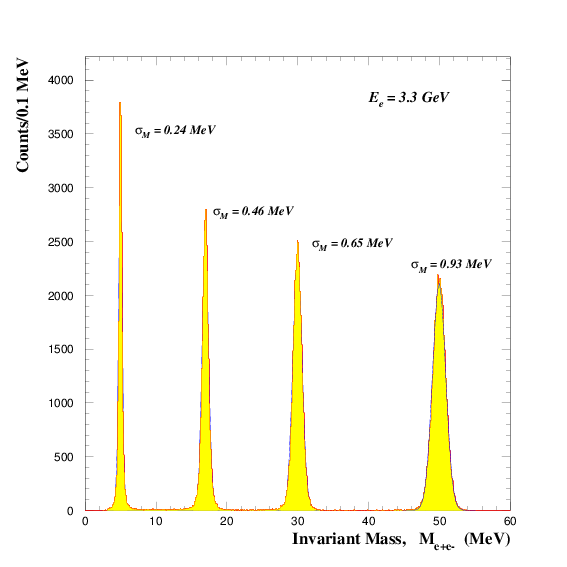}}
  \caption{The $e^{+}e^{-}$ invariant mass resolution at 2.2 GeV (left) and 3.3 GeV (right).}
  \label{fig:full_res}
\end{figure}

\subsection{Detection Efficiency and Resolutions}
\label{subsec:eff_res}
Fig.~\ref{fig:det_eff} shows the efficiency of detecting 3-cluster events in the \pbo ~calorimeter with the scattered electron having energy between (0.03 - 0.7) $\times$ E$_{beam}$ and detected along with the $e^{+}e^{-}$ pair from the $X$ decay  for E$_{beam}$ = 2.2 and 3.3 GeV. The two innermost layers of HyCal are excluded. It also includes the effect of displaced decay vertex on the geometrical acceptance (important only at the low mass and low $\epsilon^2$ range). This shows that the proposed experimental setup is sensitive to the 3 - 60 MeV mass range. At lower beam energies the detection efficiency has a steep fall-off with invariant mass. Although the 2.2 GeV detection efficiency is lower than the 3.3 GeV efficiency it will serve as a systematic check and
will also be used to boost the combined statistical significance.

\begin{figure}[h!]
  \centerline{
    \includegraphics[height=0.5\textwidth]{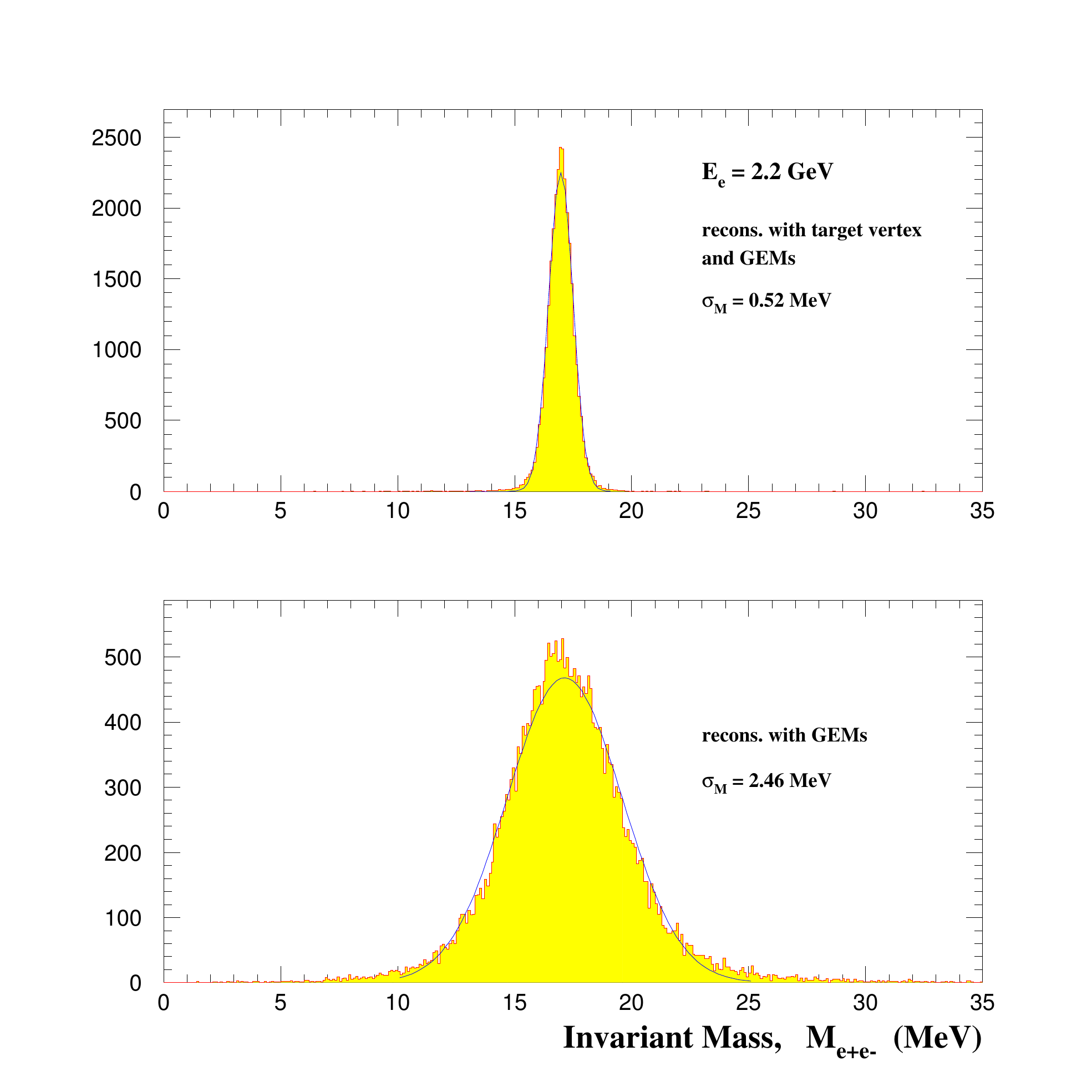}\includegraphics[height=0.5\textwidth]{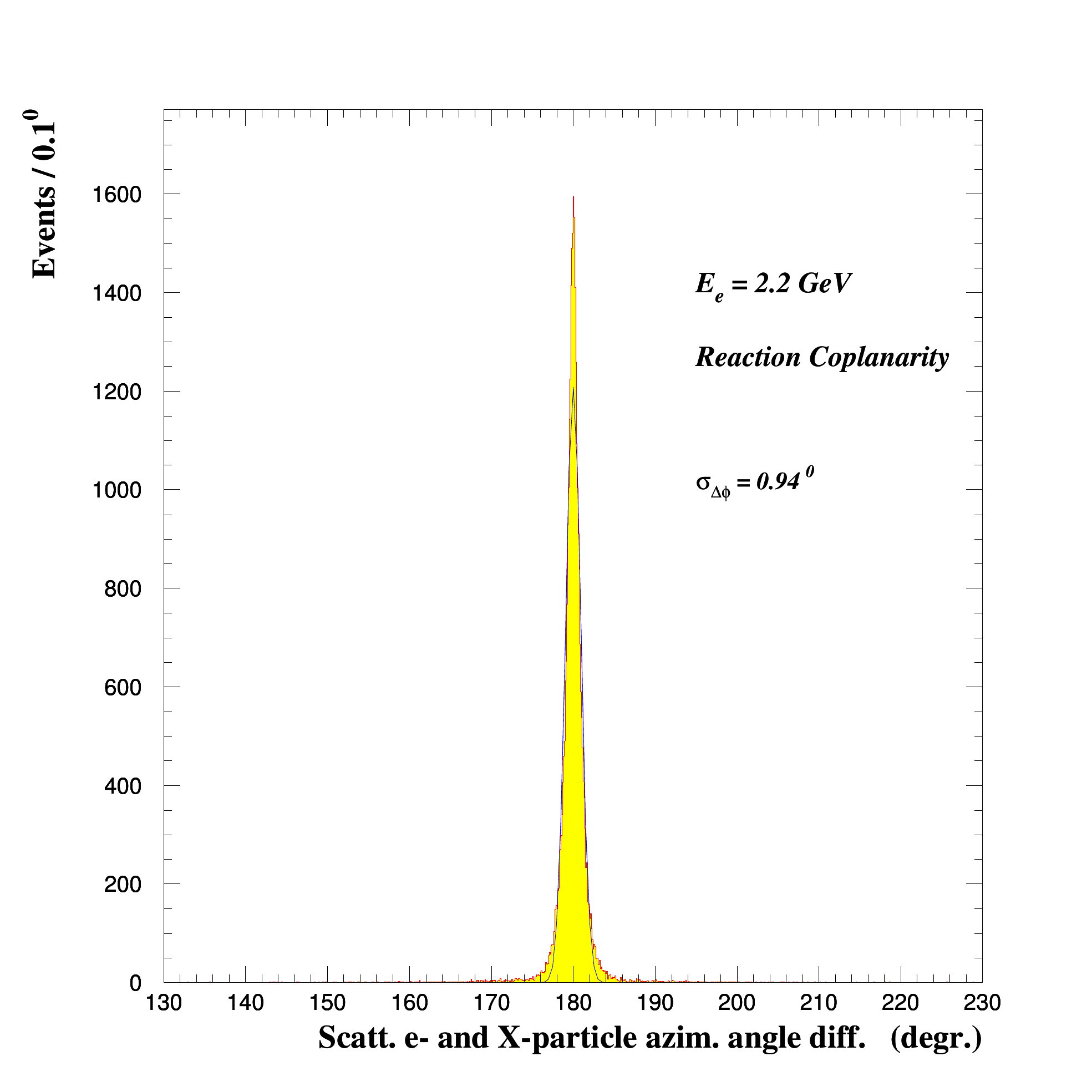}}
  \caption{(left) The invariant mass resolution at 2.2 GeV with vertex reconstruction (top) and without vertex reconstruction (bottom). (right) The difference in azimuthal angle between scattered electron and $X$ particle. The azimuthal angle resolution of $\sim$ 0.9$^{\circ}$ will be used to ensure the coplanarity of the $X$ particle production.}
  \label{fig:novertex}
\end{figure}

The invariant mass resolution for a hypothetical $X$ particle with a mass of 5, 17 and 25 MeV is shown in Fig.~\ref{fig:full_res} when using the calorimeter, GEM and vertex reconstruction with beam energy of 2.2 and 3.3 GeV is shown in Fig.~\ref{fig:full_res}. The impact of the vertex reconstruction with the HyCal and the two GEM planes is demonstrated in Fig.~\ref{fig:novertex} (left) where the invariant mass resolution with (top) and without (bottom) the vertex reconstruction is shown for 2.2 GeV beam energy. As can be seen in these plots, the invariant mass resolution is significantly improved when using the target vertex. Using the GEMs alone, however, will serve as an additional systematic check of our results. Events that originate in the target will have the same centroid for the reconstructed invariant mass using both methods. This will help to identify false peaks in our analysis. The azimuthal angle resolution is $\sim$~0.9$^{\circ}$, shown in Fig.~\ref{fig:novertex} (right). This will help ensure the coplanarity of the $X$ particle production.
 The coplanarity cut will be used to reject random uncorrelated clusters. The advantage of this cut can be seen in the analysis 
of PRad data documented in Appendix A of the proposal (Figs.~\ref{fig:IM_EC}-\ref{fig:IM_coplanar}).


\section{Monte Carlo Simulations of the Background}\label{sec:MC}

\begin{figure}[hbt!]
  \centerline{
    \includegraphics[height=0.35\textwidth]{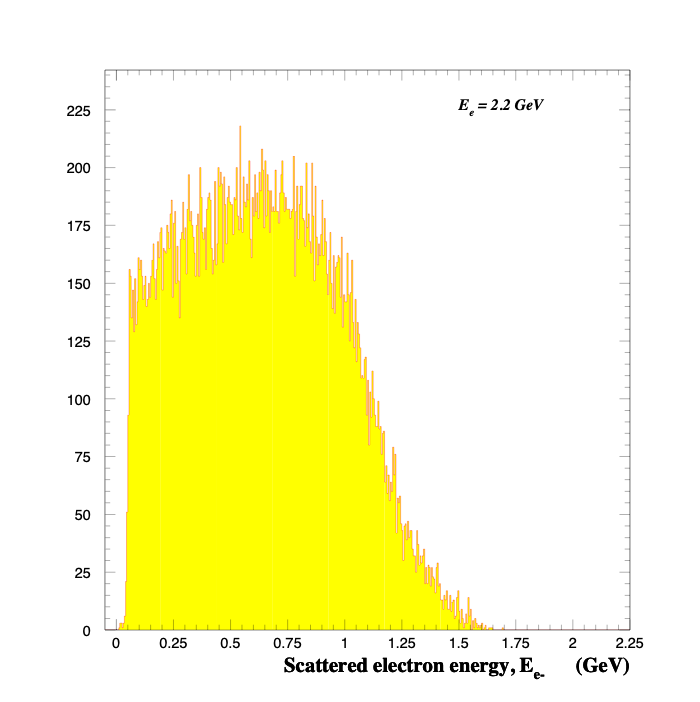}
    \includegraphics[height=0.35\textwidth]{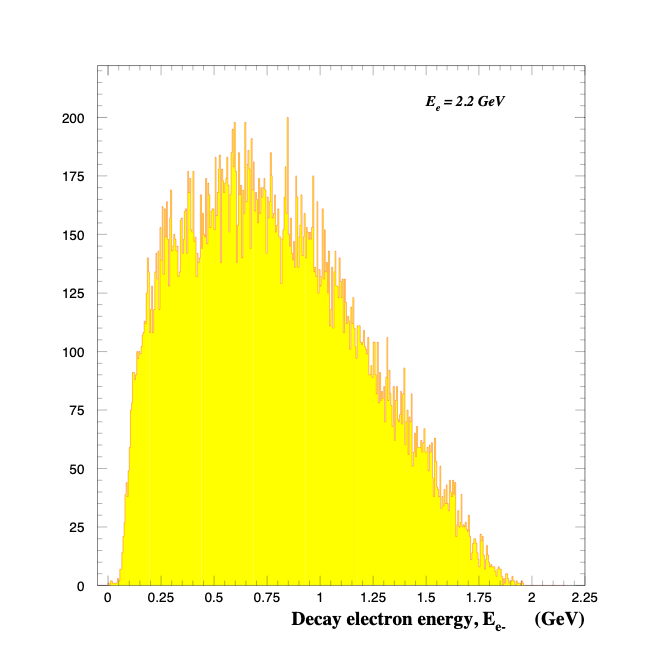}
    \includegraphics[height=0.35\textwidth]{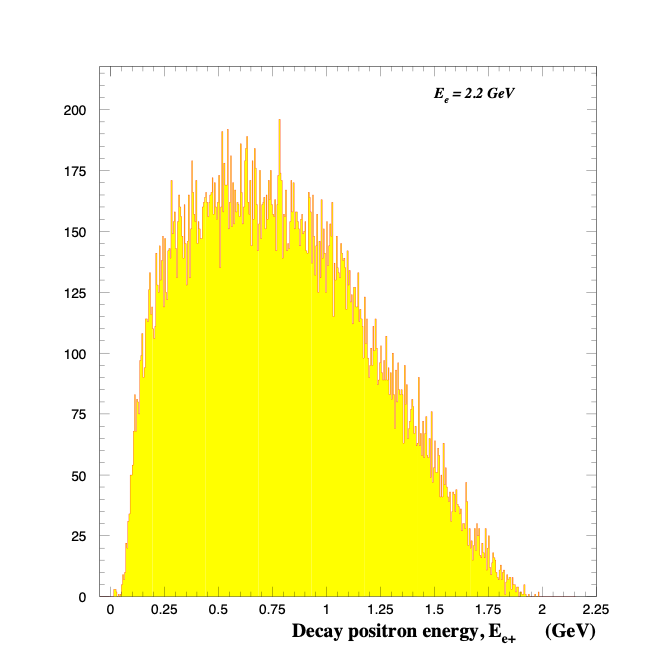}}
  \caption{The simulated scattered electron energy spectrum (left) and the energy spectrum of the electron and positron from the decay of the $X$ particle.}
  \label{fig:energy_spect}
\end{figure}
A comprehensive simulation of the experiment was carried out using the Geant3 and Geant4 simulation packages developed for the PRad experiment. This simulation takes into account realistic geometry of the experimental setup, and detector resolutions. The generated scattering events were propagated within the Geant simulation packages, which included the detector geometry and materials of the PRad setup. This enabled a proper accounting of the external Bremsstrahlung of particles passing through various materials along its path. The simulation included light propagation and digitization of the simulated events. These steps are critical for the precise reconstruction of the position and energy of each event in the \mbox{HyCal}. We consider the Geant4 simulation to be a more complete simulation of the experiment, with the Geant3 simulation serving as a crosscheck. The $X$ production and decay into $e^{+}e^{-}$ was generated according to the rate equation shown in Sec.~\ref{sec:method} with the $X$ produced along the virtual photon direction. The decay length was also taken into account. The simulated scattered electron spectrum as well as the energy spectrum of the $e^{+}e^{-}$ decay of the $X$ particle are shown in Fig.~\ref{fig:energy_spect} for the 2.2 GeV beam energy.
\begin{figure}[hbt!]
  \centerline{
    \includegraphics[height=0.35\textwidth]{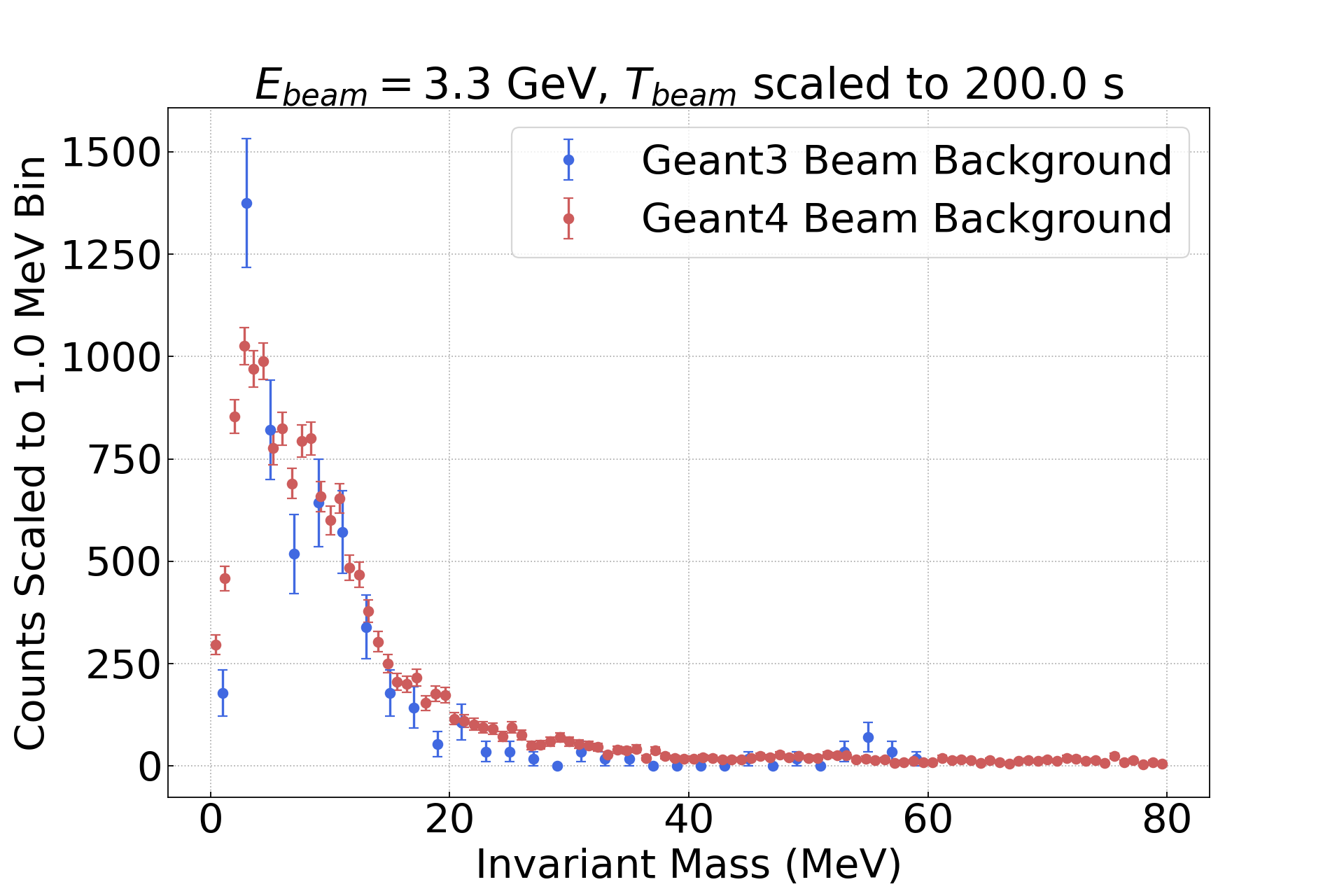}
    \includegraphics[height=0.35\textwidth]{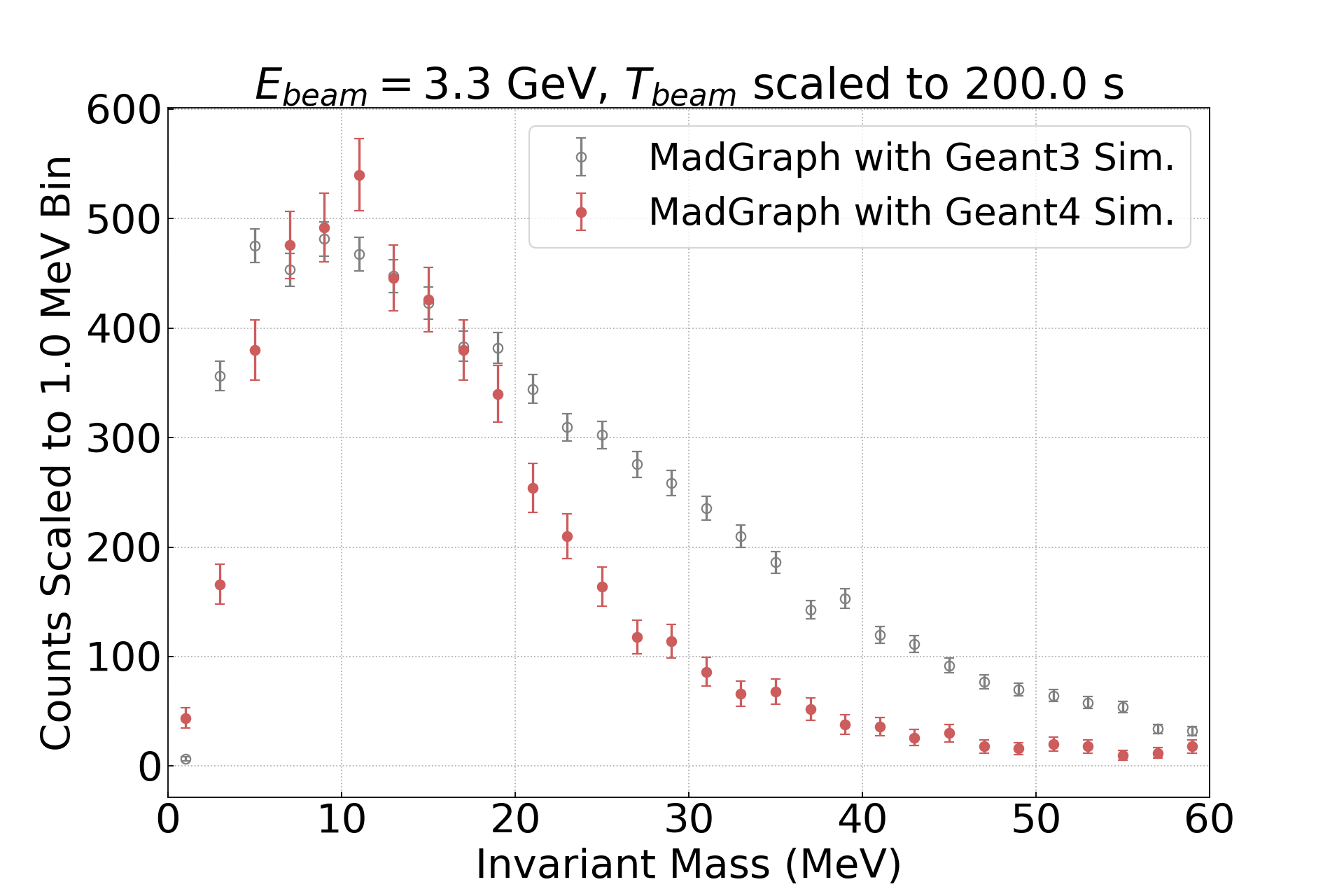}}
  \caption{(left) The simulated background using the Geant3 (blue) and Geant4 (red) background generators. The red points were generated for 200 s of 100 nA beam at 3.3 GeV (corresponding to 1.25$\times$10$^{14}$ electrons on target) while the black points were generated for 3.3 s and scaled to 200 s. (right) The simulated background using the Madgraph5 generator in the Geant3 (black) and Geant4 (red) simulation of the experiment, both for beam time equivalent to 200 s. }
  \label{fig:sim_bgd}
\end{figure}

The $ep$ elastic and $e-e$ M{\o}ller  generators developed for the PRad experiments were
used to verify that these background processes are kinematically suppressed. We have also simulated the Bethe-Heitler and the radiative background processes. The Bethe-Heitler background process is kinematically suppressed and the radiative process is the irreducible background.  The background was simulated for about 200 seconds of 3.3 GeV electron beam with a current of 100 nA (corresponding to 1.25$\times$10$^{14}$ electrons on target). The reconstructed invariant mass $M_{e^{+}e^{-}}$ spectrum for these simulated events are shown in Fig.~\ref{fig:sim_bgd} (left) , where the backgrounds from the Geant3 and Geant4 simulations are compared and found to be consistent with each other. 
An alternate simulation was performed using the MadGraph5 event  generator~\cite{Conte:2012fm}, that was used by the HPS collaboration~\cite{hps_prop2}. 

\begin{figure}[hbt!]
    \centering
    {\includegraphics[width=0.45\textwidth]{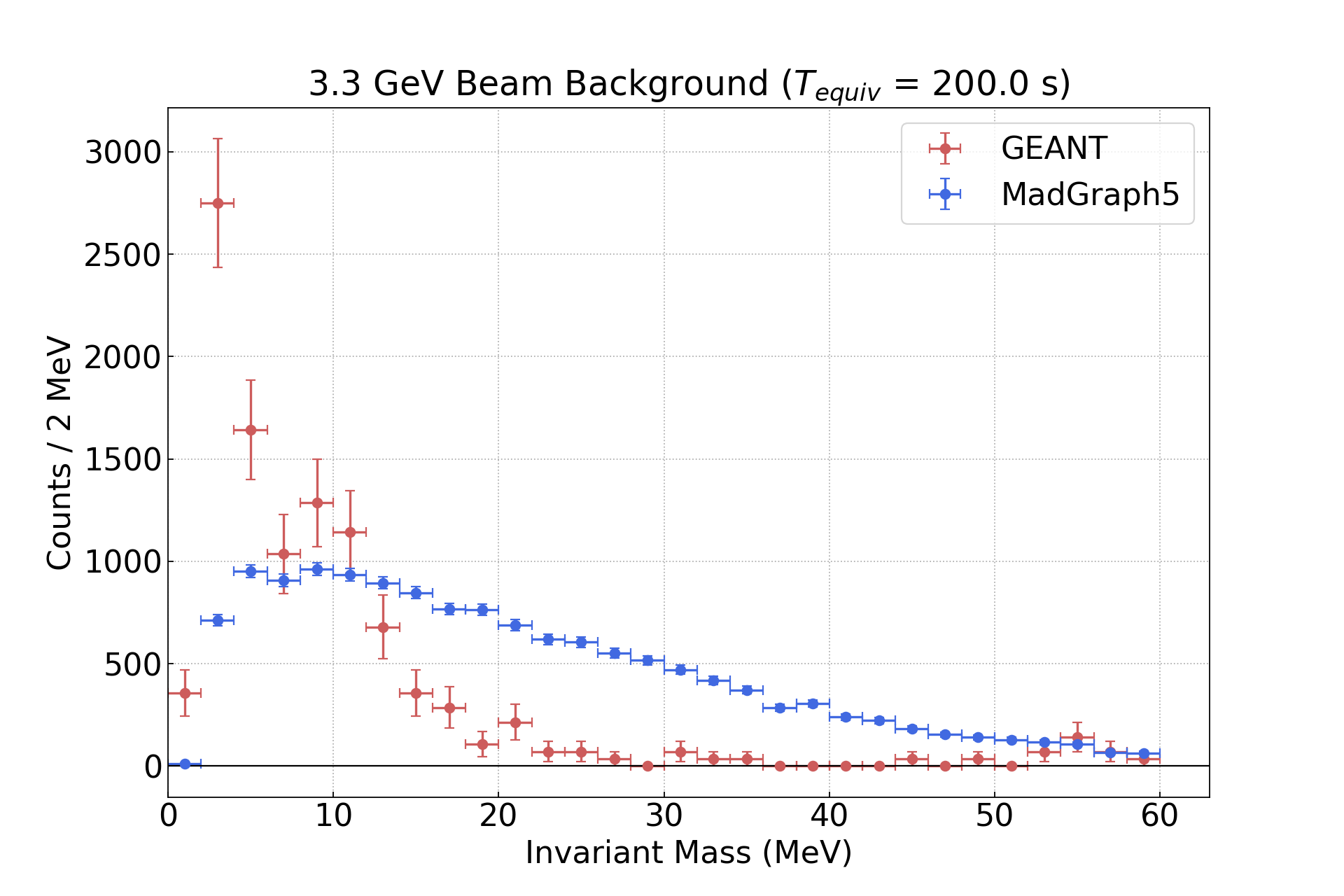}
    \includegraphics[width=0.45\textwidth]{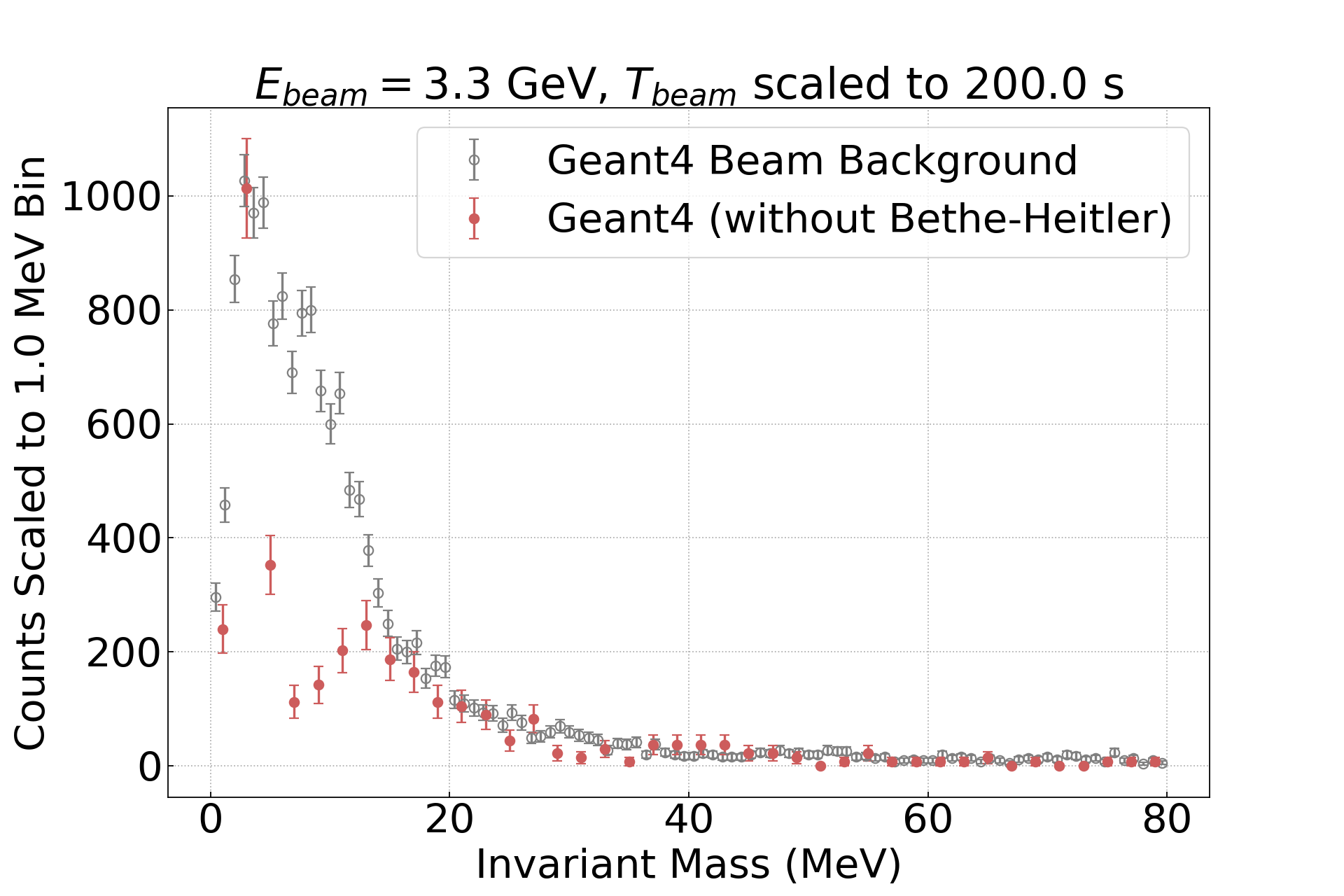}}
    \caption{(left) Comparison of the simulated background distributions performed by Geant (red points) and by MadGraph5 (blue points), they 
    are normalized to the same 200 seconds of equivalent beam time. (right) The background with the Geant4 generator with the Bethe-Heitler process turned off (red) compared the full Geant4 simulations (blue), also for 200 s of beam time.}
    \label{fig:sim_compare}
\end{figure}
\begin{figure}[hbt!]
    \centering
    {\includegraphics[width=0.45\textwidth]{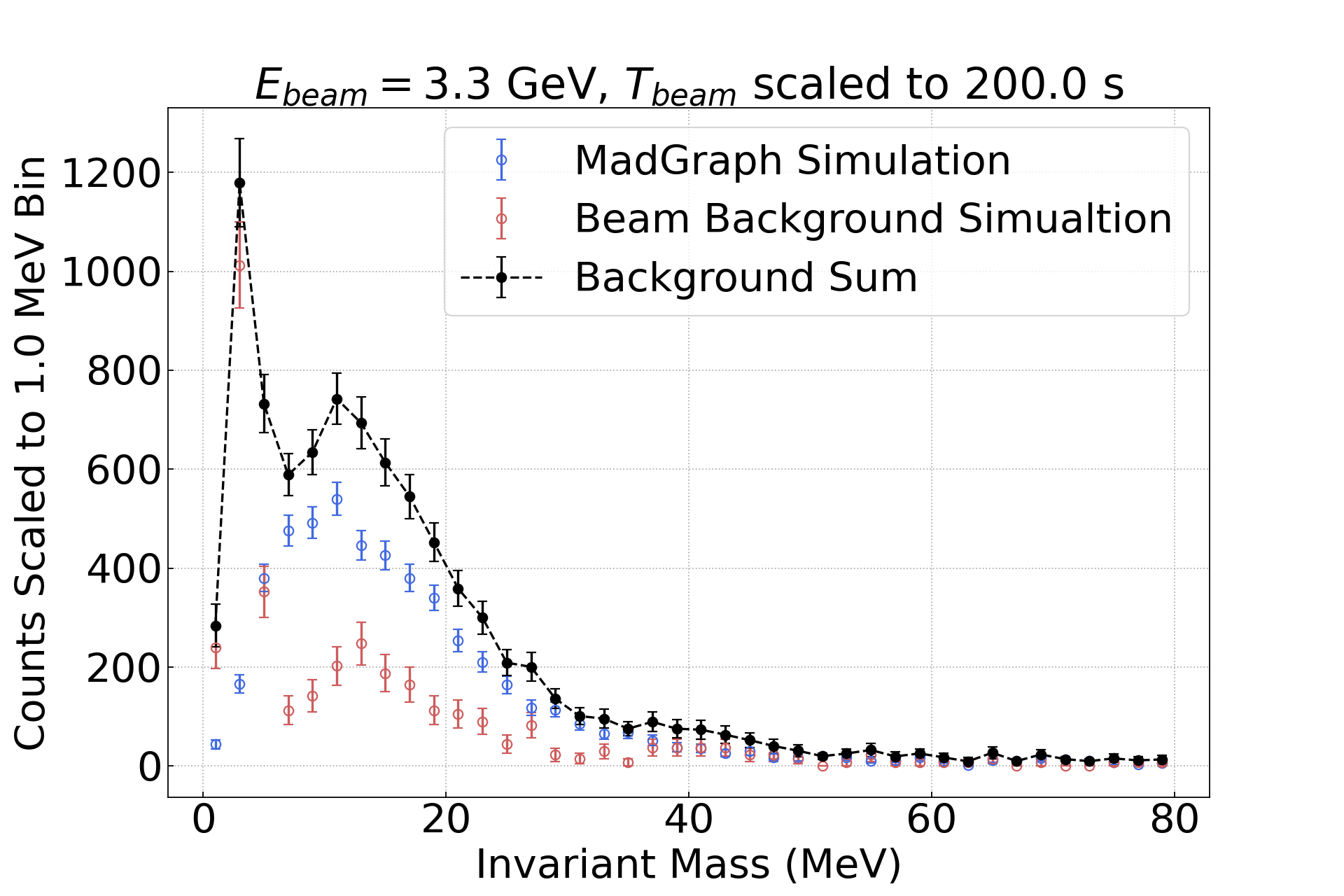}
    \includegraphics[width=0.45\textwidth]{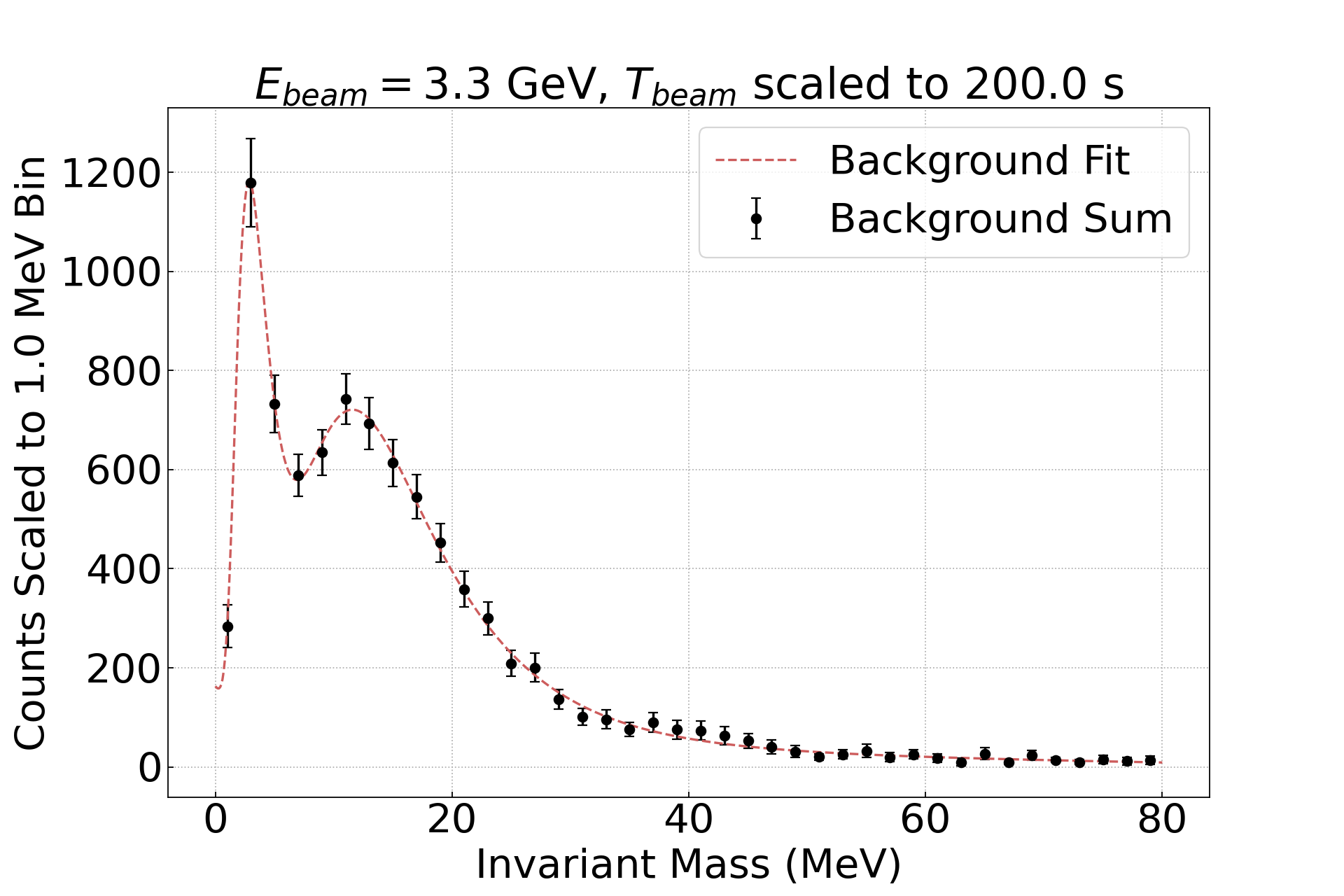}}
    \caption{(left)The background generated by the hybrid generator (black), along with the two components of the hybrid generator, Madgraph5 (blue) and Geant4 excluding Bethe-Heitler (red) for 200 s of equivalent beam time. (right) The simulated background from the hybrid generator (black points) was fit to a sum of two Landau distribution a log distribution and a constant distribution (red solid line). }
    \label{fig:sim_hyb}
\end{figure}

Using MadGraph5 we generated a large sample (2M events) of trident background events including the Bethe-Heitler, radiative, and interference 
processes. These events were fed into the same two Geant simulation 
packages that were used to trace them through the 
experimental setup and detection of events. 
The resulting simulated background for 200 seconds of equivalent beam time is shown in Fig.~\ref{fig:sim_bgd} (right), where once again the 
Geant3 and Geant4 simulations outputs are compared and found to be consistent with each other. However, as seen in Fig.~\ref{fig:sim_compare}, the Geant and Madgraph5 background generators seem to disagree in the mass range of interest. It is expected that Madgraph5 is the better generator for the Bethe-Heitler radiative trident background. Therefore we have created a hybrid generator by combining the Madgraph5  and the Geant4 (with Bethe-Heitler turned off) generators to utilize the best features of the two generators. The background generated from this hybrid generator is shown in  Fig.~\ref{fig:sim_hyb} (right) for 200 seconds of equivalent beam time.
\begin{figure}[hbt!]
    \centering
    {\includegraphics[width=0.6\textwidth]{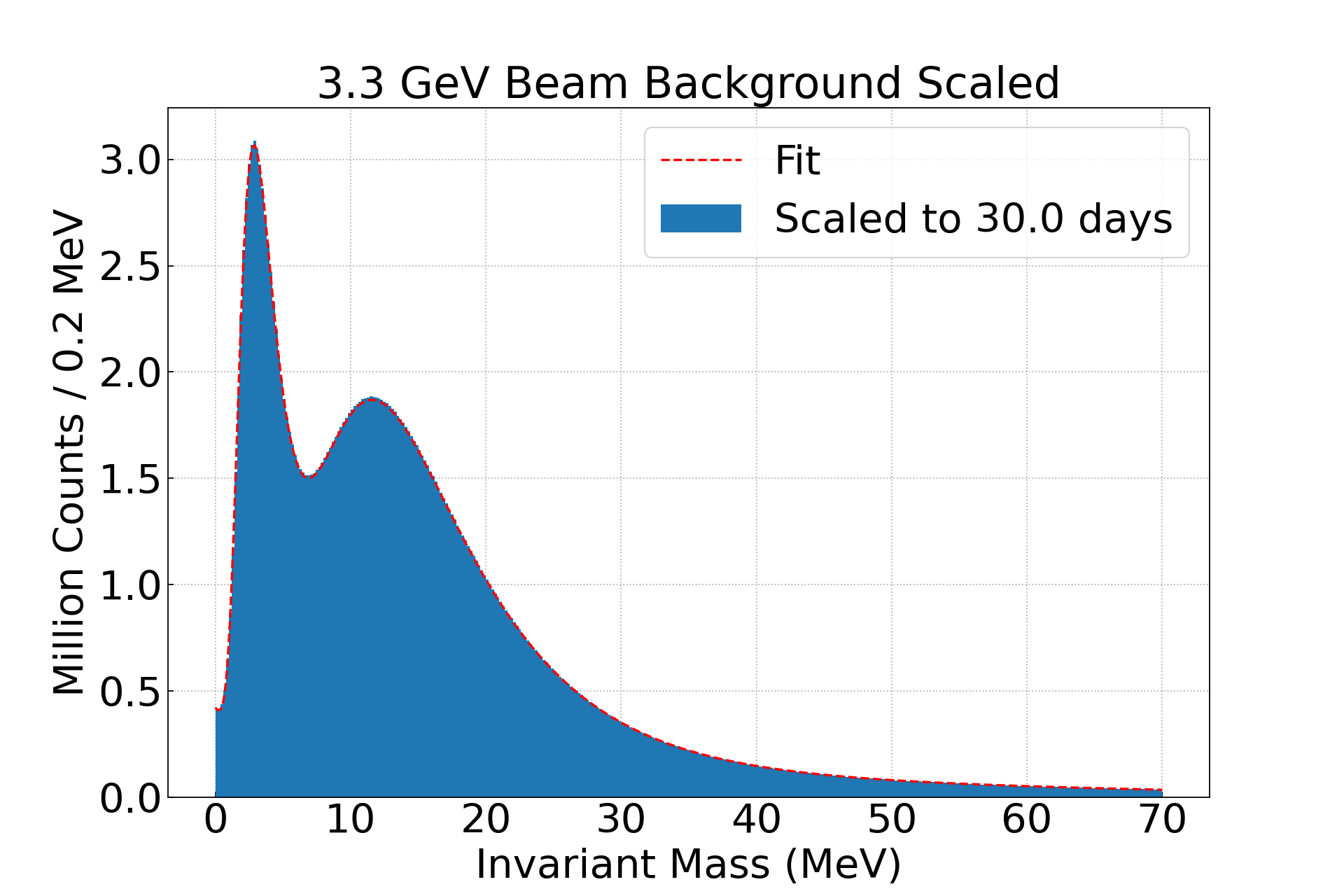}}
    \caption{The simulated background scaled for 30 days of 100 nA beam with beam energy of 3.3 GeV.}
    \label{fig:sim_hyb_scaled}
\end{figure}
The simulated background was fit to a sum of a Landau distribution a log distribution and a constant distribution. The fit was used to scale the background by sampling the number of events bin-by-bin to give the expected background for 30 days of 3.3 GeV beam at 100 nA, as shown in Fig.~\ref{fig:sim_hyb_scaled}. The shape of the background can be validated by comparing it to the background using similar event selection of PRad data (see Figs.~\ref{fig:IM_coplanar} and \ref{fig:PRadH_MTsub} in Appendix A and B respectively).


These background distributions were used to estimate the epsilon {\it vs.} mass sensitivity. 

%

The formula used to evaluate the background and estimate the $\epsilon^2$ sensitivity is the same one used in the APEX and HPS  proposals~\cite{apex, hps}. Note that it is standard practice in such search experiments to use a sensitivity limit of 2-2.4$\sigma$~\cite{apex, hps}, however, we have used a sensitivity limit of 5$\sigma$ (discovery criterion used by PDG), which make our estimates more conservative compared to previously approved proposals.
We will develop and apply a more sophisticated statistical analysis for the actual high-statistics experimental data set (if approved). \\


\section{Beam Time Request and Statistics}
\begin{figure}[hbt!]
  \centerline{
  \includegraphics[height=0.45\textwidth]{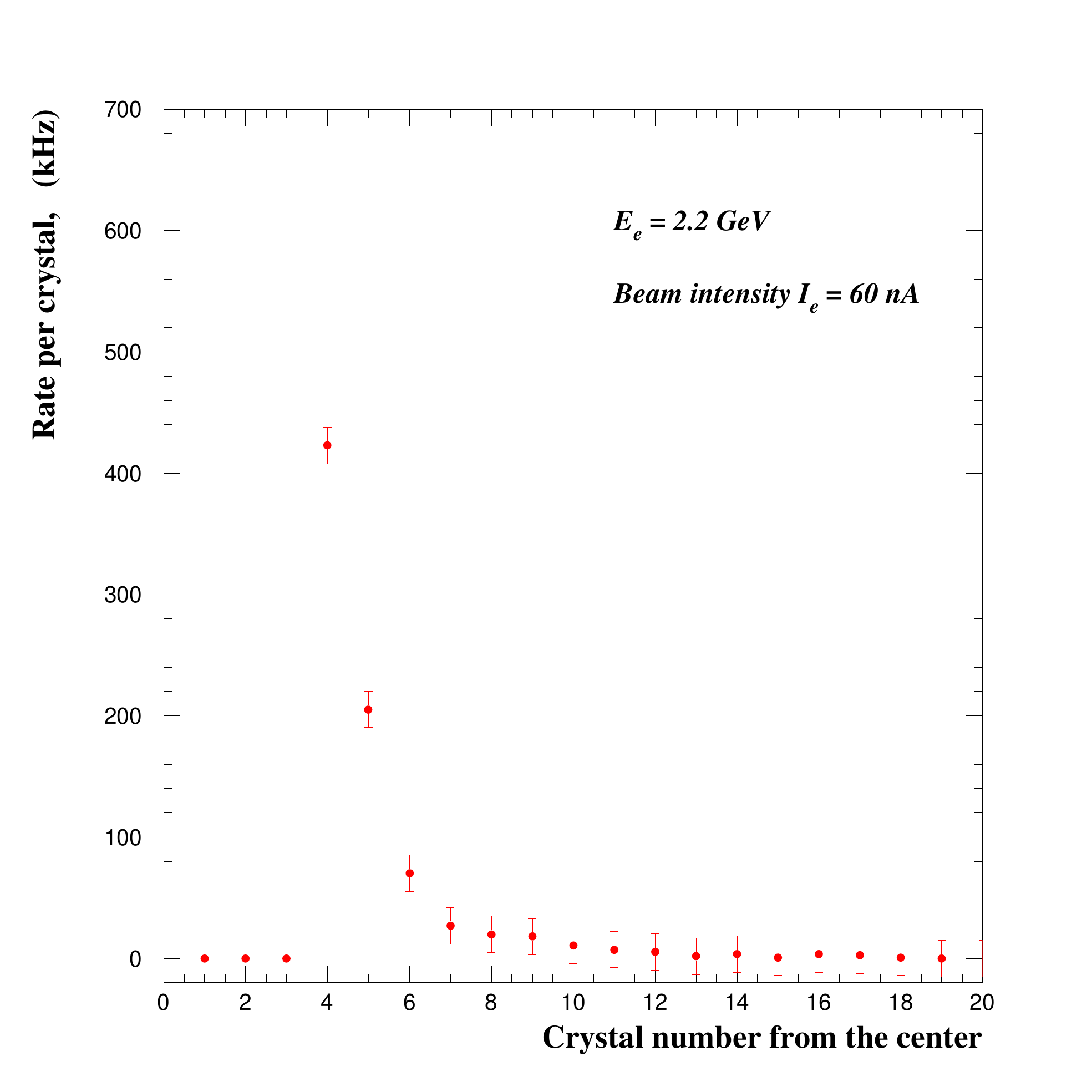}
    \includegraphics[height=0.4\textwidth]{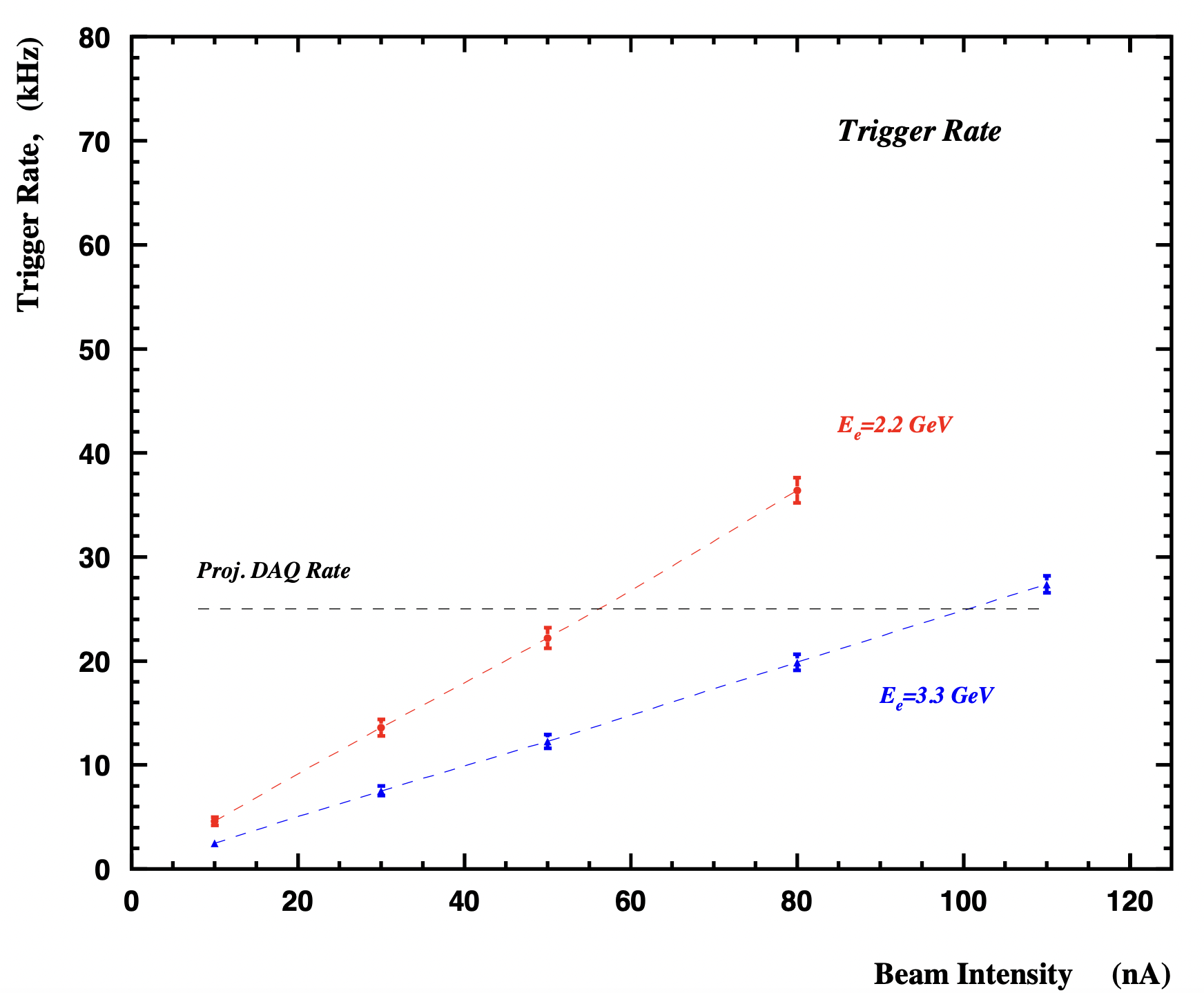}}
  \caption{(left) The rate per crystal for 60 nA of electron beam at 2.2 GeV beam energy. (right) The trigger rate {\it vs.} beam current.}
  \label{fig:daq_rate}
\end{figure}
The trigger rate is calculated for a 2.2 GeV and 3.3 GeV electron beam  incident on a 1~$\mu$m Ta target with the calorimeter placed 7.5 m from the target. We count starting from the third inner most layer of crystals (second innermost crystal after the layer covered with a W absorber),  with the condition that the total energy deposited in the \pbo{} part of the calorimeter is larger than 0.7 $\times$ E$_{beam}$. We count all three cluster events that have cluster energy within (0.02 - 0.85) $\times$ E$_{beam}$. Since the GEMs detectors are not used in the trigger, all clusters, charged or neutral, are counted. Charged events account for $\sim$ 1\% of all 3 cluster events. A cluster separation resolution of $\sqrt{2}$ crystals is assumed as demonstrated in previous experiments that have used the HyCal calorimeter.
The calculated rate per crystal for 60 nA of electron beam at 2.2 GeV beam energy is shown in  Fig.~\ref{fig:daq_rate} (left)  and the trigger rate as a function of the beam current is shown in Fig.~\ref{fig:daq_rate} (right). Given the projected maximum DAQ rate of 25 kHz, we have used 50 nA beam current at 2.2 GeV and 100 nA beam current at 3.3 GeV to calculate the beam time.
\begin{table}[hbt!]
\caption{The beam time request}
\begin{center}
\begin{tabular}{ | l | c |}
\hline
 & Time (days)\\
\hline
Setup checkout, tests and calibration & 4.0 \\
\hline
Production at 2.2~GeV @ 50 nA & 20.0 \\
\hline
Production at 3.3~GeV @ 100 nA & 30.0 \\
\hline
Energy change & 0.5 \\
\hline
No target background sampling at 2.2 \& 3.3 GeV & 5.5 \\
\hline

Total & 60.0 \\
\hline
\end{tabular}
\label{tab:beam_time}
\end{center}
\end{table}
As shown in table~\ref{tab:beam_time}, we request 20 days at 50 nA for the 2.2 GeV beam energy and 30 days at 100 nA at 3.3 GeV and total of 5.5 days for background running without a target. An additional 4.0 days are requested for calibration of the calorimeter and 0.5 days for energy change, for a total of 60 days of beam time.

\begin{figure}[hbt!]
    \centering
    \includegraphics[width=0.7\textwidth]{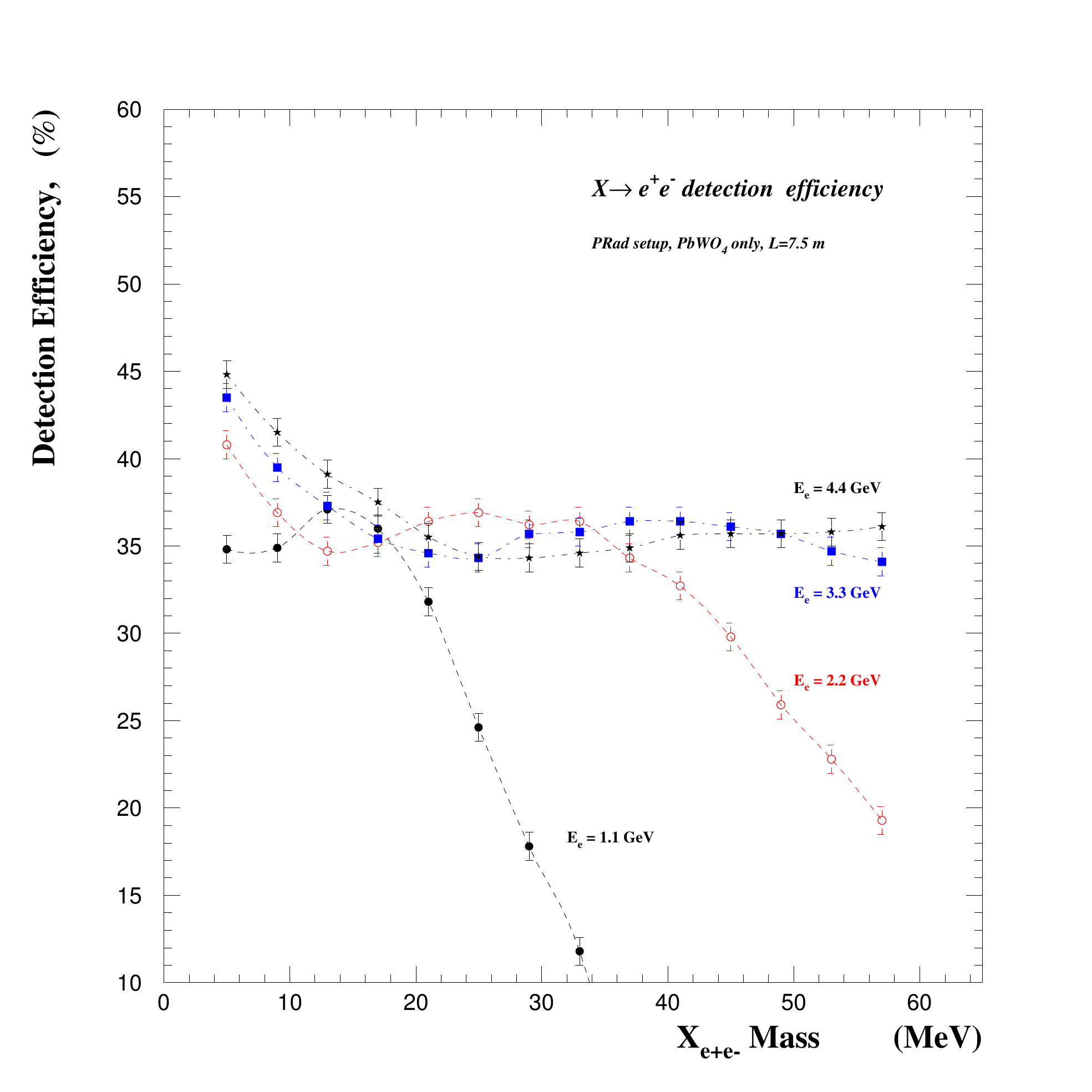}
    \caption{Detection efficiency of an $X\rightarrow e^+e^-$ decay for 1.1 GeV, 2.2 GeV, 3.3 GeV, and 4.4 GeV beam energies. 1.1 GeV was not chosen due to the rapid efficiency drop at higher masses.} 
    \label{fig:geom_acc}
\end{figure}

The use of two beam energies is critical to the success of this proposed search experiment.
The two energies were chosen as they have 
high detection efficiencies within the mass region of interest and the relative ease of scheduling during the CEBAF low-energy run periods. Other reasonable choices of beam energies are possible, as can be seen in Fig.~\ref{fig:geom_acc} where the detection efficiency is shown for four different beam energies. 
As this figure shows, this experiment (if approved) could run with 2.2 and 4.4 GeV beam energies for ease of scheduling between the different experiment Halls.

\subsection{Blinding the Data Analysis}
In order to avoid experimenter's bias during data analysis, we will adopt a suitable data blinding/un-blinding procedure for the analysis. One of the popular methods used is the hidden signal box method. In this approach, a predefined subset of the data containing the potential signal is excluded from the analysis until all aspects of the analysis are complete. The excluded region or ``signal box'' is defined in terms of any two experimental parameters, chosen to separate the signal from backgrounds. Once the event selection, background determination and signal efficiency is finalized the  hidden signal box is opened. 

Another option is the data prescaling method, where a small fraction of the data is used to tune the event selection and determine the background and efficiency. Once the analysis is finalized, it is applied without change to the full data set. We expect to adopt one of these methods for a blinded analysis of the data.

\section{Projected Results}
\label{sec:projected_results}
\begin{figure}[hbt!]
  \centerline{
  \includegraphics[width=0.5\textwidth]{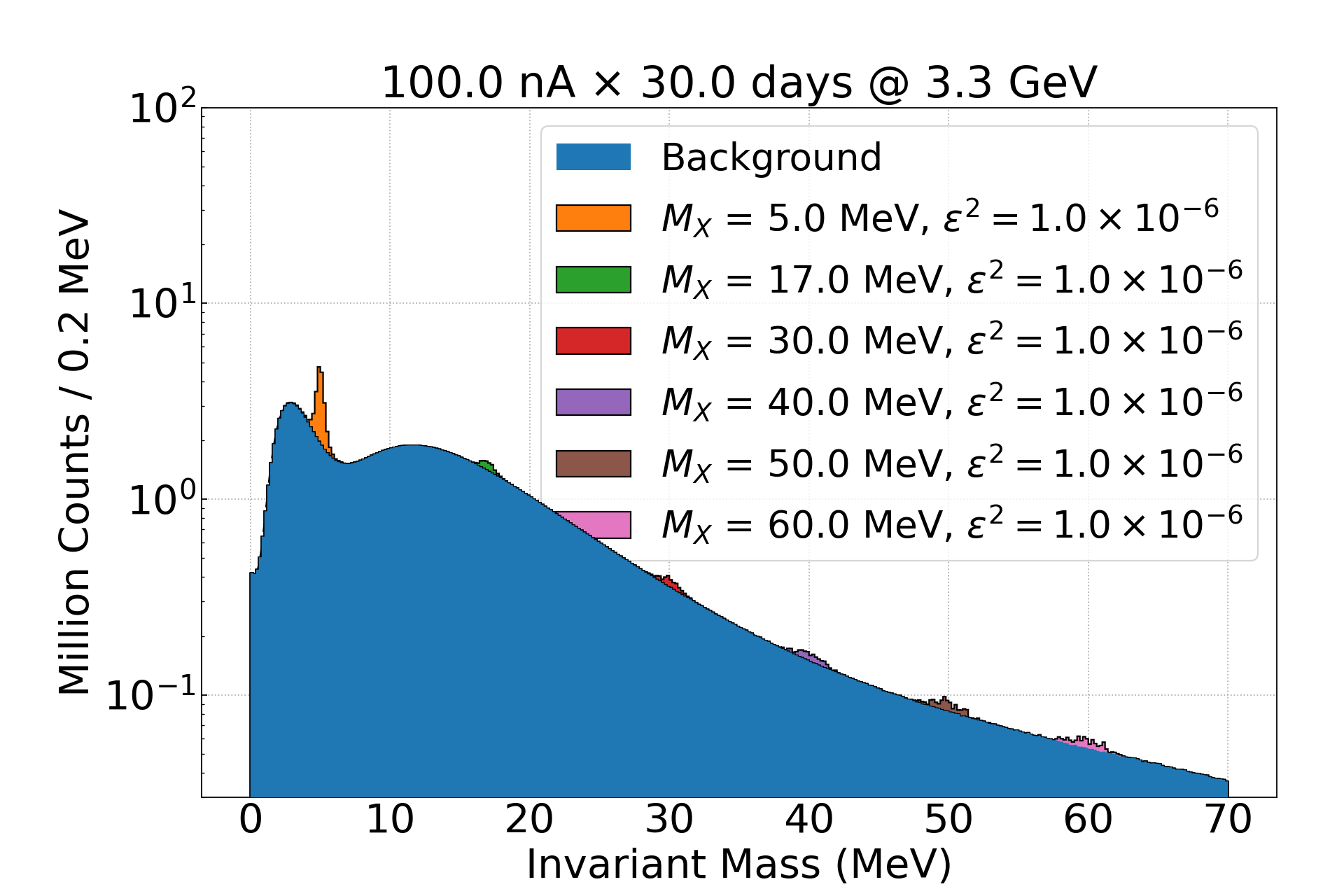}
  \includegraphics[width=0.5\textwidth]{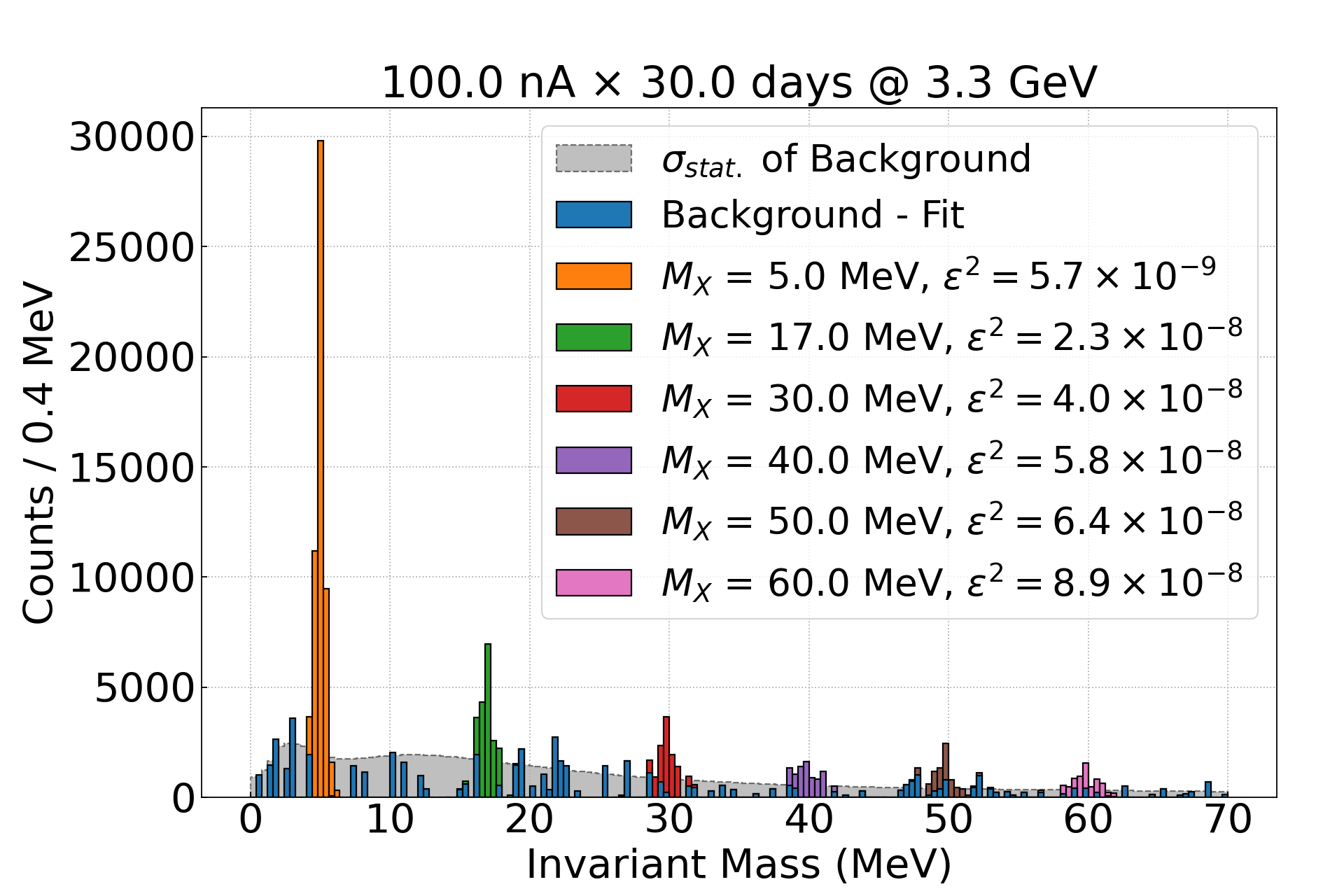}}
  \caption{Projections of 30 days of beam time with 3.3 GeV beam energy and 100 nA beam current. (left) The distribution of signals and background with various $m_X$ assuming $\epsilon^2 = 10^{-6}$ for illustration of the signal position and width. (right) The signal distribution over the background subtracted by fit, assuming $\epsilon^2$ at the threshold to reach a 5.0 significance within $\pm3\sigma$ of the signals.}
  \label{fig:3GeV_projection}
\end{figure}
\begin{figure}[b!]
  \centerline{
  \includegraphics[width=0.75\textwidth]{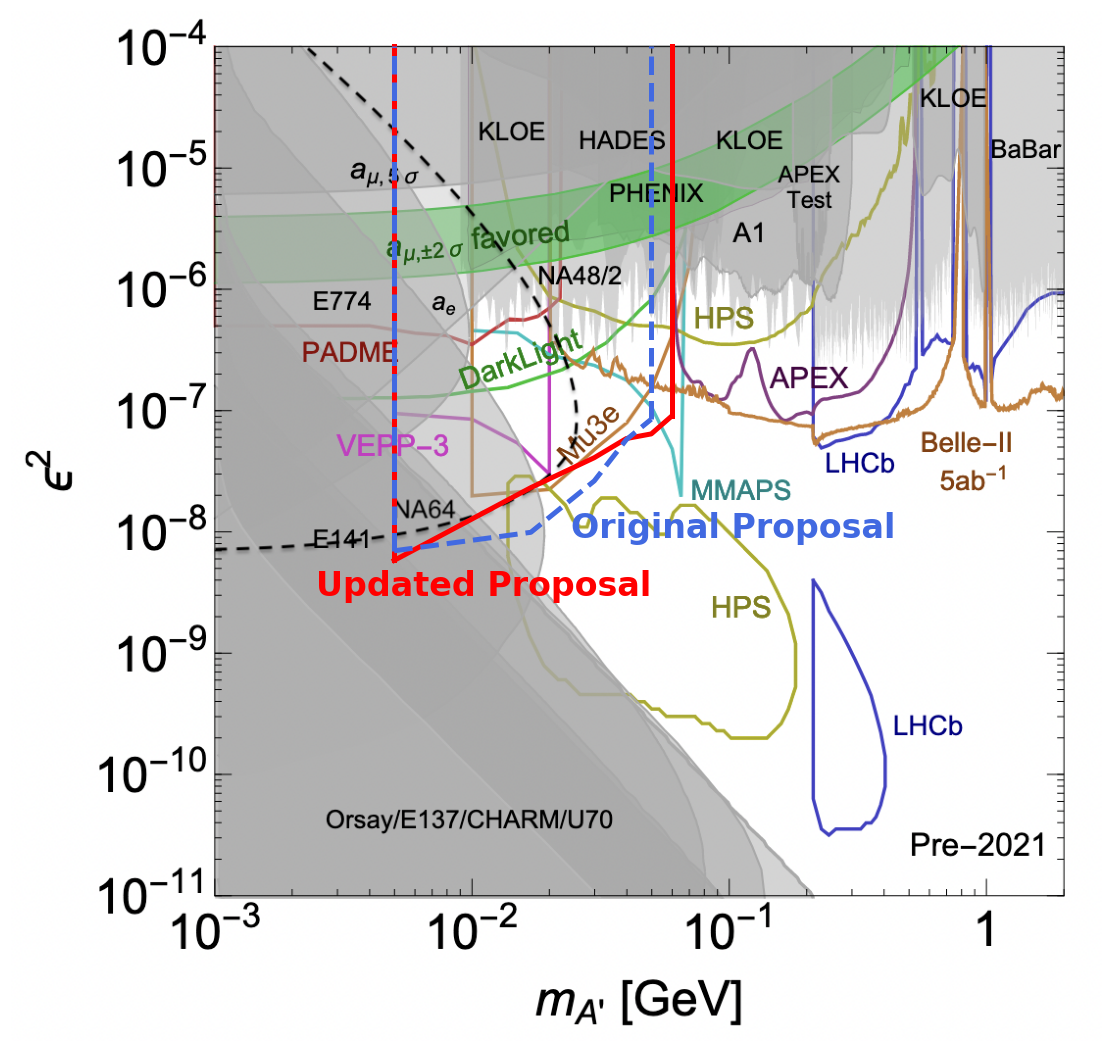}}
  \caption{Projected coverage of the square of the coupling constant ($\epsilon^2$) and mass ($m_{X}$) parameter space by this proposed experiment is shown as the thick red lines for the combined statistics of the two beam energies. The projections are superimposed on top of the constraints plot shown in Fig.~\ref{fig:cc} which was adapted from Ref.~\cite{Alexander:2016hsr}.}
  \label{fig:proj_ps}
\end{figure}

The projected sensitivity to 
$\epsilon^2$ is dominated by the 3.3 GeV run due to lower background level and better acceptance over the mass range of search. The projections for 30 days of 3.3 GeV beam at 100 nA are shown in Figure \ref{fig:3GeV_projection}. The production rates for the $X$ particle is calculated using the rate equation (Eq.~\ref{eq:rate}) described in Sec.~\ref{sec:method}. To obtain the sensitivity it is re-arranged as:
\begin{equation}
\label{eq:sensi}
\epsilon^2 = \frac{N_{X}}{{\mbox{5}} \times N_e T \frac{m_e^2}{m_X^2}},
\end{equation}
\noindent
where $N_X$ is the number of $X$ particles produced, $N_e$ is the number of incident electrons (1.62$\times$10$^{18}$ for 30 days at 100 nA), $T$ is the target thickness (2.5$\times$10$^{-4}$ r.l.), $m_X$ is the mass of the produced $X$, and $\epsilon^2$ is the square of the dimensionless coupling constant of the $X$ to SM matter. The decay length which leads to a displaced decay vertex was calculated as~\cite{Bjorken:2009mm,hps},
$$ l_{decay} = \gamma c\tau \approx \frac{3E_X}{\alpha m_{X}^2 \epsilon^2}. $$
The displaced decay vertex has an impact on the geometrical acceptance at the lowest mass and lowest $\epsilon^2$ ranges covered in this proposed experiment.
As designed this proposal does not have sufficient vertex resolution to measure the displaced decay vertex. However, a displaced vertex type experiment could be designed with additional tracking.
\begin{table}[hbt!]
\caption{Search Sensitivity}
\begin{center}
\begin{tabular}{ | c | c | c | c | c | c |}
\hline
 $m_X$ & $\sigma_{m_X}$ & Background & Signal Counts & Lowest  & Lowest\\
 MeV & MeV & Counts & (5.0 Significance) & $\epsilon^2$ & $\epsilon^2$ \\
\hline
\hline
\multicolumn{5}{| c |}{30 days of 3.3 GeV at 100 nA}& combined with signal \\
\multicolumn{5}{| c |}{}&from 20 days at 2.2 GeV\\
\hline
5.0 & 0.263 & 15.46M & 19.67k & 5.75E-09 & 4.98E-09 \\
\hline
17.0 & 0.467 & 19.58M & 22.13k & 2.29E-08 & 1.98E-08\\
\hline
30.0 & 0.692 & 7.42M & 13.63k & 4.04E-08 & 3.50E-08\\
\hline
40.0 & 0.938 & 4.24M & 10.30k & 5.82E-08 & 5.04E-08\\
\hline
50.0 & 1.009 & 2.53M & 7.96k & 6.35E-08 & 5.50E-08\\
\hline
60.0 & 1.154 & 1.87M & 6.85k & 8.90E-08 & 7.70E-08\\
\hline
\end{tabular}
\label{tab:sensitivity_table}
\end{center}
\end{table}


\noindent
Table~\ref{tab:sensitivity_table} shows the projected mass resolutions, background counts, signal, and $\epsilon^2$ sensitivity for sample signal masses $m_X=$~5,~17,~30,~40,~and~50~MeV.
The mass resolutions are taken from the simulation of signal events shown in Fig.~\ref{fig:full_res} (right).
The background event count is calculated by selecting events from the background simulation within a $\pm$~3$\sigma_{m_{X}}$ window.
The signal counts necessary for a 5$\sigma$ significance, given the background counts listed in Table~\ref{tab:sensitivity_table}, are calculated using the criteria for a 5$\sigma$ significance, i.e. the ``discovery'' criteria; 
\begin{equation}
\frac{N_{\mbox{signal}}}{\sqrt{N_{\mbox{signal}} + N_{\mbox{bgd}}}} \geq {\mbox{5}}.
\end{equation}
Finally, the lowest $\epsilon^2$ achievable is calculated using Eq.~\ref{eq:sensi} with $N_X$ obtained by scaling 
the signal counts with the detection efficiency and impact of the decay length. 

\noindent
The 2.2 GeV runs will serve as reference runs to better understand the background and the signal, so it was conservatively estimated that 50\% of their statistics can be combined into the 3.3 GeV data for the final results. Using these range of sensitivities the bounds for the $\epsilon^2 - m_{X}$ parameter space is plotting in Fig.~\ref{fig:proj_ps} for the combined projected statistics.


\section{Relationship to Other Experiments}
\label{sec:oexpt}

As described in Sec.~\ref{sec:motivation}, there are a number of motivators to search for hidden sector particles on the MeV scale.
As such, there are a number of other experiments with physics goals related to this proposal.
At Jefferson Lab, there are several experiments both approved and proposed to search for hidden sector particles: APEX, HPS, and DarkLight.
At other laboratories, there is NA64 and MAGIX.\\

\noindent
{\bf APEX}\\
The $A^\prime$ Experiment (APEX) experiment will run in Hall A at JLab.
The experiment uses a tungsten multi-foil target with an incident electron beam with a goal of producing a vector boson $A^\prime$ through a \breml{} process.
Using the two High-Resolution Spectrometers (HRSs), the $e^+e^-$ decay of the $A^\prime$ will be detected.
APEX will be sensitive to $A^\prime$ masses from 65 - 525 MeV with a coupling constant as small as $\epsilon^2>$~9$\times$10$^{-8}$~\cite{apex}.
The proposed experiment will complement the APEX experiment by searching lower masses and by being sensitive to neutral decay channels.\\

\noindent
{\bf HPS}\\
The Heavy Photon Search (HPS) is a search experiment that has began running and continues to run in Hall B at JLab.
HPS uses a tungsten target with an incident electron beam with a goal of producing a vector boson $A^\prime$ through a \breml{} process.
A magnetic spectrometer setup downstream of the CLAS detector setup will detect any charged $A^\prime$ decay products and reconstruct the displacement of the vertex.
HPS will be sensitive to $A^\prime$ masses from 20 - 1000 MeV with coupling constant as small as $\epsilon^2>$~10$^{-7}$ from the resonance search and 10$^{-8}\leq\epsilon^2\leq$~10$^{-10}$ from the displaced vertex search~\cite{hps}.
The proposed experiment will reach lower masses and will also have some overlap in the mass range (20 - 60 MeV).
In the regions of overlap, this proposed experiment will complement the HPS program by partially filling in the coupling constant gap between the resonance and displaced vertex search methods. 
Using a magnetic spectrometer free setup, this proposal will also be sensitive to neutral decay channels of hidden sector particles.\\

\noindent
{\bf DarkLight}\\
DarkLight is a set of experiments at JLab to search for MeV-scale dark photons.
An approved proposal from DarkLight focuses on a 100 MeV electron beam from the JLab energy recovery linac incident on a windowless hydrogen target.
The DarkLight setup uses a 0.5 Tesla solenoid and silicon and gas tracking detectors to detect the $e^+e^-$ decay of $A^\prime$ vector bosons in the 10 - 90 MeV mass range produced through a \breml{} process.
The DarkLight experiment will also be susceptible to the invisible decay of the produced $A^\prime$~\cite{darklight:2012pp,Balewski:2014pxa}.
A more recent DarkLight proposal focuses on a 45 MeV beam from the CEBAF injector impinging on a Tantalum target.
Two dipole spectrometers will measure the $e^+e^-$ decay of an $A^\prime$ produced through a \breml{} process.
The magnetic spectrometers will be set to focus on the 17 MeV region to quickly reach a $\epsilon^2>$~3$\times$10$^{-7}$ to test the X17 explanation of the ATOMKI anomaly discussed in Sec.~\ref{sec:atomki}~\cite{darklight:2020pp}.
This proposed experiment, in the regions of overlap (10 - 60 MeV), will reach lower coupling constants, $\epsilon^2$, and will be sensitive to the neutral decay of produced hidden sector particles in that mass range.
By using the existing PRad setup this proposal is cost-effective and essentially ready to run if approved.\\

\noindent
{\bf NA64}\\
The NA64 experiment is an active target beam dump experiment at the CERN Super Proton Synchotron.
NA64 searches for the $e^+e^-$ decay of an $A^\prime$ produced through a \breml{} process by 150 GeV electrons that strike WCAL, a calorimeter comprised of tungsten and plastic scintillator that also functions as an active target.
The most recent results exclude masses below about 25 MeV in the coupling constant range of 10$^{-8}\leq\epsilon^2\leq$~10$^{-6}$~\cite{Banerjee:2019hmi}.
An upgrade is planned that will close the coupling constant gap for the a mass of 17 MeV in an effort to test the X17 explanation of the ATOMKI anomaly~\cite{Bueno:2020dnx}.
While the exclusion from the 2020 NA64 results covers a large portion of the parameter space (3 - 60 MeV) that this proposed experiment will cover, there are some notably significant differences between the two approaches.
The NA64 experiment had 8.4$\times$10$^{10}$ electrons incident on a $\sim$ 30 - 40 (depending on the run) radiation length active target that served as a dump for recoil electrons and secondary particles produced by the beam.
This proposed experiment will have 1.62$\times$10$^{18}$ electrons incident on a $2.5\times10^{-4}$ radiation length target.
Using a thinner target will mitigate multiple scattering of the beam and the $X$ decay products.
Moreover, the use of a high resolution calorimeter and additional tracking detectors will allow us to  trigger on 3 cluster events. 
This allows for the detection of all three particles in the final state for the direct detection of the full kinematics of the $X$ production. 
The suppression of the Bethe-Heitler background in this proposal is another major difference from the NA64 experiment. 
The NA64 experiment and this proposal have very different systematics and should be considered as complementary searches. \\

\noindent
{\bf MAGIX}\\
MAGIX is an experiment proposed to run at the MESA accelerator at Mainz.
The experiment will use two magnetic spectrometers to study a varied physics program.
One of the programs is to detect the $e^+e^-$ decay of a dark photon $\gamma^\prime$ produced through a \breml{} process by 150 MeV electrons incident on a heavy nuclear supersonic gas jet target.
The projected reach covers a mass range of about 8 - 70 MeV and will reach $\epsilon^2>$~8$\times$10$^{-9}$ at low mass and $\epsilon^2>$~2$\times$10$^{-7}$ at high mass~\cite{Doria:2018sfx,Doria:2019sux}.
This proposal will cover a similar mass range as MAGIX (overlapping in 8 - 60 MeV) and reach smaller coupling constants, as well as being sensitive to neutral decay channels. Additionally, this proposal by using the existing PRad setup is essentially ready to run if approved.







\section{Summary}
We propose a direct detection search for hidden sector particles in the 3 - 60 MeV mass range using the magnetic-spectrometer-free PRad setup in Hall-B.
This experiment will exploit the well-demonstrated PRad setup to perform a ready-to-run and cost-effective search. This search experiment is timely as well as urgent given the recent confirmation of the muon $(g-2)$ anomaly and the 17-MeV particle proposed to account for the excess $e^+e^-$ pairs found in a nuclear transition in $^8$Be from one of its $1^+$ resonance to its ground state, and the electromagnetically forbidden M0 transition in $^4$He. The hidden sector particle mass range of 3 - 60 MeV remains relatively unexplored. 

The experiment will use 2.2 GeV and 3.3 GeV CW electron beams, with a current of $50 - 100$ nA, on a 1 $\mu$m Ta foil placed in front of the PRad setup. All three final state particles will be detected in the \pbo ~part of the HyCal calorimeter, and a pair of GEM chambers will be used to suppress the neutral background and the events not originating from the target. This technique will help effectively suppress the background from the radiative and Bethe-Heitler processes and provide a sensitivity of 7.2$\times$10$^{-8}$ - 5.9$\times$10$^{-9}$ to the kinetic mixing interaction coupling constant $\epsilon^2$. The experiment will help fill the void left by current, ongoing, and other planned searches, thereby helping to validate or place limits on hidden sector models. We request 60 days of beam time for this experiment.\\

\noindent
\appendix{\center{\bf \Large Appendix A: Data Mining the PRad Experiment}}\\

This proposal will use the existing PRad setup with some small modifications. 
As such, we began the preparations for this experiment by analyzing the existing PRad data. 
PRad is an experiment in Hall B at JLab that aims to precisely measure the proton radius in order to solve the proton radius puzzle.
The PRad experiment used a novel magnetic spectrometer-free design to measure $e-p$ elastic scattering at very low $Q^2$ and over a large $Q^2$ range simultaneously.
The data was recorded at 1.1 GeV and 2.2 GeV beam energies and could kinematically be able to produce low MeV $X$ candidates in the $\sim$ 3 - 50 MeV range.

\section{PRad Carbon Data}
\label{app:Carbon}

As described in Sec.~\ref{sec:method}, the \breml{} production of an $X$ scales with atomic number Z$^2$. 
The bulk of the data was taken with a windowless hydrogen gas target (Z = 1) which would have a very low $X$ production rate and has the added complication of being an extended target.
However, at each energy a short (1 hour) run was taken with a carbon foil target to better understand the associated systematics in the experiment~\cite{Mihovilovic:2016rkr}.
This led us to look at the carbon foil data since it would have a 36 times higher cross section and avoids the complications associated with an extended target.
As the production rate of an $X$ in the PRad data would be very low, this search was intended to test the experimental methods of this proposal rather than a true attempt to find an $X$. 

\begin{figure}[hbt!]
    \centering
    \includegraphics[width=0.75\textwidth]{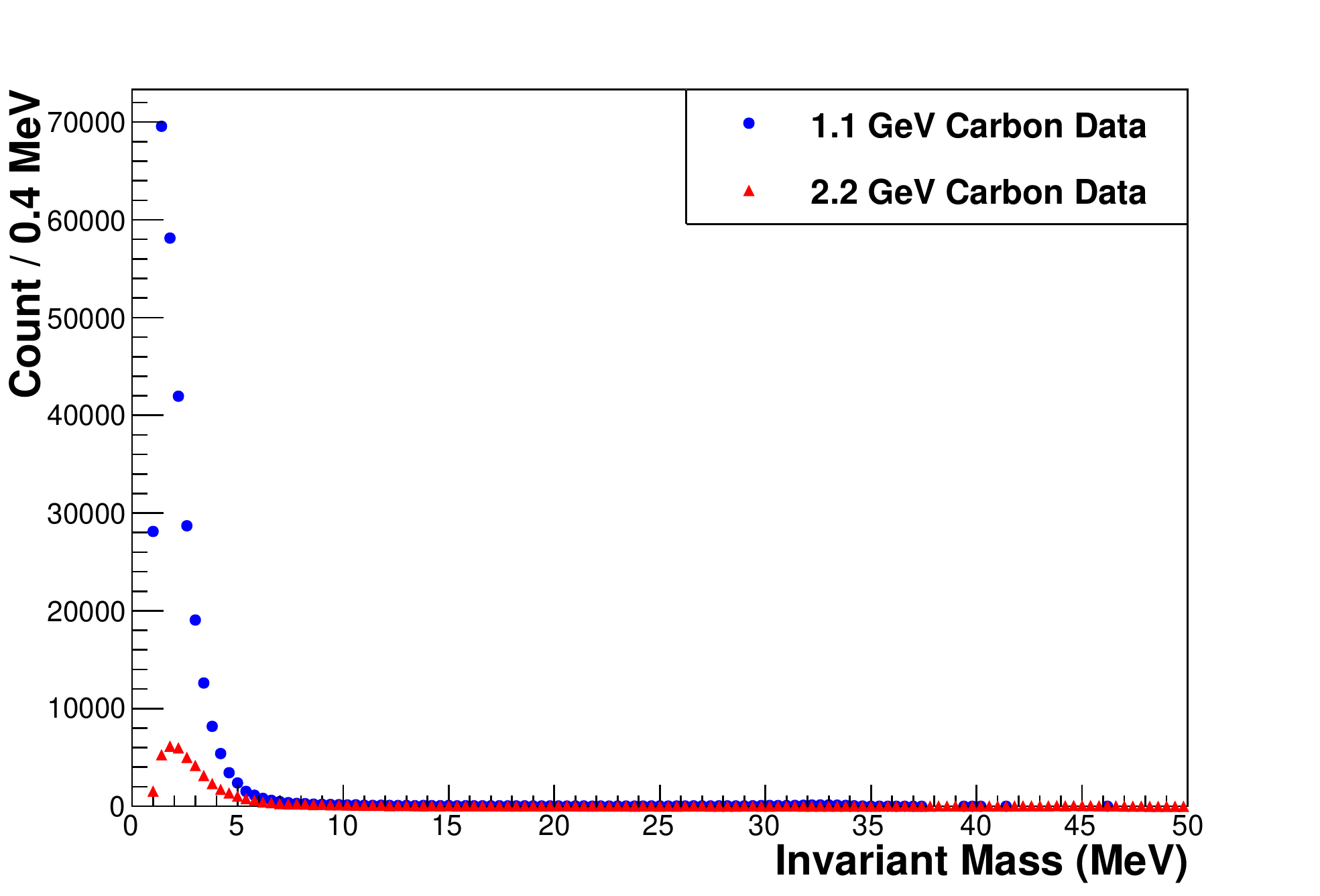}
    \caption{Invariant mass of events in PRad carbon data that pass initial event selection criteria. See the text for more information on these criteria.}
    \label{fig:IM_noEC}
\end{figure}

\begin{figure}[hbt!]
    \centering
    \includegraphics[width=0.75\textwidth]{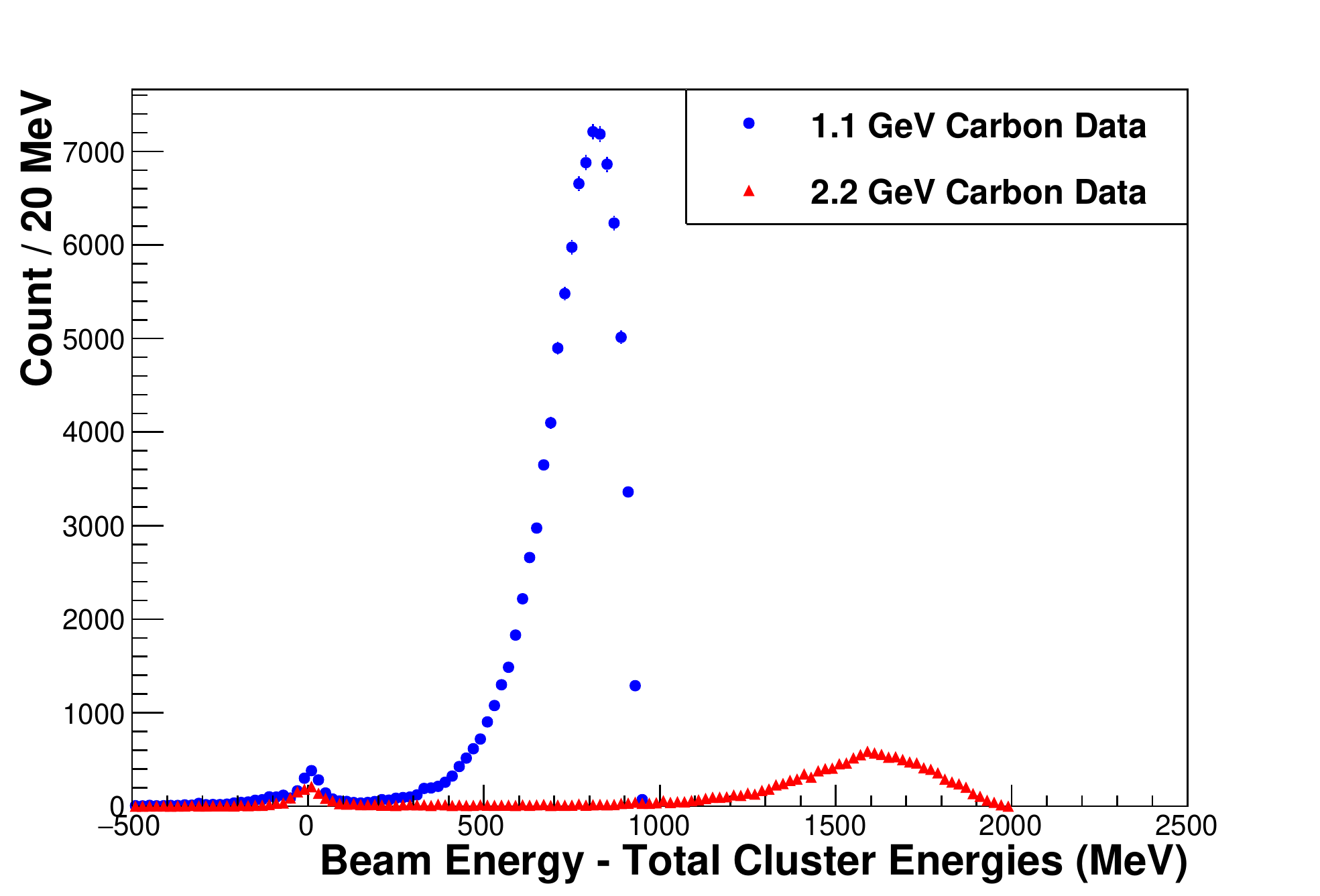}
    \caption{The difference between the beam energy and the sum of the 3 clusters in PRad carbon data that pass event selection. We can see around 0 that there is a clear peak of energy conserved events.}
    \label{fig:ediff}
\end{figure}

To begin, we selected events with exactly 3 clusters in the \pbo{} part of HyCal that have matching hits in the GEM tracker, thus explicitly looking for an $X\to e^+e^-$ event and excluding other decay channels.
Events which have extra clusters that do not match the event selection criteria are automatically vetoed.
This setup does not have the ability to distinguish which clusters are the recoil electron and which would be the $e^+e^-$ pair so all combinations are looped over, leading to three different pairings to reconstruct a potential $X$ mass.
Monte Carlo simulations have shown that this looping procedure will still yield a clear peak in the event of a true $X$ signal and that the incorrect pairings will create a smoothly distributed background.
Figure~\ref{fig:IM_noEC} shows the invariant mass and energy conservation of events that pass these selection criteria.
Figure~\ref{fig:ediff} shows the difference of the beam energy and the sum of the energy of the three cluster.
It is evident from these figures that the dominant event sources are inelastic.
The \breml{} production of an $X$ is an elastic process, so these can easily be cut.

\begin{figure}[hbt!]
    \centering
    \includegraphics[width=0.75\textwidth]{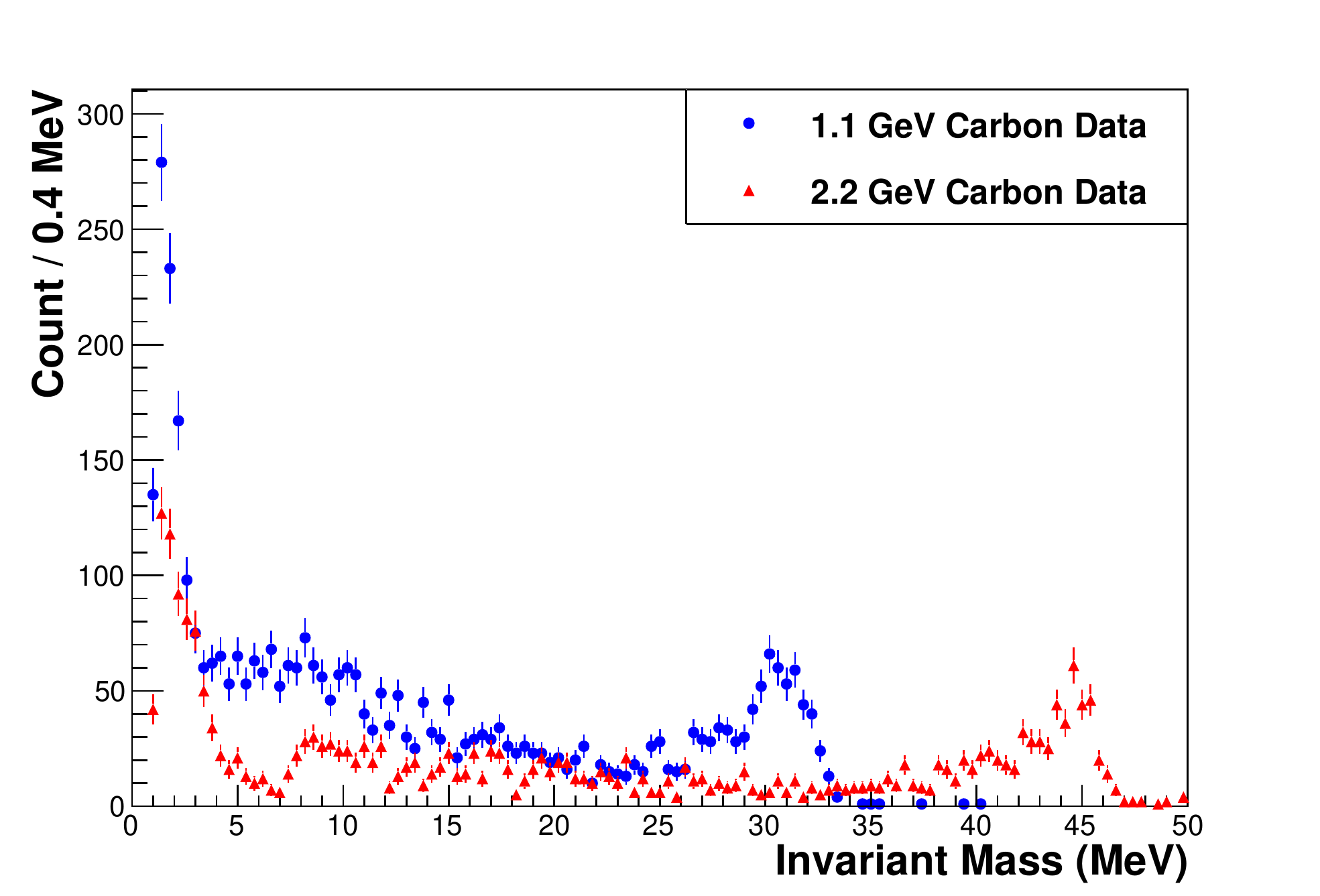}
    \caption{Invariant mass of events in PRad carbon data which pass the energy conservation cut. The peak at low invariant mass is electromagnetic background and the peak at high invariant mass is \moll{} scattering events with a low energy accidental.}
    \label{fig:IM_EC}
\end{figure}

A tight $\pm$ 50~MeV energy conservation cut is applied to these events.
With the application of this cut in Fig.~\ref{fig:IM_EC}, much of the background is removed and we see clear electromagnetic background at low invariant mass, a peak just below 10~MeV, and a peak at the high end of the invariant mass.
A Monte Carlo simulation shows that the peak at the high end of the spectrum is consistent with what would be expected from \moll{} scattering off of the target.
\moll{} scattering is a two-particle final state, so the third particle in this case is a low energy accidental.

\begin{figure}[hbt!]
    \centering
    \includegraphics[width=0.75\textwidth]{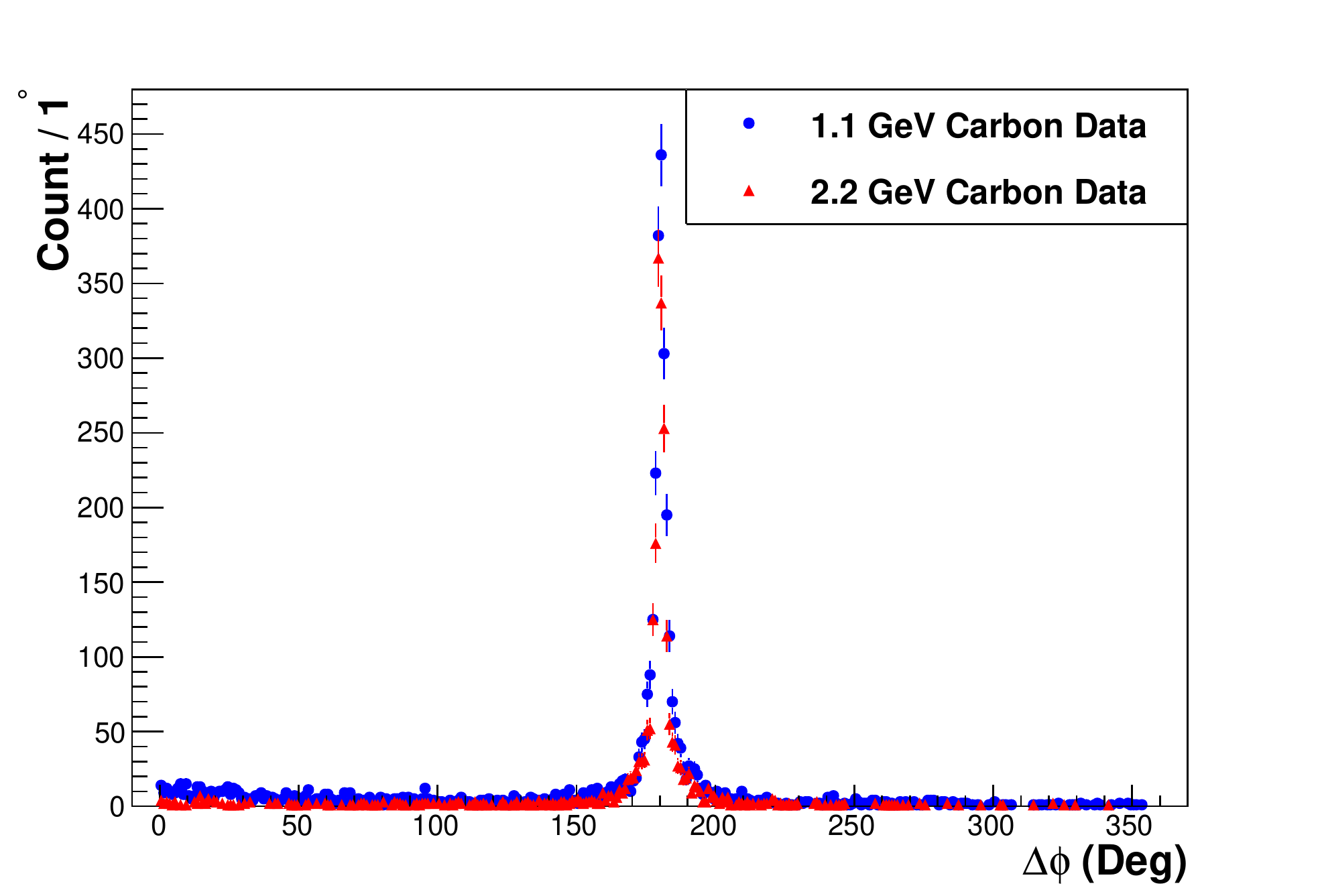}
    \caption{Coplanarity of the $X$-candidate and the recoil electron-candidate in PRad carbon data. The sum of the 4-momentum of the two clusters used to form the invariant mass and the 4-momentum of the third cluster should be coplanar if it truly the production of an $X$.}
    \label{fig:coplanar}
\end{figure}

A feature of these kinematics is that the reconstructed $X$ momentum and the recoil electron momentum are coplanar (the recoil nucleus does not receive enough energy at these kinematics to be noticeable in this test).
Figure~\ref{fig:coplanar} shows the $\Delta\phi$ distribution of the $X$-candidate and the recoil electron-candidate.
There is a clear peak of events that are coplanar (180$^\circ$) and a low background that are not.

\begin{figure}[hbt!]
    \centering
    \includegraphics[width=0.75\textwidth]{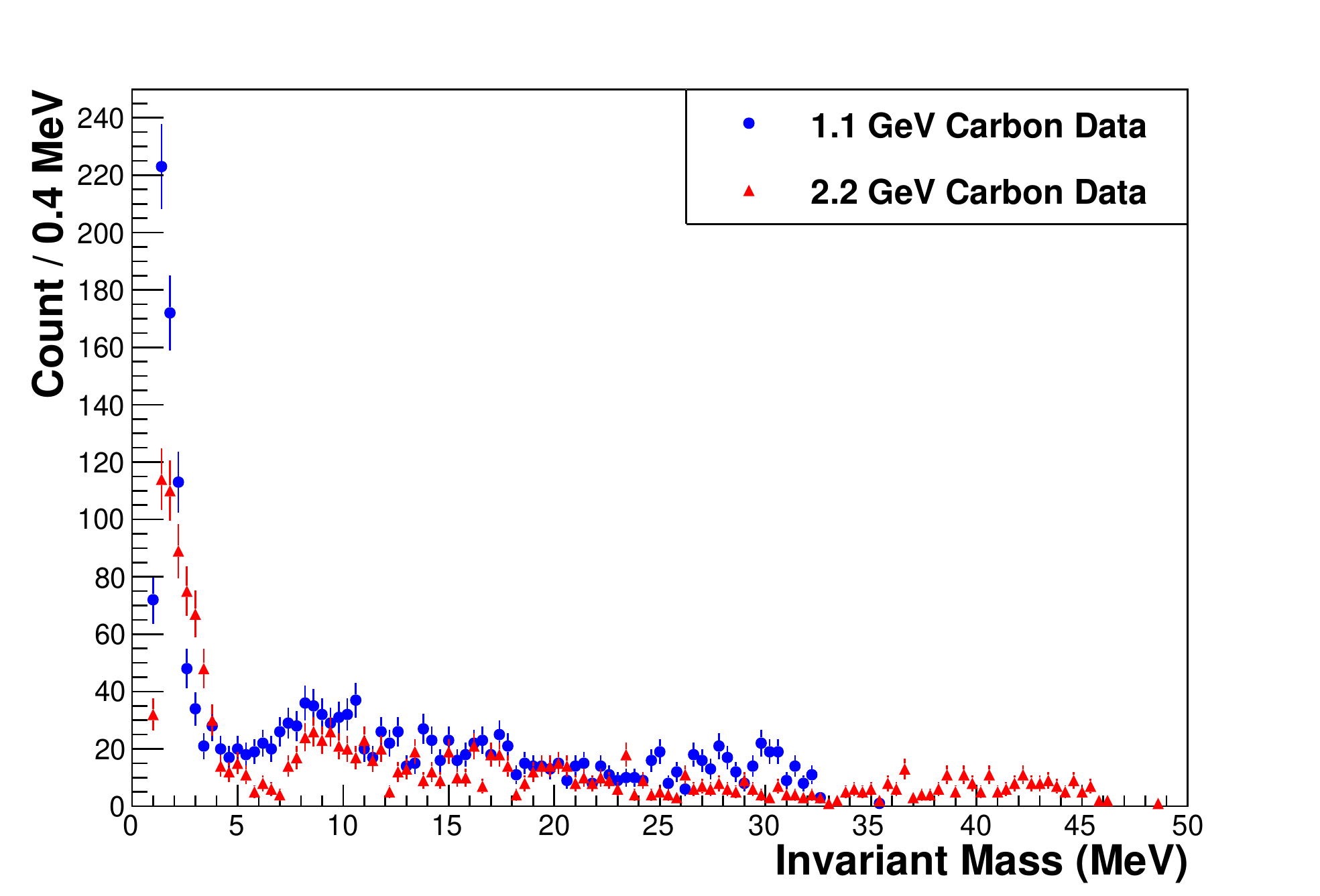}
    \caption{Invariant mass of events in PRad carbon data after applying a coplanarity cut. The \moll{} peak has been removed because the third cluster is an accidental that will, typically, not be coplanar with the clusters from the \moll{} electrons.}
    \label{fig:IM_coplanar}
\end{figure}

By placing a 180$^\circ\pm$5$^\circ$ cut on the $\Delta\phi$ distribution of these events we veto events that are simply 2-particle events that have a low-energy accidental third particle, such as \moll{} events.
In those cases, because the accidental is disconnected kinematically from the rest of the event, the $\Delta\phi$ spectrum will form a smooth background rather than a sharp peak.
We can see from Fig.~\ref{fig:IM_coplanar} that the application of this cut has completely removed the \moll{} invariant mass peak.

At this point, it is curious to note that the bump just below 10~MeV has survived all of these cuts.
However, when one looks closer it becomes evident that there is a slight shift in the peak placement between the 1.1~GeV and 2.2~GeV data.
This highlights the need for the use of multiple beam energies.
By incorporating more than one beam energy, events that do not originate from a particle decay will see a shift in their invariant mass.
This is critical to avoid misidentifying another physics process as evidence of new physics.

\begin{figure}[hbt!]
    \centering
    \includegraphics[width=0.75\textwidth]{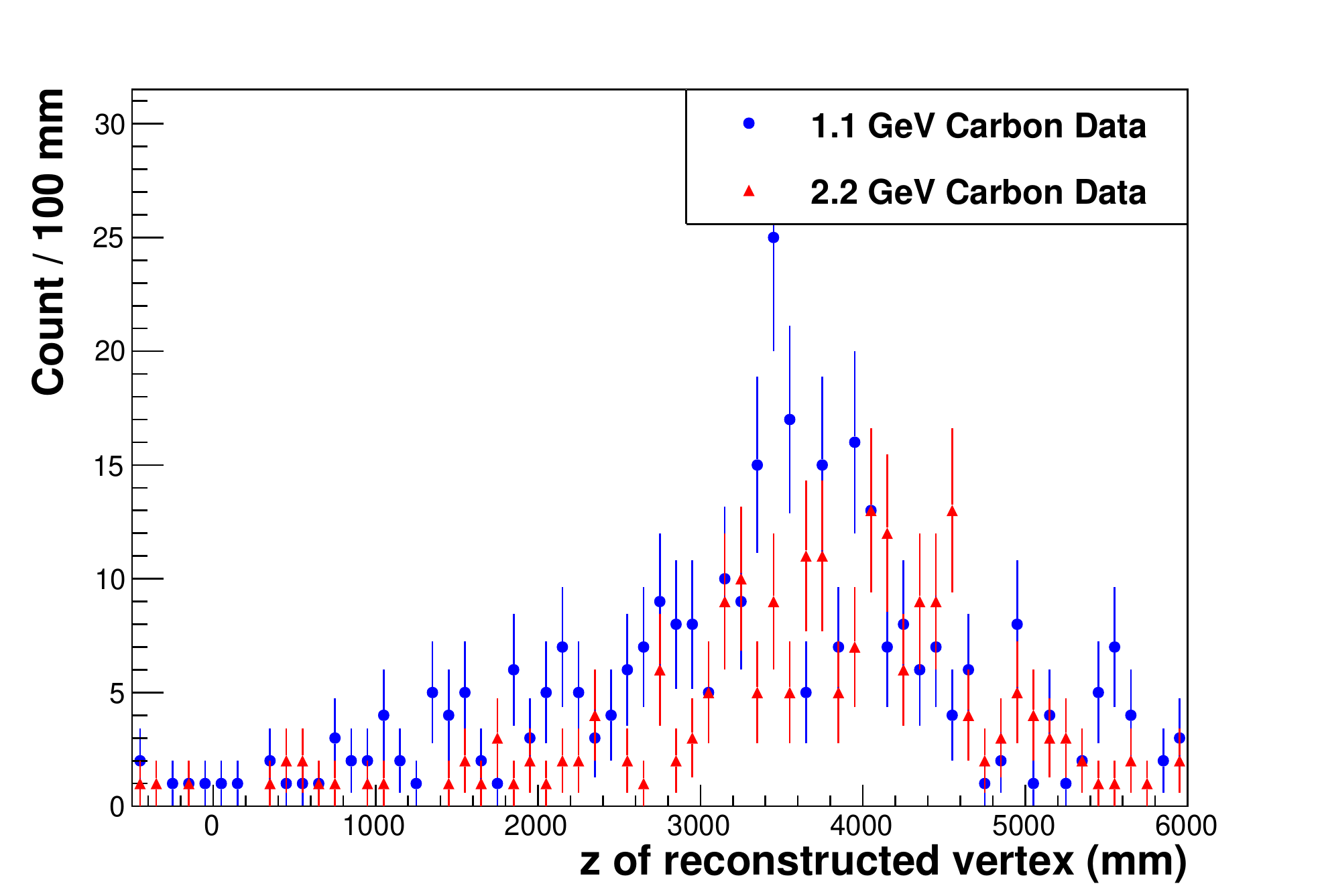}
    \caption{z-position of the reconstructed vertex of potential $X$ events from 6 - 12~MeV in PRad carbon data, where there appears to be a peak. If the events were produced in the target, we would see a build-up around 0.}
    \label{fig:z_vertex}
\end{figure}

Finally, we took a look at a very rough tracking of the events.
There is minimal distance between the GEM and HyCal which results in very poor resolution of the vertex of events.
However, we can do a very rough attempt at determine the z-vertex of these events.
Specifically, we look at the z position of closest approach of the two tracks that form the $X$ invariant mass.
Figure~\ref{fig:z_vertex} shows the reconstructed z position of the vertex of events in the bump described in the last paragraph, isolated by placing a cut on the invariant mass from 6 - 12~MeV.
Here, it can clearly be seen that these events do not originate from the target (z = 0).

This study shows that even without the optimizations we plan for the PRad experimental setup, the methodology outlined in this proposal is sound.
HyCal has excellent energy resolution for selection of elastic event and determining the invariant mass of events.
HyCal also has excellent position resolution for rejecting events that due not meet the coplanarity criteria.
We can also see that even with poor vertex reconstruction the PRad data can determine that events do not originate from the target.
The proposed experiment will have two GEM layers, separated by 10~cm.
These, paired with the third hit in HyCal will provide much improved vertex resolution, making it simpler to veto non-target events.\\

\section{PRad Hydrogen Data}
\label{app:Hydrogen}

As a follow-up, we have analyzed the PRad hydrogen data using this same method that was used for the carbon target. This is done as an additional check of the carbon results. 
As this uses the exact same setup, the final shape of the background ought to be the similar with both targets. 
The analysis procedure is identical, with the exception of an additional empty target subtraction step that is not necessary for the carbon target.

We use identical event selection criteria as in the analysis of the carbon target.
These criteria require exactly 3 clusters in the \pbo{} region of HyCal that meet the following conditions:
\begin{itemize}
    \item Have greater than 50 MeV and less than 10\% of the beam energy
    \item Have matching GEM hits (indicating a charged particle)
    \item Are not located in the two innermost rings of \pbo{} crystals
\end{itemize}

\begin{figure}[hbt!]
    \centering
    \includegraphics[width=0.75\textwidth]{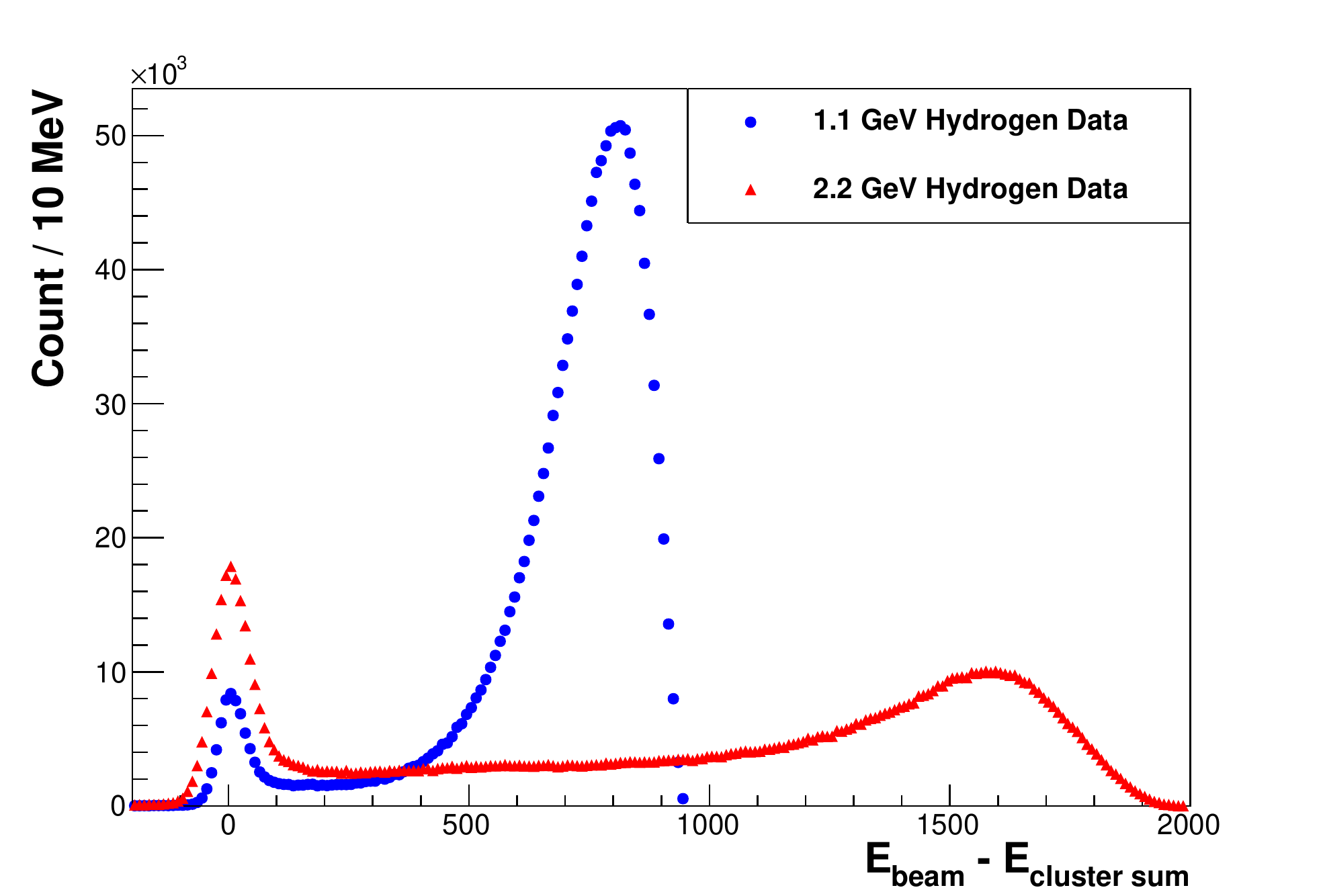}
    \caption{The difference between the beam energy and the energy of 3 clusters that meet event selection criteria in PRad hydrogen data. The peak around 0 indicates elastic events.}
    \label{fig:PRadH_EC}
\end{figure}

Events that meet these criteria are then assessed for energy conservation.
In Fig.~\ref{fig:PRadH_EC} we see, as in the carbon data, very clear separation of the elastic and inelastic peaks (though it should be noted that there is also visible inelastic background under the elastic peak that was not present in the carbon data).
Though this distribution is wider than that of the carbon data, we opt to keep the value of a $\pm$~50~MeV energy conservation cut.
This is both for consistency with the carbon analysis and to minimize the inelastic background contribution.

\begin{figure}[hbt!]
    \centering
    \includegraphics[width=0.75\textwidth]{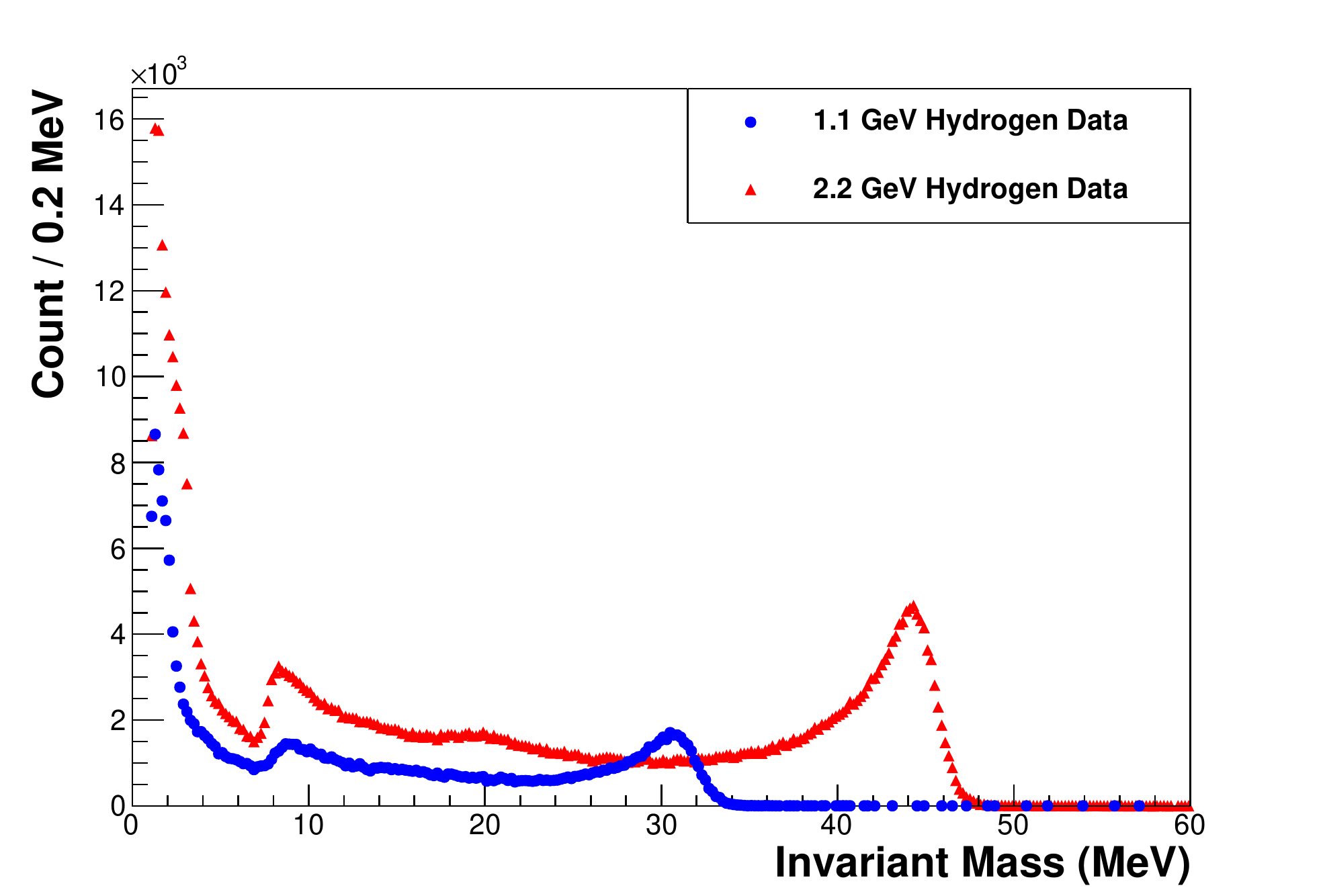}
    \caption{The invariant mass of selected events in PRad hydrogen data after a $\pm$~50~MeV energy conservation cut.}
    \label{fig:PRadH_ECIM}
\end{figure}

In Fig.~\ref{fig:PRadH_ECIM}, we see the invariant mass distributions after the energy conservation cut.
Apparent here is the same peak as carbon just below 10~MeV and the \moll{} peaks at the high energy end of the distribution.

\begin{figure}[hbt!]
    \centering
    \includegraphics[width=0.75\textwidth]{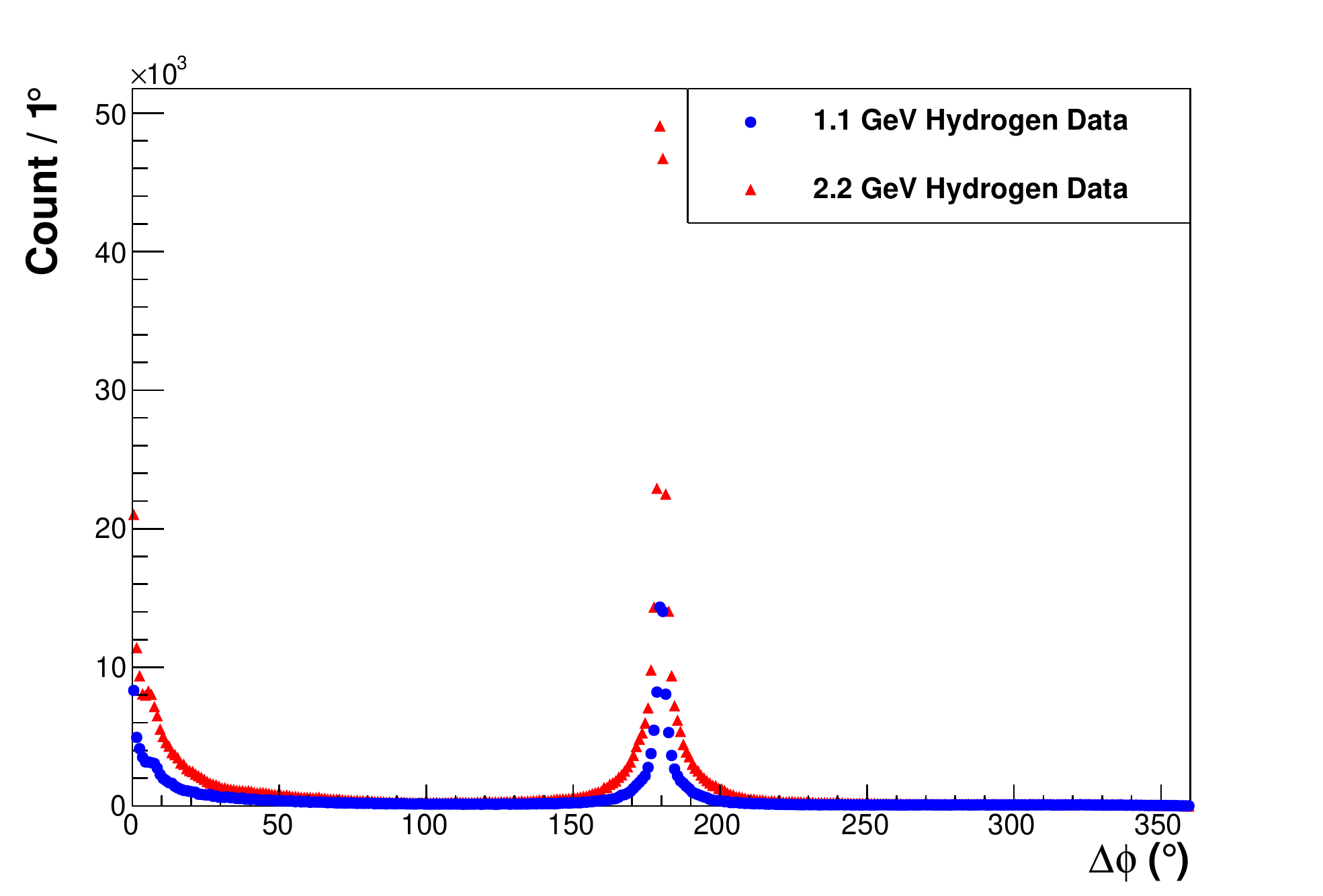}
    \caption{Distribution of $\Delta\phi$ between $X$ and $e^\prime$ candidates in PRad hydrogen data. The peak around 180$^\circ$ indicates coplanar events.}
    \label{fig:PRadH_dphi}
\end{figure}

\begin{figure}[hbt!]
    \centering
    \includegraphics[width=0.75\textwidth]{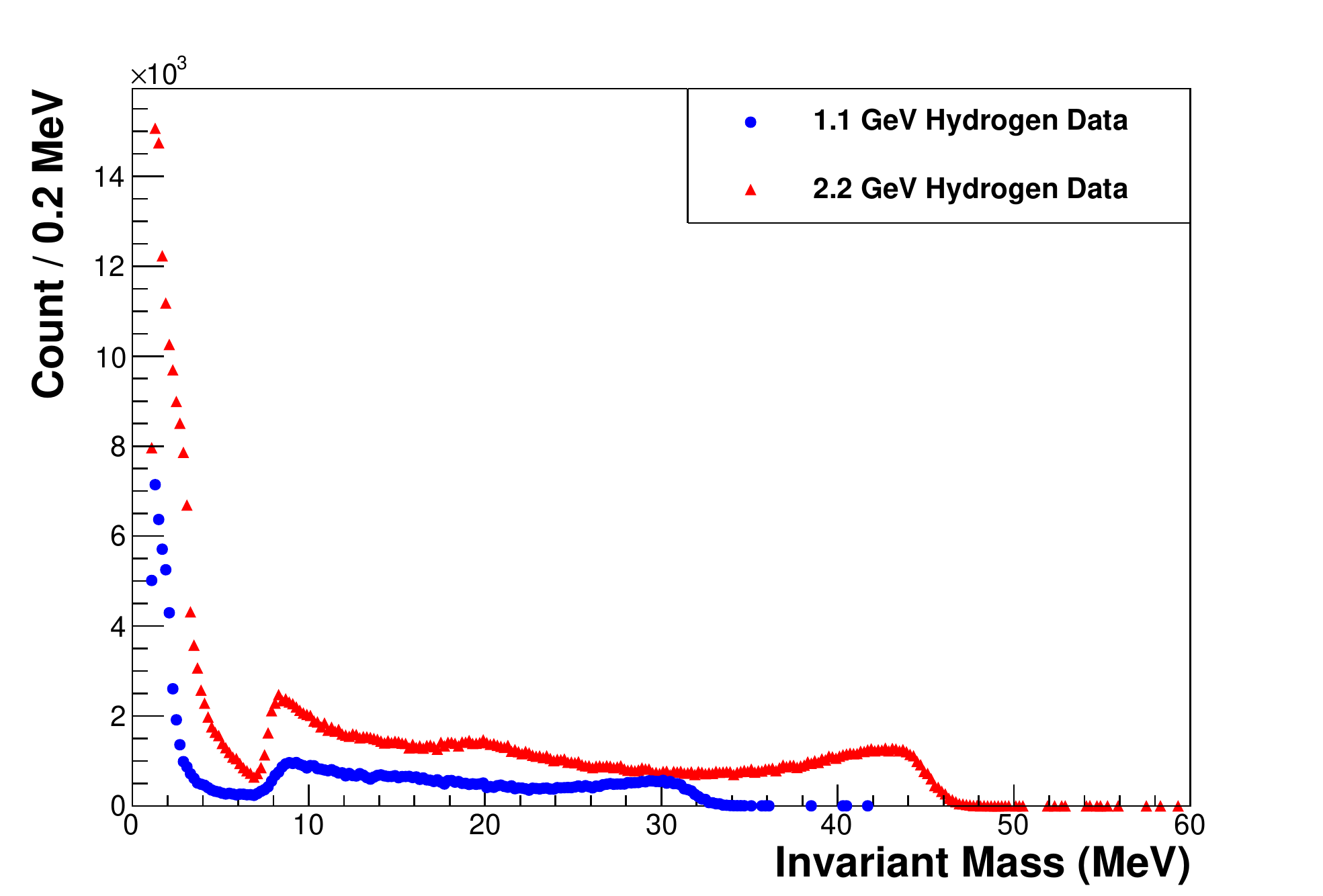}
    \caption{The invariant mass of selected events in PRad hydrogen data after the $\pm$~50~MeV energy conservation cut and 180$\pm$5$^\circ$ coplanarity cut.}
    \label{fig:PRadH_dphiIM}
\end{figure}

To clean up \moll{} scattering and accidentals, we next look at the coplanarity of the events.
Fig.~\ref{fig:PRadH_dphi} shows the distribution of $\Delta\phi$ between the $X$ and $e^\prime$ candidates.
We then place a 180$\pm$5$^\circ$ cut to isolate coplanar events, giving the invariant mass distribution shown in Fig.~\ref{fig:PRadH_dphiIM}.
This cut has slightly reduced the baseline of the background and greatly diminished the \moll{} peak.

The gaseous hydrogen target is contained in a C101 copper ``windowless'' target cell. 
This cell has a hole in each end to prevent the electron beam from interacting with the cell. 
However, there still exists the possibility of the beam interacting with the hydrogen gas that flows out of the cell and the beam halo interacting with the copper cell.
The PRad experiment recorded several runs with a low flow of hydrogen gas into the chamber surrounding the target cell to simulate the gas that exits the cell under normal operating conditions. 
In the beamline of Hall B are ``beam charge monitors'' (BCMs), which record the amount of charge that flows past them, which is exactly related to the number of electrons on target. 
When scaled up to match the total beam charge of the filled targets, subtracting the empty target data can effectively remove backgrounds from the beam interacting with non-target materials~\cite{Pierce:2021vkh}. By nature of being a gas target held in a metal cell, this additional step is added to the analysis to subtract the empty target data. 

\begin{figure}[hbt!]
    \centering
    \includegraphics[width=0.75\textwidth]{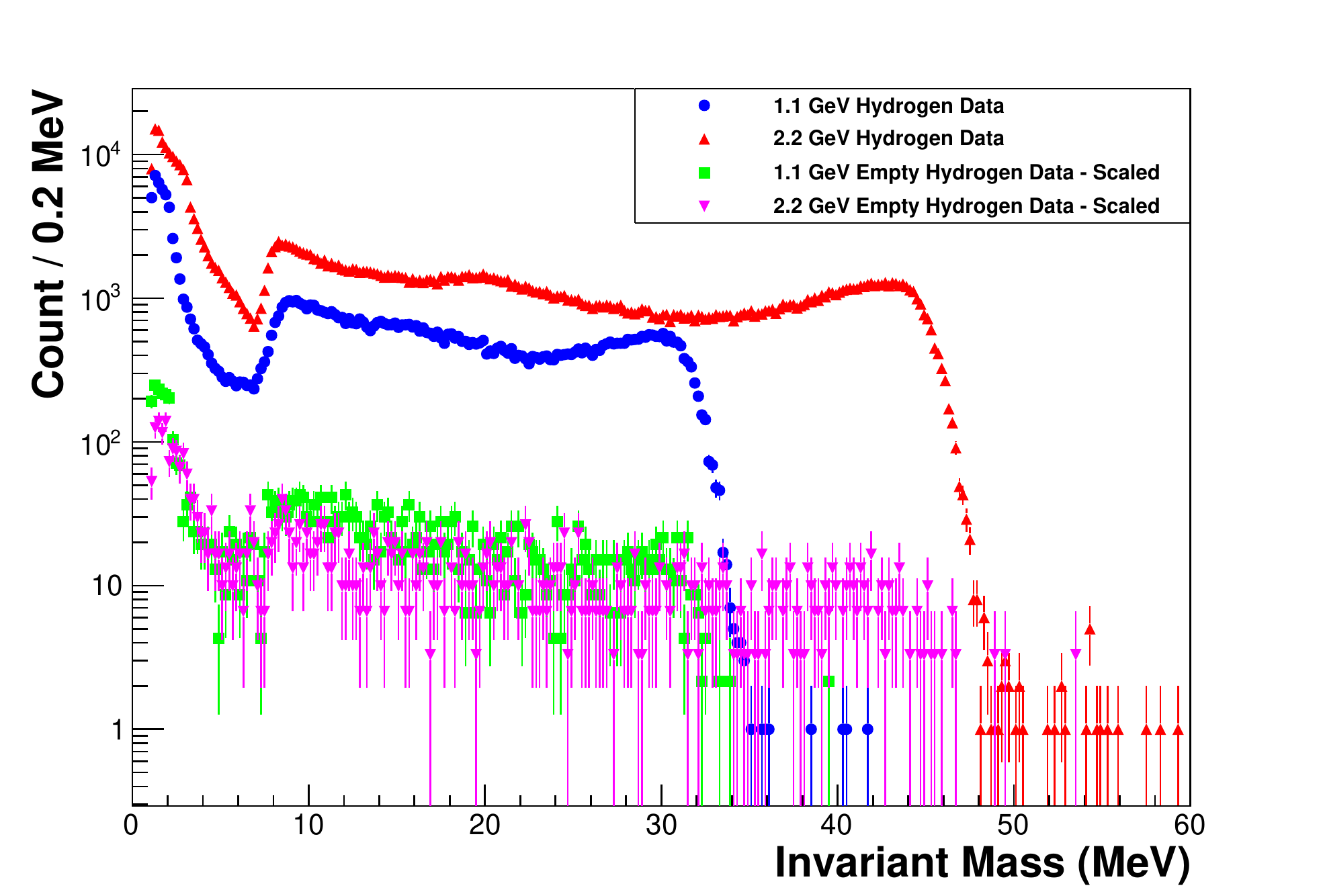}
    \caption{A comparison of the invariant mass distributions of the full PRad hydrogen target and the empty PRad hydrogen target. Note that the y-axis is set to ``log'' scale in order to discern the structure in the empty target.}
    \label{fig:PRadH_MTcomp}
\end{figure}

\begin{figure}[hbt!]
    \centering
    \includegraphics[width=0.75\textwidth]{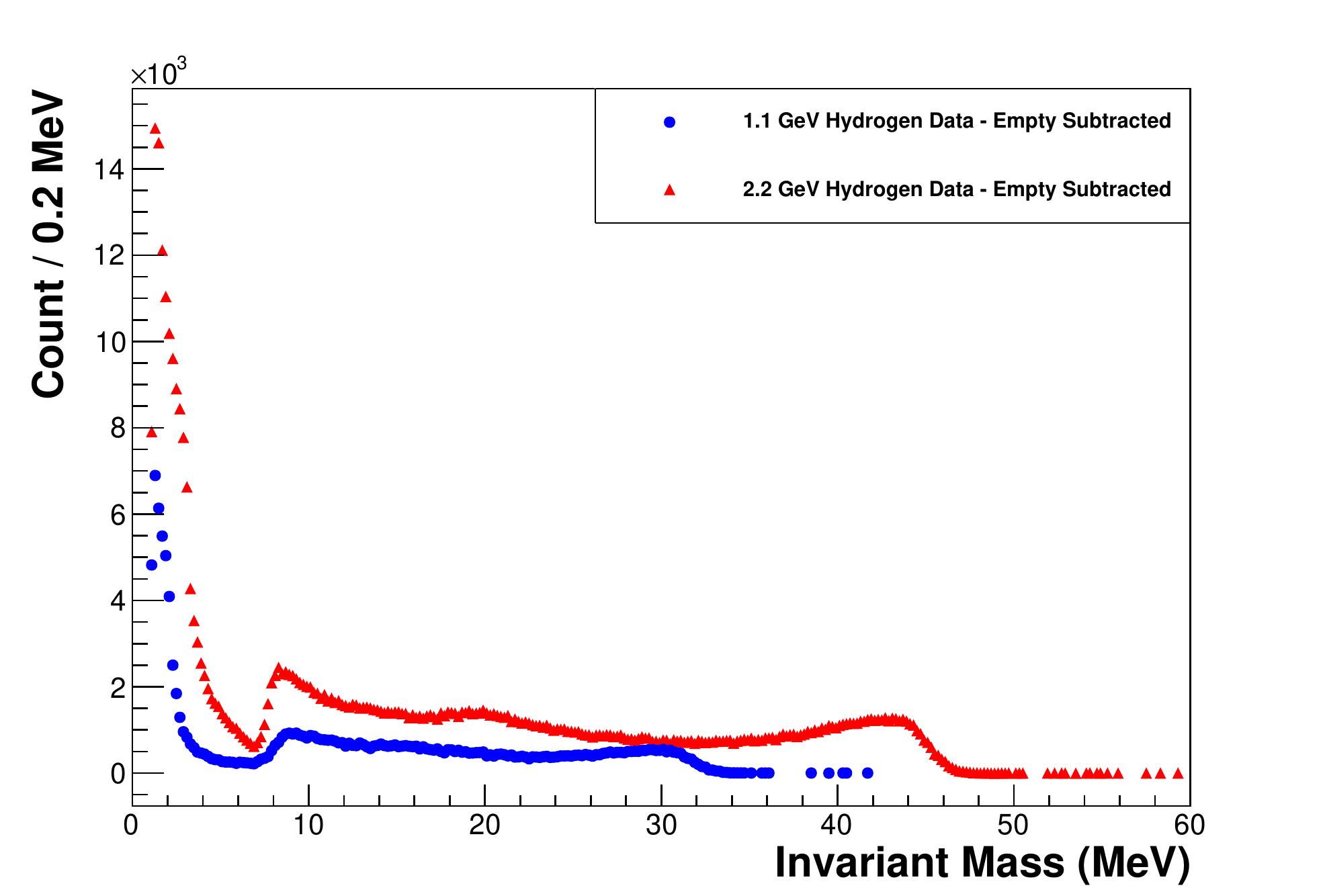}
    \caption{The invariant mass of the selected events in the PRad hydrogen target after subtracting the background assessed with the empty target.}
    \label{fig:PRadH_MTsub}
\end{figure}

Fig.~\ref{fig:PRadH_MTcomp} shows a comparison of the full targets to the empty after they have been scaled by total beam charge.
It should be noted that the figure has the y-axis set to ``log'' scale in order for any structure in the empty target distributions to be visible as they are approximately an order of magnitude lower in counts than the full targets. 
Fig.~\ref{fig:PRadH_MTsub} shows the result of subtracting the empty target.
As the empty target background is incredibly small, it is worth noting that it is nearly identical to Fig.~\ref{fig:PRadH_dphiIM}.

With the exception of the obvious difference in statistics (orders of magnitude more hydrogen data than carbon), it is notable that Fig.~\ref{fig:PRadH_MTsub} and Fig.~\ref{fig:IM_coplanar} have the same overall structure.

\newpage
\bibliographystyle{unsrt}
\bibliography{references}
\end{document}